\documentclass[twocolumn]{aastex631}
\usepackage{graphicx}
\usepackage{xfrac}
\usepackage{mathrsfs}
\usepackage{soul}

\newcommand\CI{[C\,\textsc{i}]}
\newcommand\CII{[C\,\textsc{ii}]}

\newcommand{\kms}{{\hbox {\,km\thinspace s$^{-1}$}}}

\newcommand{\hto}{{\hbox {H$_{2}$O}}}

\label{key}

\newcommand{\spt}{SPT2349$-$56}

\received{XXX}
\revised{YYY}
\accepted{ZZZ}
\submitjournal{ApJ}

\shorttitle{A radio-loud AGN in SPT2349}
\shortauthors{Chapman et al.}
\graphicspath{{./}{figures/}}

\begin{document}

\title{Brightest Cluster Galaxy Formation in the $z$=4.3 Protocluster SPT\,2349-56: Discovery of a Radio-Loud AGN.}

\correspondingauthor{Scott C. Chapman}
\email{scott.chapman@dal.ca}

\author[0000-0002-8487-3153]{Scott C.\ Chapman}
\affiliation{Department of Physics and Atmospheric Science, Dalhousie University, Halifax, NS, B3H 4R2, Canada}
\affiliation{NRC Herzberg Astronomy and Astrophysics, 5071 West Saanich Rd, Victoria, BC, V9E 2E7, Canada}
\affiliation{Department of Physics and Astronomy,  University of British Columbia, Vancouver, BC, V6T1Z1, Canada}
\affiliation{Eureka Scientific Inc, Oakland, CA 94602, USA}
\author{Ryley Hill}
\affiliation{Department of Physics and Astronomy,  University of British Columbia, Vancouver, BC, V6T1Z1, Canada}
\author[0000-0002-6290-3198]{Manuel Aravena}
\affiliation{N\'ucleo de Astronomia, Facultad de Ingenieria y Ciencias, Universidad Diego Portales, Av. Ej\'ercito 441, Santiago, Chile}
\author[0000-0002-0517-9842]{Melanie Archipley}
\affiliation{Department of Astronomy, University of Illinois, 1002 West Green St., Urbana, IL 61801, USA}
\affiliation{Center for AstroPhysical Surveys, National Center for Supercomputing Applications, 1205 West Clark Street, Urbana, IL 61801, USA}
\author[0000-0002-9181-9948]{Arif Babul}
\affiliation{Department of Physics and Astronomy, University of Victoria, Victoria, BC V8P 1A1, Canada}
\author{James Burgoyne}
\affiliation{Department of Physics and Astronomy,  University of British Columbia, Vancouver, BC,  Canada}
\author{Rebecca E. A. Canning}
\affiliation{Institute of Cosmology and Gravitation, University of Portsmouth, Dennis Sciama Building, Portsmouth, PO1 3FX, UK}
\author{Carlos De Breuck}
\affiliation{European Southern Observatory, Karl Schwarzschild Strasse 2, 85748 Garching, Germany}
\author[0000-0002-0933-8601]{Anthony H.\ Gonzalez}
\affiliation{Department of Astronomy, University of Florida, 211 Bryant Space Science Center, Gainesville, FL 32611-2055, USA}
\author[0000-0003-4073-3236]{Christopher C. Hayward}
\affiliation{Center for Computational Astrophysics, Flatiron Institute, 162 Fifth Avenue, New York, NY 10010, USA}
\author{Seon Woo Kim}
\affiliation{Department of Astronomy, University of Illinois, 1002 West Green St., Urbana, IL 61801, USA}
\author{Matt Malkan}
\affiliation{Department of Physics and Astronomy, University of California, Los Angeles, CA 90095, USA}
\author{Dan P.\ Marrone}
\affiliation{Steward Observatory, University of Arizona, 933 North Cherry Avenue, Tucson, AZ 85721, USA}
\author{Vincent McIntyre}
\affiliation{International Center for Radio Astronomy Research, Curtin University, GPO Box U1987, 6102 Perth, Australia}
\author[0000-0001-7089-7325]{Eric Murphy}
\affiliation{NRAO National Radio Astronomy Observatory 520 Edgemont Road Charlottesville, VA 22903}
\author{Emily Pass}
\affiliation{Harvard-Smithsonian Center for Astrophysics, 60 Garden Street, Cambridge, MA 02138, USA}
\affiliation{Department of Physics and Atmospheric Science, Dalhousie University, Halifax, NS, B3H 4R2, Canada}
\author{Ryan W.\ Perry}
\affiliation{Department of Physics and Atmospheric Science, Dalhousie University, Halifax, NS, B3H 4R2, Canada}
\author[0000-0001-7946-557X]{Kedar A. Phadke}
\affiliation{Department of Astronomy, University of Illinois, 1002 West Green St., Urbana, IL 61801, USA}
\affiliation{Center for AstroPhysical Surveys, National Center for Supercomputing Applications, 1205 West Clark Street,
Urbana, IL 61801, USA}
\author{Douglas Rennehan}
\affiliation{Center for Computational Astrophysics, Flatiron Institute, 162 Fifth Avenue, New York, NY 10010, USA}
\author{Cassie Reuter}
\affiliation{Department of Astronomy, University of Illinois, 1002 West Green St., Urbana, IL 61801, USA}
\author[0000-0002-9181-9948]{Kaja M.\ Rotermund}
\affiliation{Lawrence Berkeley National Laboratory, Berkeley, CA 94720, USA}
\author{Douglas Scott}
\affiliation{Department of Physics and Astronomy,  University of British Columbia, Vancouver, BC, V6T1Z1, Canada}
\author[0000-0002-9181-9948]{Nick Seymour}
\affiliation{International Center for Radio Astronomy Research, Curtin University, GPO Box U1987, 6102 Perth, Australia}
\author{Manuel Solimano}
\affiliation{N\'ucleo de Astronomia, Facultad de Ingenieria y Ciencias, Universidad Diego Portales, Av. Ej\'ercito 441, Santiago, Chile}
\author[0000-0003-3256-5615]{Justin Spilker}
\affiliation{Department of Physics and Astronomy and George P. and Cynthia Woods Mitchell Institute for Fundamental Physics and Astronomy, Texas A\&M University, 4242 TAMU, College Station, TX 77843-4242, US}
\author[0000-0002-2718-9996]{Anthony A.\ Stark}
\affiliation{Harvard-Smithsonian Center for Astrophysics, 60 Garden Street, Cambridge, MA 02138, USA}
\author[0000-0002-9181-9948]{Nikolaus Sulzenauer}
\affiliation{Max-Planck-Institut für Radioastronomie, Auf dem Hugel 69, Bonn, D-53121, Germany}
\author[0000-0002-9931-5162]{Nick Tothill}
\affiliation{School of Science, Western Sydney University, Locked Bag 1797, Penrith NSW 2751, Australia}
\author[0000-0001-7192-3871]{Joaquin D. Vieira}
\affiliation{Department of Astronomy, University of Illinois, 1002 West Green St., Urbana, IL 61801, USA}
\affiliation{Center for AstroPhysical Surveys, National Center for Supercomputing Applications, 1205 West Clark Street,
Urbana, IL 61801, USA}
\author{David Vizgan}
\affiliation{Department of Astronomy, University of Illinois, 1002 West Green St., Urbana, IL 61801, USA}
\author{George Wang}
\affiliation{Department of Physics and Astronomy,  University of British Columbia, Vancouver, BC, V6T1Z1, Canada}
\author[0000-0003-4678-3939]{Axel Weiss}
\affiliation{Max-Planck-Institut für Radioastronomie, Auf dem Hugel 69, Bonn, D-53121, Germany}

\begin{abstract}
We have observed the $z\,{=}\,$4.3 protocluster \spt\ with the Australia Telescope Compact Array (ATCA)  with the aim of detecting radio-loud 
active galactic nuclei (AGN) amongst the ${\sim}\,$30 submillimeter (submm) galaxies (SMGs) identified in the structure. 
We detect the central complex of submm sources 
at 2.2\,GHz 
with a luminosity of 
$L_{2.2}\,{=}\,(4.42\pm$0.56)$\,{\times}\,10^{25}$\,W\,Hz$^{-1}$. 
The Australian Square Kilometre Array Pathfinder (ASKAP) 
also detects the source at 888\,MHz, constraining the radio spectral index to $\alpha\,{=}\,-$1.6$\pm$0.3, consistent with ATCA non-detections at 5.5 and 9\,GHz, and implying 
$L_{1.4,\,{\rm rest}}\,{=}\,(2.4\pm$0.3)$\,{\times}\,10^{26}$\,W\,Hz$^{-1}$. 
This radio luminosity is about 100 times higher than expected from star formation, assuming the usual far-infrared (FIR)-radio correlation, which is a clear indication of an AGN driven by 
a forming brightest cluster galaxy (BCG). None of the SMGs in \spt\ show signs of AGN in any other diagnostics available to us (notably $^{12}$CO out to $J\,{=}\,$16, OH$_{\rm 163\mu m}$, \CII/IR, and optical spectra), 
highlighting the radio continuum as a powerful probe of obscured AGN in high-$z$ 
protoclusters.
No other significant radio detections are found amongst the cluster members, with stacking on either all members or just the ten most luminous members yielding non-detections consistent with the FIR-radio correlation for star-forming galaxies.
We compare these results to field samples of radio sources and SMGs, along with the 22 SPT-SMG gravitational lenses also observed in the ATCA program, as well as powerful radio galaxies at high redshifts.
Our results allow us to better understand the effects of this gas-rich, overdense environment on early supermassive black hole (SMBH) growth and cluster feedback. 
We estimate that $(3.3\pm0.7)\,{\times}\,10^{38}$\,W of power are injected into the growing intra-cluster medium (ICM) by the radio-loud AGN, whose energy over 100\,Myr is comparable to the binding energy of the gas mass of the central halo. The AGN power is also comparable to the instantaneous energy injection from supernova feedback from the 23 catalogued SMGs in the core region of 120\,kpc projected radius. The \spt\ radio-loud AGN may be providing strong feedback on a nascent ICM. 
\end{abstract}



\section{Introduction}
\label{sec:intro}

Submillimeter (submm) galaxies (SMGs) are important sites of stellar mass build-up at cosmic noon and earlier (e.g.,\ \citealt{Chapman03N,Chapman05,smail2004}), 
with star-formation rates (SFRs) as high as hundreds to thousands of solar masses per year. 
Several studies have also suggested that SMGs 
may be good tracers of dark matter halos at early cosmic time \citep[e.g.,][]{blain04, Chen16, Dud20}. 
Simulations conducted by \citet{Miller15} found that while many dark matter halos at $z\,{=}\,$2--4 do not contain any SMGs, large and rare associations of five or more SMGs do trace massive overdensities of dark matter that have the potential of evolving into present-day massive clusters. Supporting this, in the recent past,  several high-redshift protoclusters have been identified entirely through their submm emission 
\citep[e.g.,][]{Chapman09,Daddi09a,Capak11,Casey15,Miller,Oteo18,GomezGuijarro19,wang2021}. 

AGN and star-formation processes in galaxy evolution are clearly related \citep[e.g.][]{kormendy2013}. 
Enhanced AGN activity relative to the field environment has been found in massive protoclusters at $z\,{=}\,$2--3 \citep[e.g.,][]{pentericci2002,lehmer2009,digby-north2010}, 
which is likely related to the enhancement of star formation in galaxy protocluster members \citep[e.g.][]{elbaz2007,Chapman09,brodwin013,Casey15,gilli2019}. 
%
The suppression of star formation in galaxy clusters requires mechanical and radiative feedback, which is naturally generated by AGN.
Extended X-ray emission has been detected in clusters, showing empty regions or cavities in the hot gas \citep[e.g.][]{fabian2012}, which can naturally be explained as shocked gas from the feedback.
The Clusters Around Radio-Loud AGN (CARLA) survey of around 400 high-redshift radio galaxies (HzRGs) 
from $z\,{=}\,$1--3 \citep{wyz2013} showed that in the majority of cases, the radio AGN is located near the center of the galaxy overdensity as traced by their stellar mass ({\it Spitzer}-IRAC emission). This is strong evidence that  radio galaxy feedback in a growing brighest cluster galaxy (BCG) is 
important for the evolution of massive galaxy clusters.

Galaxy overdensities in the high redshift Universe have likely not yet virialized (e.g., \citealt{overzier2013}). They have abundant reservoirs of cold gas to supply star formation, while the ongoing mergers between galaxies expected in the dense environments provide triggers for star formation.
Mergers can also provide the tidal torques necessary for the gas to overcome its angular momentum and fall to accretion disk of the supermassive black hole (SMBH). AGN require this nuclear accretion as a power source.
This is in contrast to low redshifts, where structures are virialized, and both AGN and star formation are largely suppressed in cluster galaxies
\citep{Ehlert14,Rasmussen12,vanbreukelen2009,Martini06,Kauffmann04}.
Studies of AGN in protoclusters has recently become a viable endeavor, with relatively deep {\it Chandra} and XMM-{\it Newton} observations at $z=1.5$--$3$ (e.g., \citealt{digby-north2010,wang2013,Travascio2020}). Continuing to study the rich variety of protoclusters and extending these studies to earlier times 
can inform how host galaxies are affected by their SMBHs, as well as the connection to the surrounding environment. In the $z\,{=}\,3.09$ SSA22 protocluster, 50\% of the SMGs were found to host X-ray luminous AGN \citep{Umehata19} 
-- a clear excess over the 15\% found for field SMGs \citep[e.g.][]{wang2013}.
At larger distances, an overdensity of ten SMGs found by the {\it Hershel Space Telescope} at $z\,{=}\,4.0$ \citep{Oteo18} has been studied by {\it Chandra} in the X-ray \citep{vito2020} 
and  in the radio \citep{Oteo18}, revealing no significant excess of AGN activity in the system over field SMGs (22\% versus 15\%, respectively).

The 2,500\,deg$^2$ survey conducted by the South Pole Telescope (SPT -- \citealt{Vieira10,Everett20})  at 3.0\,mm, 2.0\,mm and 1.4\,mm has uncovered a small population of nine millimeter sources ranging from $z\,{=}\,$3--7, 
which are extremely luminous, yet apparently not gravitationally lensed \citep[e.g.][]{Spilker,Reuter20,wang2021}.
A well characterized example of this is \spt, a  protocluster system at $z\,{=}\,$4.303 \citep{Miller}.
Observations at $870\,\mu$m using the Large APEX BOlometer CAmera (LABOCA; \citealt{LABOCA}) on the Atacama Pathfinder Experiment (APEX; \citealt{APEX}) telescope (with a 19-arcsec beam size) first revealed an extended structure with two distinct lobes connected by a bridge with a combined flux density of $S_{870\,\mu\rm{m}}\,{=}\,(106\pm8)$\,mJy \citep{Miller,wang2021}.
Follow-up observations with the Atacama Large Millimeter-submillimeter Array \citep[ALMA;][]{ALMA} measured the redshift of its brightest central source through $^{12}$CO lines \citep{Strandet},  
and then resolved the structure into over 30 submm-luminous sources \citep{Miller,Hill20,Rotermund21}, with a velocity dispersion suggesting a central halo mass of around 10$^{13}$\,M$_\odot$. 
A VLT/MUSE observation reveals the presence of a Ly$\alpha$ blob (LAB), with a linear size of about 60\,kpc, close to the core of \spt\ (Y.\ Apostolovski et al. in prep.). 
None of the other protocluster SMGs were detected as Ly$\alpha$ emitters (LAEs) in the MUSE data. 
Ly$\alpha$ halos are commonplace in most HzRG protoclusters 
\citep{venemans2007}.
Similar objects are often found in protoclusters identified through other means, for example optical galaxy overdensities \citep{Overzier},  and indicate the presence of significant amounts of neutral gas in the assembling cluster.

\begin{table}
 \centering{
  \caption{ATCA-selected sources within a radius of 1\,Mpc in projection (140$^{\prime\prime}$) of \spt.}
\label{table:atca_obs}
\begin{tabular}{lcccccc}
\hline
ID & RA & Dec & Freq. & Flux   \\ 
{} & {} & {} & (GHz) & ($\mu$Jy) \\
\hline
ID1 & {--} & {--} & 8.98 & $<$ 159$^{\dagger\dagger}$ \\
ID1 & {--} & {--} &5.47   & $<$ 120$^{\dagger\dagger}$  \\
ID1 & 23:49:42.760 & $-$56:38:25.05 & 2.17   & 214$\pm$27 \\
ID1$^\dagger$ & 23:49:42.55 & $-$56:38:19.4 & 0.888   & 867$\pm$189 \\
\hline
ID2 & 23:49:38.838 & $-$56:37:09.63 & 2.17 &  547$\pm$36\\
ID2$^\dagger$ & 23:49:38.750 & $-$56:37:06.09 & 0.888  & 1324$\pm$182 \\
ID3 & 23:49:43.692 & $-$56:38:01.82 & 2.17 & 135$\pm$26 \\
\hline
\hline
\end{tabular}
}\\
\flushleft{
$\dagger$ ASKAP measurement\\
$\dagger\dagger$ 3$\sigma$ ATCA limit
}
\end{table}

\begin{table}
 \centering{
  \caption{ALMA observing programs used for follow-up analysis. Details on additional ALMA Band-7 observations used in this paper can be found in \citep{Hill20}. Here the frequency is the central frequency between the upper and lower sidebands, the continuum sensitivity is calculated at the center of the primary beam and averaged over the upper and lower sidebands, and the beam is the average circular synthesized beam FWHM.}
\label{table:alma_obs}
\begin{tabular}{lccccc}
\hline
ID & Date & Freq. & $\sigma_{{\rm ctm}}$ & Beam & Array   \\ 
{} & {} & (GHz) & ($\mu$Jy) & ($^{\prime\prime}$) & \\
\hline
2015.1.01543.T & 03/20/16 & 148.3 & 10 & 0.88 & C36-2/3 \\
2018.1.00058.S & 10/03/18 & 146.8 & 12   & 0.28 & C43-6 \\
2021.1.01313.S & 07/27/22 & 146.3 & 21 & 0.27 & C-6\\
2021.1.01313.S & 09/01/22 & 231.9 &  31 & 0.47 & C-4  \\
\hline
\hline
\end{tabular}
}\\
\flushleft{

}
\end{table}

This paper presents a search for radio detections of members of the \spt\ cluster.
Section~\ref{sec:data} describes the radio and (sub)millimeter observations. Section~\ref{sec:results} presents the results derived from the source extraction and analysis. In Section~\ref{sec:discussion} we discuss the detected central radio source, the energy injected into a growing ICM, and the implications for radio-loud AGN in protoclusters. We conclude in Section~\ref{sec:conclusion}. Throughout our analysis, a Hubble constant of $H_0\,{=}\,70$\ km\,s$^{-1}$\,Mpc$^{-1}$ and density parameters of $\Omega_{\Lambda}\,{=}\,0.7$ and $\Omega_{\rm m}\,{=}\,0.3$ are assumed, resulting in a proper angular scale of 6.88\,kpc/$''$ at $z=4.3$.

\section{Data} \label{sec:data}

\begin{figure*}
    \hspace*{-1.2cm}
   \includegraphics[width=0.99\linewidth]{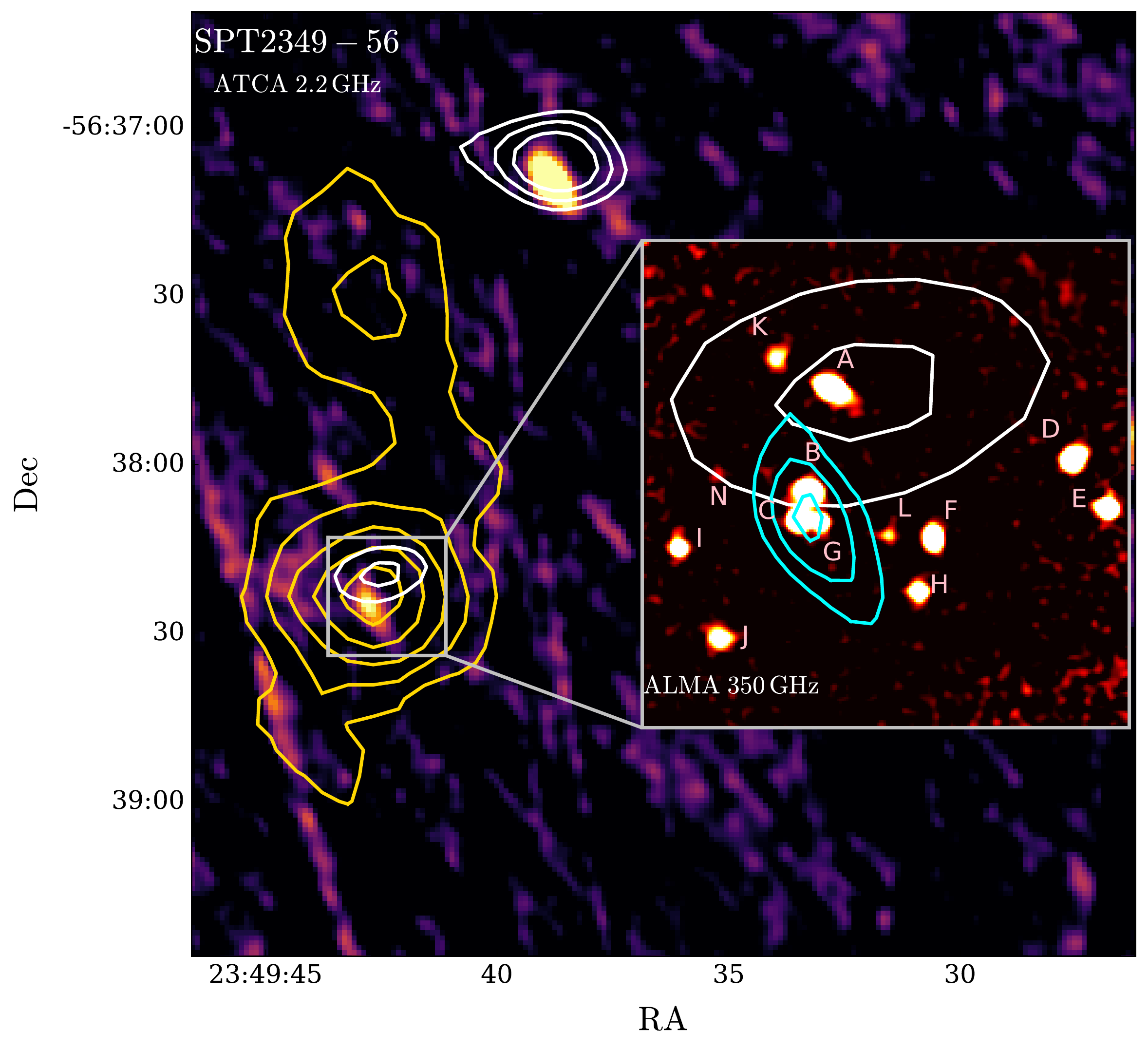}
    \caption{
{\bf Background:} 
ATCA 2.2\,GHz imaging of the \spt\ region, with gold contours highlighting the 106\,mJy extended LABOCA source at 870$\mu$m.
    The linear ATCA feature east of \spt\ is from the synthesized beam structure of a bright 20\,mJy source to the south (see Appendix \ref{sec:appendixA}).  
An ATCA radio source is identified near the LABOCA core, with an ASKAP source (white contours) overlapping.
The bright ATCA+ASKAP source to the northwest is identified with a Milky Way star (ID2 in Table~\ref{table:atca_obs}).
{\bf Inset:} 
A $20''\times20''$ zoom-in of 
{\it ALMA} 350\,GHz continuum imaging 
\citep{hill2022} with overlays of ATCA 2.2\,GHz (cyan), ASKAP  888\,MHz (white).
ATCA contours start at 3.7$\sigma$ revealing the FWHM of the source ($4^{\prime\prime}\,{\times}\,8^{\prime\prime}$). White contours (ASKAP) start at 3$\sigma$, and the FWHM of the source is $16^{\prime\prime}\,{\times}\,25^{\prime\prime}$. 
ALMA sources are named from \citet{Miller} in order of their 850\,$\mu$m flux density.  
The ATCA radio detection of the B-C-G complex of galaxies is evident.
}
    \label{fig:field}
\end{figure*}

\begin{figure}
    \centering
   \includegraphics[width=1.03\linewidth]{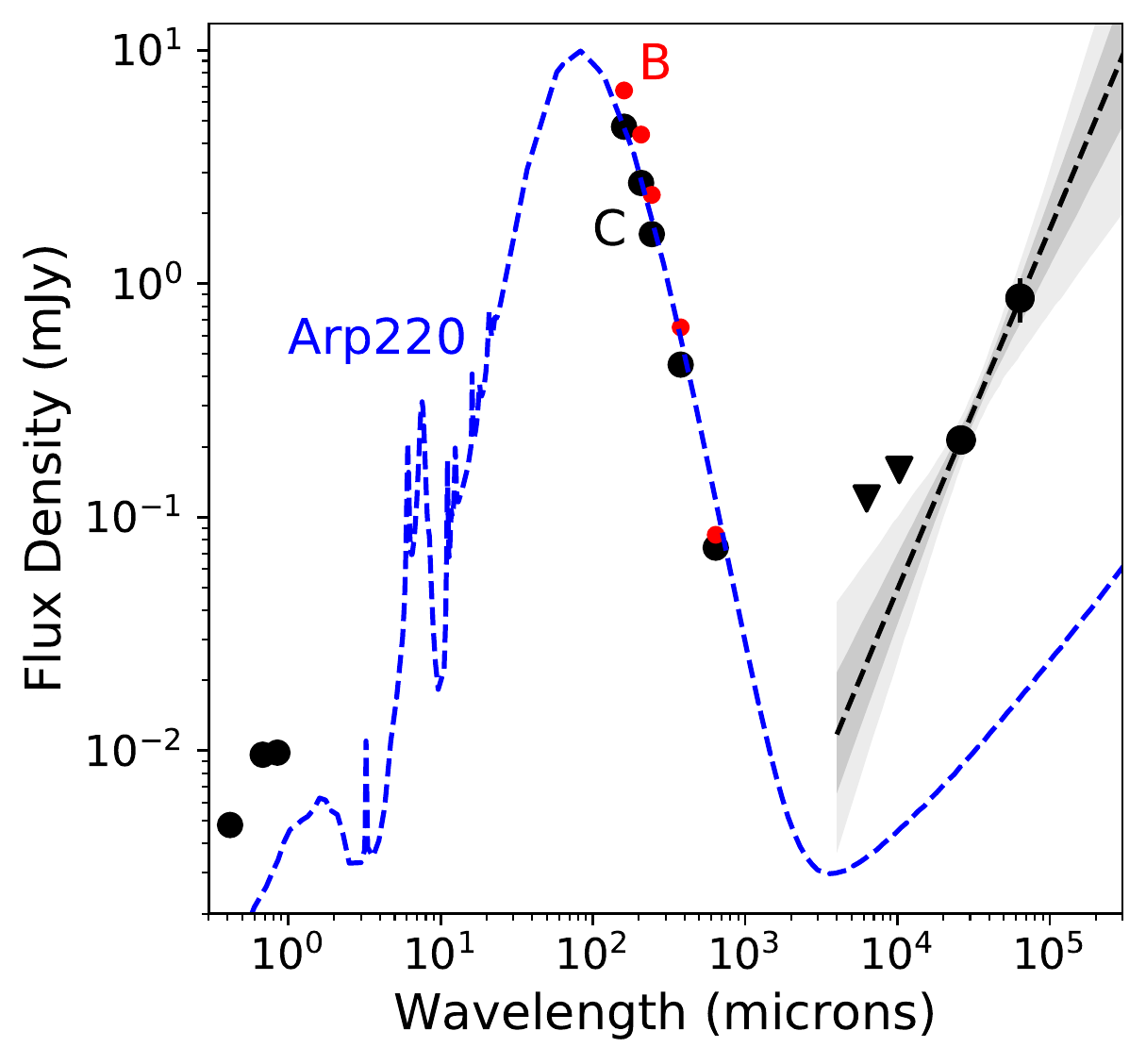}
    \caption{ 
Spectral energy distribution showing the ATCA and ASKAP radio detections at rest wavelengths, and the flux densities of the brightest two of the three central SMGs (B -- red circles; C -- black circles). Also shown is the rest-frame optical photometry of ALMA source C (as in Fig.~\ref{fig:field}), which was modelled with a $3\times10^{11}$\,M$_\odot$ stellar mass fit \citep{Rotermund21}. Source B is undetected at these wavelengths. The Arp220 SED (dashed blue line) is normalized to the submm photometry, revealing that the SPT2349 BCG galaxy complex has significant excess in radio above the far-infrared (FIR)-radio correlation for star-forming galaxies.
The 5.5\,GHz and 9\,GHz limits are shown at 3$\sigma$.
The radio spectral index is constrained to $\alpha\,{=}\,-1.58\pm0.31$ (fitted line), primarily by the ASKAP detection, and consistent with the upper limits.
The grey shadings show the 1 and 2$\sigma$ uncertainties in the fit (Appendix~\ref{sec:appendixC}); an Arp220 $\alpha\,{=}\,-0.8$ spectral index 
is ruled out at the 3$\sigma$ level by the 5.5\,GHz non-detection.
    }
    \label{fig:sed}
\end{figure}

\subsection{ATCA observations}

\spt\ was observed by the Australia Telescope Compact Array (ATCA) at 2.2, 5.5, and 9.0\,GHz between 2012, January 23 to 27,
as part of a program (C1563) to observe 23 SPT-SMGs (described in Appendix~\ref{sec:appendixC}). We used the Compact Array Broadband Backend (CABB) configured in the 1M-0.5k mode, which leads to a bandwidth of 2\,GHz per correlator window with 1\,MHz per channel of spectral resolution. The observations were performed in the most extended ATCA configuration, 6A, with six working 22\,m antennas. The on source time was 34\,min, which was typical for all SPT-SMGs observed (see Table~5). The data were edited, calibrated, and imaged using the {\tt Miriad} package. Data affected by known radio frequency interference (RFI) or with bad visibility ranges were flagged accordingly. We estimate an absolute calibration uncertainty of $5\%$ at 2.2 and 5.5\,GHz, and 10\% at 9.0\,GHz.  We inverted the visibilities using natural weighting, leading to beam sizes of $7.7^{\prime\prime}\,{\times}\,4.2^{\prime\prime}$, $3.2^{\prime\prime}\,{\times}\,2.1^{\prime\prime}$, and $2.0^{\prime\prime}\,{\times}\,1.3^{\prime\prime}$ at 2.2, 5.5, and 9.0 GHz, respectively, with associated 
RMS noise values  of 27, 40, and 53\,$\mu$Jy\,beam$^{-1}$, respectively.
Figure~\ref{fig:field} displays the ATCA\,2.2\,GHz map surrounding \spt, revealing a well-detected (8$\sigma$) source  near the core of \spt. No sources at 5.5 or 9.0\,GHz are found in the vicinity of \spt. 
ATCA sources surrounding \spt\ out to 1\,Mpc in projection are listed in Table~\ref{table:atca_obs}, and the wider-field ATCA map is shown in Appendix \ref{sec:appendixA}.

The shortest baseline is 30\,m and the images should be sensitive to emission on angular scales up to a few arcminutes. In principle, these data should not be missing any flux  on the scales covering both the ATCA and ASKAP (see below) sources, although the ATCA data will be less sensitive to lower surface brightness emission.
However the short 34\,min integration, with quite limited SNR, may still be missing some structure due to sparse uv coverage. Similar issues were discussed in an ATCA snapshot survey of distant HzRGs \citep{debreuck2000} and are elaborated in section~4. 

\subsection{ASKAP observations}

The Australian Square Kilometre Array Pathfinder (ASKAP) comprises 36 twelve-metre dishes located in the Inyarrimanha Ilgari Bundra\footnote{The name means `shared skies and stars' in the local indigenous language, Wajarri Yamatji.} at the CSIRO Murchison Radio-astronomy Observatory (MRO) in Western Australia, observing between 700\,MHz and 1.8\,GHz, with an instantaneous bandwidth of up to 288\,MHz. 
ASKAP is equipped with phased-array feeds (PAF; \citealt{hotan2014,mcconnell2016}), capable of simultaneously forming up to 36 independent beams, covering some 30\,deg$^2$.

\spt, along with all 22 of the lensed SPT-SMGs in the ATCA program, were observed by the Rapid ASKAP Continuum Survey (RACS, \citealt{mcconnell2020}), 
covering the sky south of $+41$\,deg declination at a central frequency of 887.5\,MHz, using 903 individual pointings with 15-minute observations. The beam size at the location of \spt\ is $24^{\prime\prime}\,{\times}\,13^{\prime\prime}$. We retrieved the ASKAP image surrounding \spt\ using the cutout server. At the declination of \spt\ the achieved RMS sensitivity is 189\,$\mu$Jy. 
The RMS is similar in the ASKAP images around the other 22 lensed SPT-SMGs, although the actual sensitivity depends on proximity to other nearby bright radio sources (see Appendix~\ref{sec:appendixC}).
The \spt\ ATCA-detected source is not cataloged in the RACS, but we find a 4.6$\sigma$ peak approximately 5$^{\prime\prime}$ from the ATCA source (shown in Fig.~\ref{fig:field}).

\subsection{ALMA observations}

Extensive ALMA properties of \spt\ sources B, C, and G have already been published \citep{Miller,Hill20,Rotermund21}.
Here we present several new ALMA observations (Table~\ref{table:alma_obs}), supporting our measurements of line emission in the context of searching for AGN.

ALMA Band-4 imaging (150\,GHz) was obtained under three different programs in Cycles 3, 6, and 8, all targeting the brightest peak of the LABOCA source, and tuned to place CO(7--6) ($\nu_{\rm rest}\,{=}\,$806.652\,GHz) and \CI(2--1) ($\nu_{\rm rest}\,{=}\,$809.34\,GHz) in the upper sideband, and para-H$_2$O(2$_{11}$--2$_{02}$) ($\nu_{\rm rest}\,{=}\,$752.033\,GHz) in the lower sideband.

The Cycle 3 program 2015.1.01543.T (PI: K.\ Lacaille) was observed on March 20, 2016. The array was in the C36-2/3 configuration with baselines ranging from 15 to 460\,m, and provided a naturally-weighted synthesized beam size of $0.88^{\prime\prime}$. Pallas and J2343$-$5626 were used to calibrate the amplitude and phase, respectively. The Cycle 6 program (2018.1.00058.S; PI: S.\ Chapman) observations were obtained on 2018, October 3$^{\rm rd}$ in the C43-6 array configuration with baseline lengths of 15 to 2500\,m, giving a naturally-weighted synthesized beam size of 0.28$^{\prime\prime}$. J2056$-$4714 was used to calibrate the amplitude, while J2357$-$5311 was used to calibrate the phase. Lastly, the Cycle 8 program (2021.1.01313.S; PI: R.\ Canning) observations were obtained on 2022, July 27. These observations used the C-6 array configuration with baselines of 15 to 2500\,m, giving a naturally-weighted synthesized beam size of 0.27$^{\prime\prime}$. J2357$-$5311 was used to calibrate the amplitude, while J2336$-$5236 was used to calibrate the phase.
 
 The Cycle 8 program (2021.1.01313.S) also observed CO(11--10) ($\nu_{\rm rest}\,{=}\,$1267.01\,GHz) and continuum at about 230\,GHz in Band 6.
 These observations, carried out on 2022, September 1, used the C-4 array configuration with baselines of 15 to 784\,m, giving a naturally-weighted synthesized beam size of 0.47$^{\prime\prime}$. J2357$-$5311 and J2258$-$2758 were used to calibrate the amplitude, while J2357$-$5311 and J2336$-$5236 were used to calibrate the phase.
 
 We also make use of previously-published Band 7 (345\,GHz) ALMA Cycle 5 and 6 observations \citep{Hill20}. The deep 0.5$^{\prime\prime}$-resolution (i.e. synthesized beam) Cycle 5 data contain the CO(16--15) line ($\nu_{\rm rest}\,{=}\,$1841.35\,GHz) and an OH doublet; each of the doublets is actually composed of a triplet whose frequencies are about 0.01\,GHz separated, which is completely unresolved by our spectral resolution, so we consider the OH line to be a doublet. The mean frequencies of the doublet are $\nu_{\rm rest}\,{=}\,$1837.80\,GHz and $\nu_{\rm rest}\,{=}\,$1834.74\,GHz). These lines are present in the upper sideband, which was not previously analyzed or published. The high-resolution Cycle 6 data described by \citep{Hill20} has a synthesized beam of about 0.2$^{\prime\prime}$ and is here used to further analyze kinematics through a moment analysis of the \CII\ line (Section \ref{sec:results:3.3}). 
 
All the data were calibrated using the standard observatory-supplied calibration script. Imaging was done using the {\tt CASA} task {\tt tclean}, using Briggs weighting with a robust parameter of 0.5, and in all cases channel widths were averaged down to a common 15.625\,MHz. The Cycle 6 and 8 observations covering the CO(7--6), \CI(2--1), and H$_2$O lines were combined in $uv$ space and then imaged together, while the Cycle 3 observation was imaged separately. We chose this approach as the two data sets did not overlap entirely in frequency, which led to artefacts in the imaging step. The higher-resolution Cycle 6 and 8 data cubes were then convolved to match the resolution of the Cycle 3 data (about 0.88$^{\prime\prime}$). The continuum was subtracted using the task {\tt imcontsub} after flagging all channels expected to contain line emission based on previously-detected \CII\ lines given in \citet{Hill20}. At each spatial pixel, {\tt imcontsub} extracts a one-dimensional spectrum and calculates the average over all channels not flagged by the user, then subtracts this average and returns a continuum-subtracted data cube.

The same apertures used by \citet{Hill20} to extract \CII\ line strengths and 350-GHz continuum flux densities were applied to sources B, C, and G in order to extract one-dimensional spectra for each line. The Cycle 3 and Cycles 6+8 CO(7--6), \CI(2--1), and H$_2$O spectra were averaged to produce a final spectrum. Details on how line strengths and continuum flux densities (including our procedure for deblending lines) are given in Appendix \ref{sec:appendixB}, and the spectra are shown in Figs.~\ref{fig:line_cutouts1} -- \ref{fig:line_cutouts3}. All new continuum flux densities and line strengths are listed in Table~\ref{table:lines}, and the new continuum measurements are also shown in Fig.~\ref{fig:sed}. 

\section{Results}\label{sec:results}

\subsection{Identifying and characterizing radio sources}\label{sec:results:3.1}

We first searched for radio sources at the positions of known ALMA and optically-identified members of the \spt\ protocluster.
 There is one strong radio detection at 2.2\,GHz ($S_{2.2}\,{=}\,214\,\mu$Jy) found near the \spt\ core with ATCA (detected at 8$\sigma$), which corresponds to a less robust ($4.6\sigma$) detection with ASKAP at 888\,MHz (Fig.~\ref{fig:field} and Table~\ref{table:atca_obs}). The ATCA source with a much smaller beam encompasses the bright central ALMA sources, named B, C, and G based on their rank-ordered 850\,$\mu$m flux densities \citet{Miller}.\footnote{These three sources are named C3, C6, and C13 in \citet{Hill20} based on their rank-ordered \CII\ line strength.}
It is unclear from positional uncertainty and beam size whether the emission comes from all three galaxies or just a single source. Irrespective of this, the strong radio emission would be in excess from that expected from the far-infrared (FIR)-radio correlation \citep{helou1985}. We analyse these issues in detail in Section \ref{sec:results:3.2}.



\begin{figure*}
    \centering
   \includegraphics[width=0.495\linewidth]{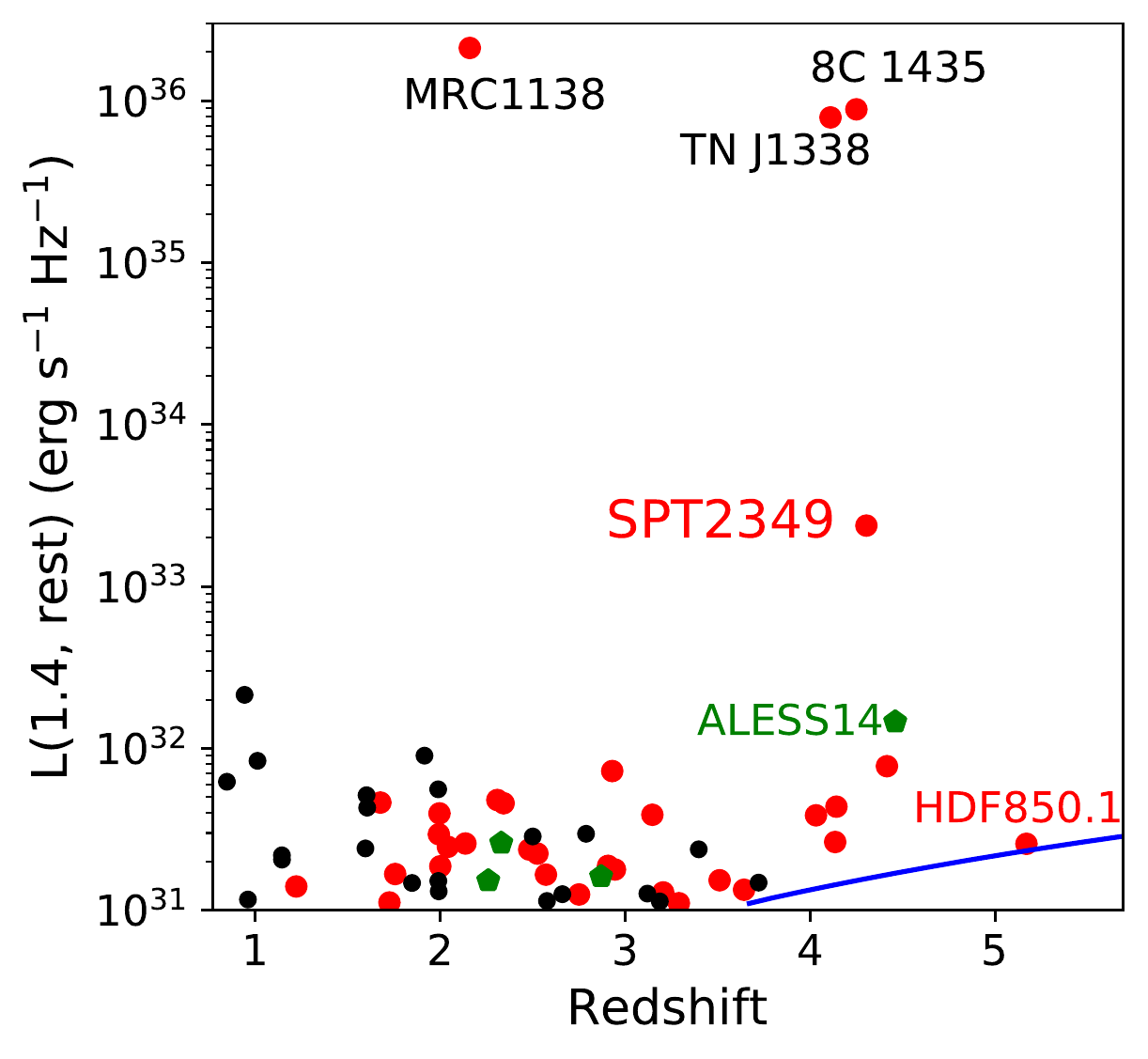}
   \includegraphics[width=0.495\linewidth]{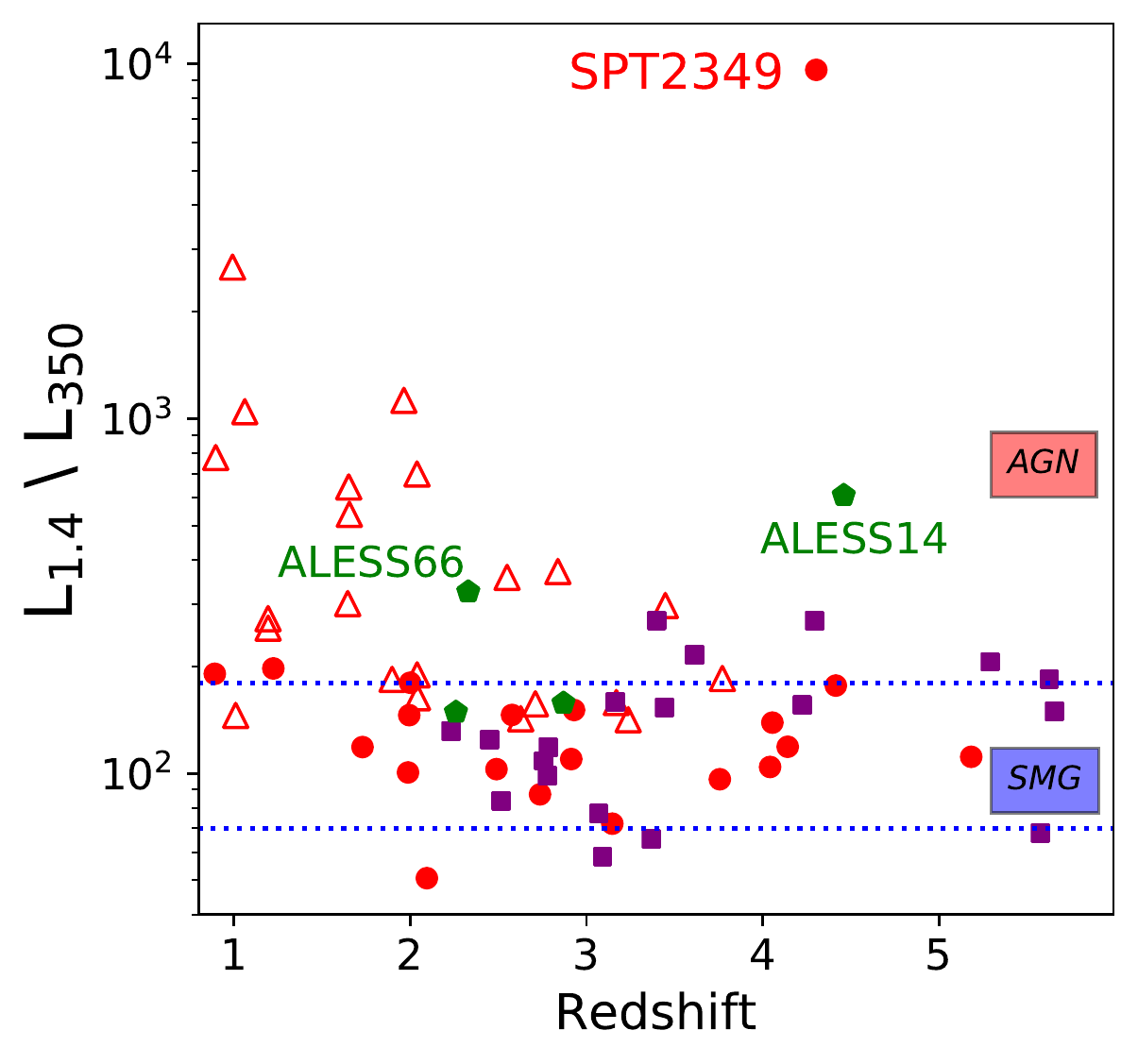}
    \caption{ 
    {\bf Left:} Redshift vs.\ 1.4\,GHz radio power using the GOODS-N sample \citep{barger2017} and radio-excess candidates from the ALESS sample \citep{thomson2014}.  The detection threshold for the GOODS-N radio sample is shown with the blue curve. Red circles show sources detected above the 3$\sigma$ level at 850\,$\mu$m, while black circles show sources not detected at this level.
    \spt\ has about 10 times more radio power than any radio source found in GOODS-N, although it has more than 500 times lower radio power than other well-studied radio-loud galaxies
    that were used to identify high redshift 
    protoclusters.
{\bf Right:} 1.4\,GHz luminosity over 350\,GHz luminosity vs.\ redshift for the submm sources with spectroscopic redshifts in GOODS-N (red circles), and lower limits on radio sources undetected in the submm (red triangles). Also shown is the ATCA survey of lensed SPT-SMGs described in Appendix \ref{sec:appendixC} (purple squares). The blue dashed line region shows where the submillimeter luminosities and radio luminosities produce consistent estimates of SFRs. None of the GOODS-N SMGs show any excess radio emission over the FIR-radio relation, while two of the ALESS sources do have a clear excess. 
Some of the higher redshift SPT-SMGs show a marginal radio excess, discussed in Appendix~\ref{sec:appendixC}.
\spt\ is about 100 times higher than the median relation at rest 1.4\,GHz. The HzRGs lie about 500 times higher with their measured S$_{850}=6-12$\,mJy.
    }
    \label{fig:distribution}
\end{figure*}

There are no other significant ($>$3$\sigma$) ATCA or ASKAP detections of any known protocluster members (Fig.~\ref{fig:field}).
The FIR-radio correlation for star-forming galaxies \citep{helou1985,ivison2010} 
would imply $S_{2.2}\,{\approx}\,12\,\mu$Jy for a $S_{850}\,{=}\,5$\,mJy source at $z\,{=}\,$4.3. 
The ten brightest \spt\ SMGs (excluding B, C, and G) span 0.8--15\,mJy, with an average of 4.7\,mJy. Thus even the brightest SMGs would only be expected to be at the $1\sigma$ level in our ATCA map.
A radio stacking analysis on these remaining ten brightest SMGs finds (11.0$\pm$10.0)$\,\mu$Jy,
which is completely consistent with the average 2.2\,GHz emission expected from the FIR-radio correlation, $\langle S_{2.2} \rangle\,{=}\,$12\,$\mu$Jy. 
Stacking on all 40 known cluster members yields $-$5.0$\pm$5.8$\,\mu$Jy.

We then consider if there might be other radio sources in \spt\ that could be 
cluster members. We searched for robustly-detected radio sources in the surroundings of \spt\ out to 1\,Mpc in projection (140$^{\prime\prime}$ in radius) from the core, roughly the region studied with ALMA by \citet{Hill20}. We find two ATCA sources above $5\sigma$, ID2 and ID3 in Table~\ref{table:atca_obs}.
ID2 is identified to a bright star, and is also detected by ASKAP.
ID3 has a clear optical counterpart, which does not have properties (especially non-detections in the $g$-band) of optical sources likely to be near $z\,{=}\,4.3$ \citep{Rotermund21}. 

We thus focus on the properties of the central ID1 radio source, starting with the positional uncertainty, $\Delta\alpha$.
From \citet{condon1997},  
we can derive the synthesized beam positional uncertainty for the ATCA and ASKAP detections, assuming that the beam is a single 2D
Gaussian with an RMS `width' $\sigma\,{=}\,$FWHM/2.354 in each coordinate. In the limit where centroiding uncertainty dominates over systematic astrometry errors and for uncorrelated Gaussian noise, we have
$\Delta\alpha\,{=}\,0.6\,({\rm SNR})^{-1}\,{\rm FWHM}$.
For both the ATCA and ASKAP sources in \spt, we have confirmed that the source size and position angle is indistinguishable from other brighter, unresolved  sources in the field, in agreement with the synthesized beam. We conclude that the \spt\ radio source is unresolved with our current data.

For the ATCA source (ID1) 
detected at SNR=7.9 
and a beam size of $4^{\prime\prime}\,{\times}\,8^{\prime\prime}$ (PA$\,{=}\,$27\,deg east of north),
the positional uncertainty is therefore 0.3$^{\prime\prime}\,{\times}\,0.6^{\prime\prime}$.
For the ASKAP source 
detected with SNR of 4.6 and a 
beam size of $24^{\prime\prime}\,{\times}\,13^{\prime\prime}$ (PA$\,{=}\,$89\,deg east of north)
the positional uncertainty is therefore 3.0$^{\prime\prime}\,{\times}\,1.7^{\prime\prime}$.
There is a 5.1$^{\prime\prime}$ roughly northern offset between the ATCA and ASKAP sources, which is consistent at the joint $2\sigma$ level. The ASKAP centroid is most consistent with the ALMA source A.
Comparison of our wider field ATCA map and the ASKAP RACS map 
reveals that the majority of the sources show excellent astrometric alignment, but we also identify a few other ATCA sources with ASKAP counterparts with several arcsecond offsets (see Appendix \ref{sec:appendixA}). In two cases, there is a robust association of the ATCA position to other cataloged objects (from 2MASS), suggesting the offset to the ASKAP position is likely due to measurement error.
For ID1, the more robust ATCA position and association to the B, C, and G galaxies in \spt\ is the most likely interpretation, with the ASKAP source being assumed to be entirely related to the ATCA source for the purposes of deriving a radio spectral index.
The 5$^{\prime\prime}$ offset is not entirely unexpected, but may be significant enough to require a physical interpretation rather than just measurement error (Appendix \ref{sec:appendixA}). It could for instance be related to a radio core-jet morphology. 
However, as noted, the 30m minimum baselines of ATCA would not resolve out flux on scales smaller than several arcmin. 
While it's not clear why the sources are offset, it appears more likely to be instrumental than physical based on the analysis in Appendix \ref{sec:appendixA}.

\subsection{Physical interpretation of ID1}\label{sec:results:3.2}

We first constrain the radio spectral index to estimate and compare luminosities between sources.
The radio source ID1 has a steep spectrum with an index of 
$\alpha\,{=}\,-1.58\pm0.28$, constrained by the ASKAP 888\,MHz detection, and the non-detections at 5 and 9\,GHz. 
The uncertainty 
can be estimated by propagation of errors on the two frequencies as follows:

\begin{equation} 
\Delta\alpha\,=\,\frac{\sqrt{\mathrm{SNR}_{2.2}^{-2} + \mathrm{SNR}_{888}^{-2}}}{\ln(2.2/0.89)}.
\end{equation} 
In Appendix~\ref{sec:appendixC}, we describe a MCMC method to assess the uncertainty for any number of spectral measurements, and show this distribution in Figure~\ref{fig:sed}. 
The spectrum is too steep to be consistent with synchrotron radiation due to shock acceleration of cosmic ray electrons from supernovae (i.e., star formation), where \cite{thomson2014} recently constrained $\alpha\,{=}\,-0.79\pm0.06$   specifically for high-$z$ SMGs.  The steep \spt\ spectrum seems to demand 
an AGN interpretation.
%


The radio luminosity can then be assessed by assuming it is associated with one of the central \spt\ galaxies at $z\,{=}\,4.3$.
With a specific luminosity of L$_{\rm 2.2}\,{=}\,$(4.4$\pm0.3)\,{\times}\,$10$^{25}$\,W\,Hz$^{-1}$, it is far larger than  expected from star formation through the FIR-radio correlation.
For reference, the FIR-radio correlation for star-forming galaxies
\citep{ivison2010}  would imply 
L$_{\rm 2.2}\,{=}\,2.4\,{\times}\,$10$^{24}$\,W\,Hz$^{-1}$ 
for a similar $S_{850}$\,{=}\,$5$\,mJy source at $z\,{=}\,$4.3.
Adopting the measured spectral index above, the radio excess increases to greater than a factor $100$ 
at a rest-frame of 1.4\,GHz, with
L$_{1.4,\,{\rm rest}}\,{=}\,(2.4\pm$0.3)$\,{\times}\,10^{26}$\,W\,Hz$^{-1}$. 
This strong radio excess suggests the presence of an AGN (e.g. \citealt{guidetti2017}); 
however, the radio emission is still distinctly less luminous than powerful radio galaxies, like those residing in other structures  studied at these redshifts, by a few orders of magnitude (Fig.~\ref{fig:distribution}).
MRC\,1138, for instance, is almost 1000 times more powerful in radio, and it is also hosted by the obvious BCG of the protocluster (e.g. \citealt{hatch2009}). 

We then compare \spt\ to  radio sources from the literature. In Fig.~\ref{fig:distribution}, the redshift versus radio power is shown using the 0.3\,deg$^2$ GOODS-N VLA sample \citep{barger2017},
which is highly complete in spectroscopic redshift. 
We compute the rest-frame radio luminosity using the equation
\begin{equation} 
L_{1.4}=\left( 4\pi d_{L}^2\, S_{1.4} / 10^{29} \right) (1+z)^{-(\alpha+1)}\, {\rm erg\,s^{-1}\,Hz^{-1}}, 
\end{equation} 
where $d_{L}$ is the luminosity distance (in cm) and $S_{1.4}$ is the  flux density in units of $\mu$Jy observed at 1.4\,GHz. This equation assumes $S_\nu\,{\propto}\,\nu^{\alpha}$, and we adopt a radio spectral index of $\alpha\,{=}\,-$0.8 \citep{ibar2010} 
for the GOODS-N sources, and the measured $\alpha$ for \spt\ and the literature HzRG sources (in fact all very close to $-$1.6). 
Shown for comparison are several well-studied HzRGs that were used as beacons to uncover massive galaxy overdensities: MRC\,1138 \citep{large1981,seymour2012}; 
TN\,J1338 \citep{debreuck1999}; 
and 8C\,1435 \citep{lacy1994}. 
\spt\ has around 10 times more radio power than any radio source found in GOODS-N, but it has less than 500 times the radio power of these HzRGs.

\begin{figure*}
    \centering
   \includegraphics[width=1.0\linewidth]{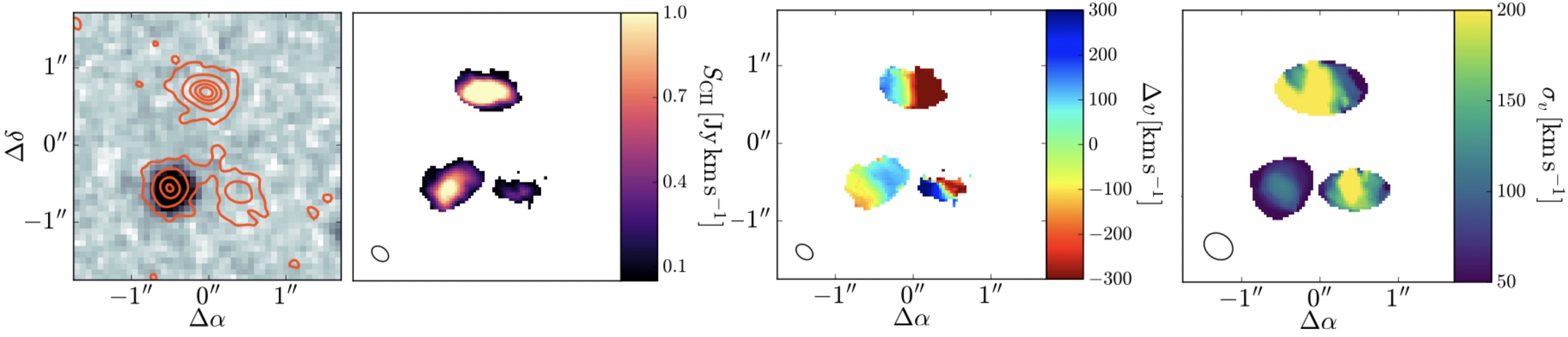}
    \caption{ 
Moment maps of B, C, and G. {\bf Left:} 
850-$\mu$m continuum (red contours) from high-resolution (0.3$^{\prime\prime}$) combined Cycle 5 and Cycle 6 ALMA data \citep{Hill20}, shown overlaid over 3-orbit {\it HST} F160W imaging \citep{hill2022}.
{\bf Mid-left:} 
\CII\ moment-0 maps from Cycle 6 high-resolution ALMA data.
{\bf Mid-right:} 
\CII\ moment-1 maps (in velocity units) from Cycle 6  ALMA data, with the zero velocity centered at the peak of each galaxy's \CII\ emission. All three sources show a clear velocity gradient (listed in Table~\ref{table:properties}, along with dynamical mass comparisons).
{\bf Right:} 
\CII\ moment-2 maps (in velocity units) are shown from lower-resolution data to increase the SNR, revealing centrally concentrated dispersions. In all panels, the synthesized beam FWHM is shown in the bottom-left corner.
}
    \label{fig:moments}
\end{figure*}

Figure~\ref{fig:distribution} also directly assesses the departure of \spt\ from the radio-FIR correlation by plotting the luminosity ratio of 1.4\,GHz to 350\,GHz versus redshift for all GOODS-N submm sources with spectroscopic redshifts (\citealt{barger2014}; A.\ Huber in prep.).
All of the submm sources in GOODS-N are radio-detected, even at $z\,{=}\,5.2$, and the submm luminosity and radio luminosity produce consistent estimates of the SFRs for all sources -- there is no sign of AGN from their radio emission.
A similar analysis of the subset of gravitationally lensed SPT-SMGs also observed from ATCA in this program (Appendix~\ref{sec:appendixC}) suggests the majority (87\%) also follow this relation; however, there are three very significant outliers in this sample which is most likely attributed to an AGN contribution from the foreground lensing galaxy (discussed further in Appendix~\ref{sec:appendixC}). 
%
\spt\ is an outlier by a factor of about $100$ from this envelope (assuming the radio emission is coming exclusively from ALMA source C). The HzRGs shown in the left panel of figure~\ref{fig:distribution} have comparable S$_{850}=6-12$\,mJy to other SMGs shown (e.g., \citealt{dannerbauer2014,debreuck1999}), and would remain about 500 times above \spt\ in the radio/submm ratio plot in the right panel.
By contrast, the GOODS-N radio sources without submm detection rise significantly above this envelope, into the AGN regime.

\cite{thomson2014} have used deep JVLA (1.4\,GHz) and GMRT (610\,MHz) to study the 76 ALMA-identified SMGs in the CDFS field (the ALESS survey -- e.g., \citealt{Simpson14}).
They find four SMGs whose radio-FIR values are $>2\sigma$ above  the sample median, which they classify as potential AGN. The most robust of these (ALESS 066.1) is a strong X-ray source with an inverted radio spectrum ($\alpha>0.51$). 
Of the remaining three, one (ALESS 014.1) has a flat radio spectrum ($\alpha>-0.1$) and an obviously high radio luminosity, 
while the other two (ALESS 094.1 and ALESS 118.1) have spectral index limits consistent with star formation ($\alpha\sim-0.8$).
%
We show these four SMGs in figure~\ref{fig:distribution}, where it is clear that none are comparable to \spt\ in radio luminosity or departure from the FIR-radio relation. In fact, two of the four are not at all unusual in their properties relative to the other samples.

Radio emission provides an extinction-free probe of AGN (which even X-ray cannot claim, since practical sensitivity limits preclude the detection of the most obscured, Compton-thick AGN with 
$N_{\rm H}>10^{24}$\,cm$^{-2}$). Traditionally radio AGN are divided into two subsets \citep{padovani2017}: {\it (i)} radio-loud AGN L$_{1.4}>10^{24}$\,W\,Hz$^{-1}$, which exhibit steep spectrum radio jets and lobes on kpc scales \citep{yun1999}; and {\it(ii)} radio-quiet AGN, with 
 flat-spectrum, lower luminosity radio emission, typically contained within a compact, several pc, core  \citep{Blundell2007}. 
\spt\ is solidly a radio-loud AGN, whereas most of the other candidate AGN found in the surveys described above (GOODS-N and ALESS)   cannot clearly be defined as such. 

    %


\begin{table*}
 \centering{
  \caption{Continuum and line properties of B, C, and G. $S_{147}$ and $S_{231}$ are the continuum flux densities at 147 and 231\,GHz, respectively, while the other columns provide various line strengths (the line is indicated by the subscript). The OH doublet arises from blended hyper-fine triplets centered at 1835 and 1838\,GHz, and the H$_2$O line is the para-$2_{11}$--$2_{02}$ line.}
\label{table:lines}
\begin{tabular}{lcccccccccc}
\hline
ID & RA,Dec & 
$S_{147}$ & $S_{231}$ & $F_{\rm CO(16-15)}$  & $F_{\rm CO(11-10)}$  &  $F_{\rm CO(7-6)}$ &  $F_{\rm H_{2}O}$ &    $F_{\rm OH}$  & $F_{\rm [CI](2-1)}$\\ 
{} & {} & 
{$\mu$Jy} & {$\mu$Jy} & {Jy\,km\,s$^{-1}$} & {Jy\,km\,s$^{-1}$}  & {Jy\,km\,s$^{-1}$} & {Jy\,km\,s$^{-1}$} & {Jy\,km\,s$^{-1}$} & {Jy\,km\,s$^{-1}$}   \\
\hline
B  &  23:49:42.79,    -56:38:24.0 & 589$\pm$15 & 3322$\pm$156 & 0.12$\pm$0.05 &  0.26$\pm$0.09 & 0.73$\pm$0.05 & 0.22$\pm$0.02 &  0.93$\pm$0.11 & 0.46$\pm$0.03 \\ 
C  & 23:49:42.84,     -56:38:25.1 &  336$\pm$11 & 1810$\pm$118 & 0.10$\pm$0.06  & 0.17$\pm$0.05 & 0.61$\pm$0.03 & 0.15$\pm$0.01 & 0.83$\pm$0.08 & 0.35$\pm$0.03\\ 
G  &  23:49:42.74,    -56:38:25.1  &181$\pm$23 & 136$\pm$9 & 0.10$\pm$0.05 &  0.12$\pm$0.05  & 0.18$\pm$0.02 & 0.02$\pm$0.01 &  0.22$\pm$0.06 & 0.10$\pm$0.02 \\
\hline
\hline
\end{tabular}
}\\
\flushleft{

}
\end{table*}

\begin{table*}
 \centering{
  \caption{Physical properties of B, C, and G. SFR$_{\rm H_2O}$ is the SFR estimated using Eq.~\ref{eq:h2o}, $S_{850}$/$F_{\rm H_2O}$ is the ratio of 850\,$\mu$m continuum flux density (from \citealt{Hill20}) to H$_2$O line strength, and SFR$_{\rm LIR}$/SFR$_{\rm H_2O}$ is the ratio of the FIR-derived SFR (from \citealt{Hill20}) to the H$_2$O-derived SFR. $V_{\rm p-p}$ is the peak-to-peak velocity from moment-1 maps, while FWHM$_{\rm cen}$ is the central velocity dispersion (multiplied by $2\sqrt{2\ln2}$) from moment-2 maps (see Fig.~\ref{fig:moments}), and FWHM$_{\rm int}$ is the width of the [C{\sc ii}] line after fitting a single Gaussian to the lines shown in Fig.~\ref{fig:line_cutouts1}--\ref{fig:line_cutouts3}. $M_{\rm dyn,\,disk}$ is the dynamical mass derived using $V_{\rm p-p}$ and a disk model (Eq.~\ref{eq:dyn_mass_disk}), while $M_{\rm dyn,\,cen}$ and $M_{\rm dyn,\,int}$ are dynamical masses derived using the velocity dispersion measurements and Eq.~\ref{eq:dyn_mass_vir}.}
\label{table:properties}
\begin{tabular}{lccccccccc}
\hline
ID & SFR$_{\rm H_2O}$ & $S_{850}$/$F_{\rm H_2O}$ & SFR$_{\rm LIR}$/SFR$_{\rm H_2O}$ & $V_{\rm p-p}$ & $M_{\rm dyn,\,disk}$ & FWHM$_{\rm cen}$ & $M_{\rm dyn,\,cen}$ & FWHM$_{\rm int}$ & $M_{\rm dyn,\,int}$  \\ 
{} & M$_\odot$ yr$^{-1}$ & 10$^{-3}$ {km$^{-1}$\,s} & {}& {km\,s$^{-1}$} & 10$^{10}$ {M$_\odot$ } & {km\,s$^{-1}$} & 10$^{10}$ {M$_\odot$ } & km\,s$^{-1}$ & 10$^{10}$ {M$_\odot$} \\
\hline
B & 1100$\pm$410 & 31$\pm$3 & 0.8$^{+0.5}_{-0.4}$ & 600$\pm$50 & 18.2$\pm$2.2 & 540$\pm$20 & 11.0$\pm$0.8 & 612$\pm$10 & 14.0$\pm$0.5 \\ 
C & 750$\pm$280 & 31$\pm$2 & 0.8$^{+0.5}_{-0.4}$ & 240$\pm$50 & 2.9$\pm$0.8 & 280$\pm$20 & 2.9$\pm$0.4 & 358$\pm$5 & 4.7$\pm$0.2  \\ 
G & 80$\pm$60 & 65$\pm$23 & 2.3$^{+2.0}_{-1.8}$ & 690$\pm$50  & 18.1$\pm$2.5 & 520$\pm$20 & 7.6$\pm$0.8 & 901$\pm$54 & 22.8$\pm$2.8 \\ 
\hline
\hline
\end{tabular}
}\\
\flushleft{
{ \ \ }\\
{ \ \ }\\
}
\end{table*}

\subsection{Resolved properties of the `BCG' sources}\label{sec:results:3.3}

Given the radio detection in \spt, it is of interest  to assess the properties of the B, C, and G ALMA sources, and to compare them to other protocluster members. As noted, these are three very submm-luminous sources in the core region ($S_{850}\,{=}\,$6.7\,mJy for B, 4.7\,mJy for C, and 1.3\,mJy for G), with only source A being brighter, although two even more luminous sources are present in the northern extension (sources N1 and N2; \citealt{Hill20}).

The most distinguishing features of this trio (beyond their flux-ordered source names serendipitously spelling out `BCG') are their locations near the center-of-mass of the cluster core, and their immediate environment. They are very close neighbours (they lie within an arcsecond of each other), and are likely to be interacting. Further, there is a notable arc seen in \CII\ surrounding the three galaxies (\citealt{Hill20}; N.\ Sulzanauer, in prep). Source C does distinguish itself with an anomalously narrow \CII\ and CO(4--3) line width for its luminosity \citep{Hill20}. 
\citet{Rotermund21} identified C as a significant outlier from the \spt\ galaxy sample in its $M_{\rm dyn}/M_{\rm gas}$ ratio inferred from the narrow CO(4--3) line width and large luminosity, similar to many high-$z$ QSOs (e.g., \citealt{Narayanan2008,walter2009,hill2019}), where selection effects favoring face-on orientation offer viable explanations.
It is also noteworthy that source C has by far the largest stellar mass of any cluster member (${>}\,10^{11}$\,M$_\odot$ \citealt{Rotermund21,hill2022}). It is associated with a bright and very compact {\it HST} F160W source \citep{hill2022}, as shown in Figs.~\ref{fig:field} and \ref{fig:moments}, and it has been suggested to be the seed of a growing BCG galaxy in this ongoing mega-merger \citep{Rennehan}.

\subsubsection{\CII\ kinematics}

We consider here a more detailed analysis of the kinematic properties  of the B, C, and G galaxies. Using high-resolution Cycle 6 \CII\ data \citep{Hill20}, which has a synthesized beam of about 0.2$^{\prime\prime}$, we construct moment 0, 1, and 2 maps of the B, C, and G sources and analyze the resolved velocity and dispersion fields. We use the {\tt CASA} task {\tt immoments}, focusing on channels between $\pm3\sigma$ of the best-fit \CII\ line, and masking pixels ${<}\,4$ times the RMS per channel. Since second moments are particularly sensitive to noise (being a squared term), we use $uv$-combined Cycle 5 and 6 data cubes (described in \citealt{Hill20}) to calculate the moment 2 maps; for reference, the resolution of the combined data is about 0.3$^{\prime\prime}$. The results (moments 0, 1, and 2) are shown in Fig.~\ref{fig:moments}.

All three sources show a clear velocity gradient and resolved, centrally-concentrated dispersion, characteristic of rotationally-supported disks. 
From these velocity gradients and velocity dispersion maps we extract peak-to-peak velocities, $V_{\rm p-p},$ and central velocity dispersions, FWHM$_{\rm cen}$. We draw a line along the semi-major axis of each galaxy, then from the moment 1 map calculate the velocity difference between the two ends, and from the moment 2 map extract the velocity dispersion at the midpoint of the line. We find that moving the position angle of the line by $\pm10\,$deg and moving the midpoint of the line by 5 pixels results in a peak-to-peak velocity change of $\pm$50\,km\,s$^{-1}$ and a central velocity dispersion change of  $\pm$10\,km\,s$^{-1}$ ($\pm$20\,km\,s$^{-1}$ in FWHM), so we quote these as our uncertainties. The results are given in Table \ref{table:properties}, multiplied by a factor of $2\sqrt{2\ln2}$ to estimate a FWHM. 

We use these peak-to-peak velocities and central dispersions to estimate masses assuming a disk model, 
%
with the enclosed dynamical mass given by 
\begin{equation} \label{eq:dyn_mass_disk}
M_{\mathrm{dyn,\,disk}} [\mathrm{M_\odot}] = 2.35 \times 10^5\, \left[ V_{\rm p-p}/\langle\sin(i)\rangle \right]^2 R,
\end{equation}
where $V_{\rm p-p}$ is the peak-to-peak velocity in km\,s$^{-1}$, $R$ is the radius in kpc, and $i$ is the inclination angle of the galaxy. We adopt a mean inclination suitable for a collection of randomly oriented disks of $\langle \sin(i) \rangle\,{=}\,\pi\,/\,4\,{\simeq}\,0.79$ (see \citealt{2009ApJ...697.2057L}), and we use the half-light radii from \citet{hill2022}, estimated by fitting S{\'e}rsic profiles to the high-resolution ALMA \CII\ moment 0 images. The results are given in Table \ref{table:properties}.

The dynamical masses were derived previously \citep{Rotermund21} from the unresolved velocity dispersions, using  the width of the integrated \CII\ lines shown in Figs.~\ref{fig:line_cutouts1}--\ref{fig:line_cutouts2}, with an assumption about the structure of the source based on the virial theorem, using the relation
\begin{equation} \label{eq:dyn_mass_vir}
M_{\mathrm{dyn}} [\mathrm{M_{\odot}}] = 2.81 \times 10^{5} \, \mathrm{FWHM}^{2} R,
\end{equation}
where FWHM is a one-dimensional velocity dispersion (multiplied by a factor of $2\sqrt{2\ln2}$) in km\,s$^{-1}$, and $R$ is the radius of the virialised structure. First, we use the resolved central velocity dispersion, FWHM$_{\rm cen}$, adopting the \CII\ size measurements from \citet{hill2022} and the central resolved velocity dispersions from the moment 2 maps (Table \ref{table:properties}). Next we use the width of the integrated \CII\ line, FWHM$_{\rm int}$, obtained by fitting a single Gaussian model to the \CII\ spectra shown in Figs.~\ref{fig:line_cutouts1}--\ref{fig:line_cutouts3} and given in Table \ref{table:properties}, again using Eq.~\ref{eq:dyn_mass_vir} and the same \CII\ size measurements. The resulting dynamical masses are provided in Table \ref{table:properties}.

Considered in the context of a disk model, source C does show a similar dynamical mass comparing  both its central and integrated velocity dispersion (Table~\ref{table:properties}, and \citealt{Rotermund21}); however, it still appears to have substantially lower mass (six times lower) than B from any kinematics  analysis. Inclination is reasonably constrained, since the aspect ratio of these galaxies is resolved by ALMA. While it remains an uncertainty in any mass modelling, the aspect ratios of B and C are similar at $\sim$1.8 (major to minor axis).

Sources B and G have similarly large inferred disk masses (18$\,{\times}\,$10$^{11}$\,M$_\odot$).
However, the distinct double-horned profile of source G (Appendix \ref{sec:appendixB}) is direct evidence for a rotating disk or bar-like structure at high inclination (explaining the broad velocity profile), while the profile for B is
possibly due to a tidal torque in response to the interaction with C. 
Source G also has a higher aspect ratio (2.3, major/minor axes) in moment-1 than B and C, suggesting the disk is seen closer to edge-on. Explicitly using this higher implied inclination in Eq.~\ref{eq:dyn_mass_disk} brings down the disk mass estimate by 25\%, more consistent with the much lower gas mass of G compared with B and C.


It is 
noteworthy that in projection at least, B is counter-rotating relative to C. 
Several studies have predicted that mergers configured with counter-rotating gas disks should lead to the most intense starbursts, and conditions for fueling the SMBHs (e.g., \citealt{1994ApJ...431L...9M, 1996ApJ...464..641M}; 
\citealt{2007A&A...468...61D}; 
\citealt{2012A&A...545A..57S}
). 

\begin{figure*}
    \centering
   \includegraphics[width=0.485\linewidth]{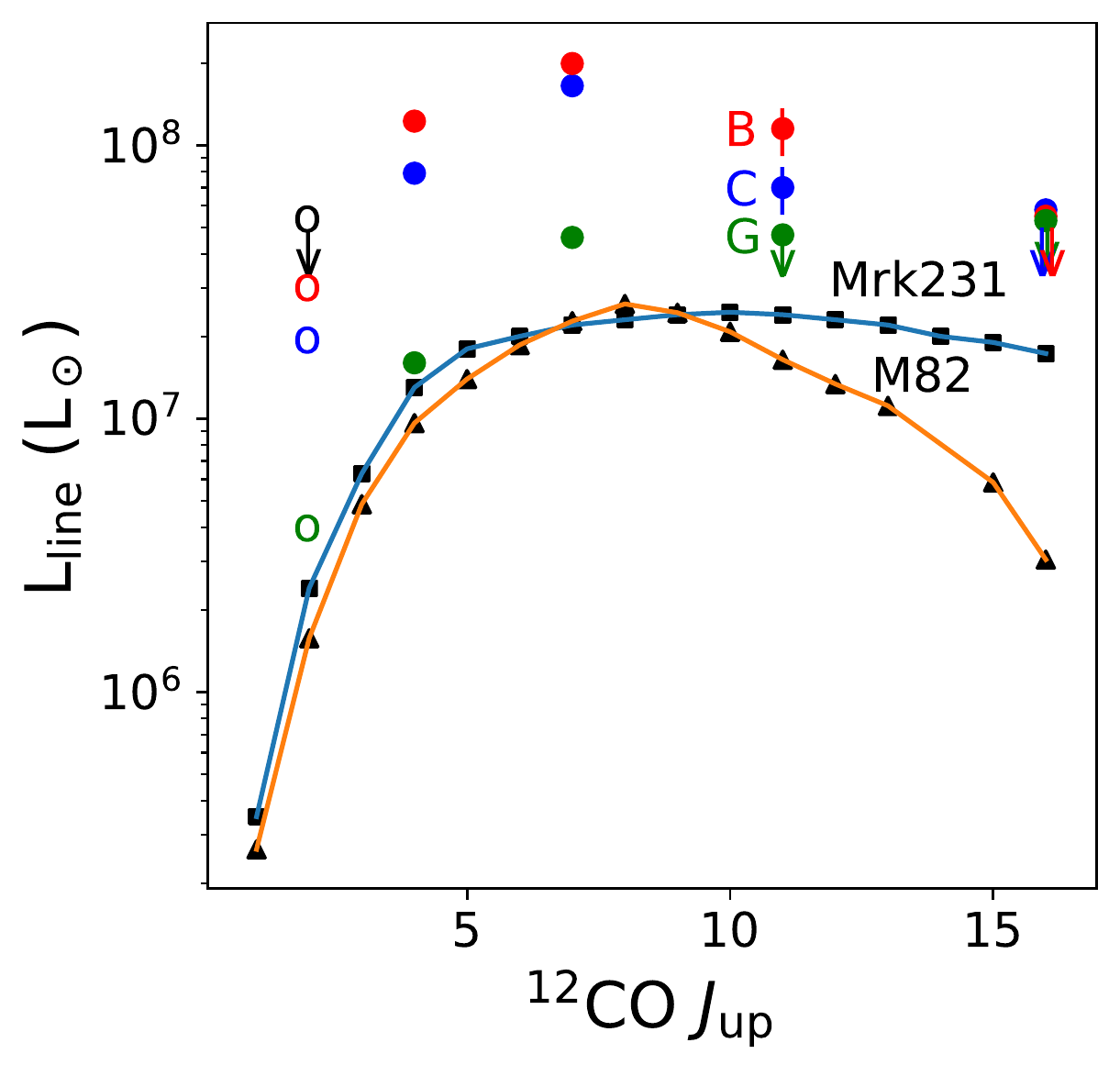}
   \includegraphics[width=0.495\linewidth]{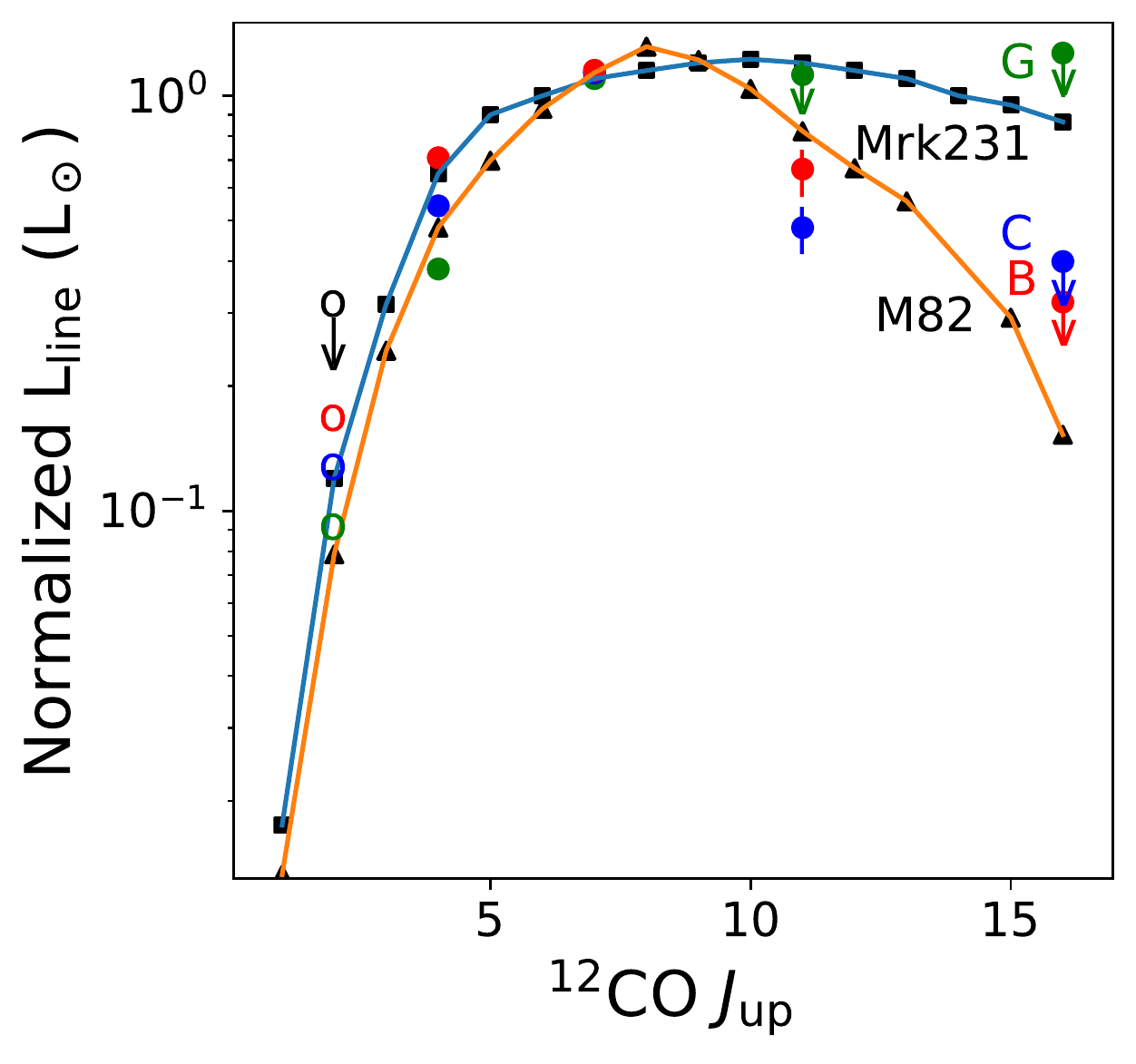}
    \caption{ 
  SLED for B, C, and G in the CO $J_{\rm upper}\,{=}\,$2, 4, 7, 11, and 16 transitions (new data for the $J_{\rm upper}\,{=}\,$7, 11, 16 lines are shown in Appendix \ref{sec:appendixB}). The ATCA detection of CO(2--1) \citep{Miller} is shown as an upper limit as it is an unresolved measurement over the core region including at least B, C, and G. Open circles illustrate the $J=2$ division if the luminosities scale from $J=4$. We compare to the $L_{\rm FIR}\,{=}\,3\,{\times}\,10^{12}$\,L$_\odot$ AGN-dominated galaxy Mrk231 \citep{vanderWerf},
  and the $L_{\rm FIR}\,{=}\,3\,{\times}\,10^{10}$\,L$_\odot$ starburst M82 \citep{kamenetzky2012}, here normalized to Mrk231 at CO(7--6).
  In the right panel the \spt\ galaxies are also normalized to the CO(7--6) luminosity of Mrk231 for comparison of the excitation curves. 
Source G is undetected 
in the $J_{\rm upper}\,{=}\,$11 and 16 lines, while B and C (shown as 2$\sigma$ upper limits) are marginally detected in CO(16--15).
}
    \label{fig:sled}
{ \ }\\
\end{figure*}

\subsubsection{Submm line properties}

We then consider line diagnostics to elucidate which of the three might be most likely to host the radio-AGN. 
We first assess the \CII/FIR ratio, which has been shown to highlight AGN with a deficit compared with star-forming galaxies (e.g. \citealt{Stacey10}).  However at high luminosities, both AGN and SMGs (without obvious AGN) exhibit similar deficits in the ratio. \citet{Hill20} have shown that all three of  B, C, and G are `deficit sources' in \CII/FIR, inhabiting similar regions in the \CII/FIR-to-FIR plot as many luminous AGN. However, this work also showed that all 12 of the most luminous SMGs in \spt\ have comparable \CII/FIR ratios, and none of these are obviously AGN from any available diagnostics.

One possibility to consider is that the FIR estimates  
are being affected by an AGN in one of B, C, or G. Since the shortest wavelength measured by ALMA is 160$\mu$m in the rest frame, the peak of the SED is not sampled, and there is little constraint on whether the dust might be substantially hotter than the T$_{\rm d}\,{\approx}40$\,K estimated in \cite{Hill20}.
To test this we make use of the para-H$_2$O(2$_{11}$--2$_{02}$) lines observed in the ALMA Band 4 dataset (Table \ref{table:lines} and Figs.~\ref{fig:line_cutouts1}--\ref{fig:line_cutouts3}). H$_2$O is strongly coupled to the FIR radiation field whether it is being produced by star-formation or AGN \citep{Omont}. \citet{Jarugula2021} compiled a sample of low- and high-$z$ submm galaxies with para-H$_2$O(2$_{11}$--2$_{02}$) measurements (including two sources, SPT0346$-$52 and SPT0311$-$58, from the same parent sample as \spt), and found that a simple single-parameter scaling relation described the correlation between $L_{\rm H_20}$ and SFR (derived from FIR) of the form
\begin{equation}\label{eq:h2o}
    \mathrm{SFR\,[M_{\odot}\,yr^{-1}]} = (2.07 \pm 0.75)\,{\times}\,10^{-5} L_{\rm H_2O}\,\mathrm{[L_{\odot}]}.
\end{equation}.

A simple test is to first take the ratio of 850$\mu$m continuum flux density to H$_2$O line strength,  where measurement errors are mostly small. These values are given in Table \ref{table:properties}, where we have used $S_{850}$ values from \citet{Hill20}. Using the same modified blackbody SED as in \citet{Hill20} to model the continuum flux density emission, a dust temperature of 40\,K at a redshift of 4.3 means that $S_{850}\,{=}\,1$\,mJy corresponds to 115\,M$_{\odot}$\,yr$^{-1}$, and so Eq.~\ref{eq:h2o} implies $S_{850}/F_{\rm H_2O}\,{=}\,$(42$\,{\times}\,10^{-3}$)\,km$^{-1}$\,s; B and C sit significantly below this value, implying they might have higher FIR than currently estimated. In contrast, G is significantly above the relation, which could imply cooler dust and lower FIR than previously estimated. This would make G less of a deficit source in \CII/FIR, and less likely to be considered an AGN by this criterion.

Using Eq.~\ref{eq:h2o}, we can also estimate SFRs directly for B, C, and G using our measured para-H$_2$O(2$_{11}$--2$_{02}$) line strengths (Table \ref{table:properties}), and compare these with the SFRs from the FIR (taken from \citealt{Hill20}). These relations show the same behaviour as our ratio of measurements above, that B and C both may have higher FIR (from hotter dust) than estimated from our current ALMA data. We note that systematic errors in converting to these physical quantities are large (as listed in \ref{table:properties}).


We next consider the CO spectral line energy distribution (SLED), which can distinguish AGN with high excitation lines driven by X-ray dominated regions (XDRs, e.g. \citealt{vanderWerf}). Here we present higher-$J$ CO transitions (Section~\ref{sec:data}) than have previously been published in \cite{Miller}. All of B, C, and G are well-detected in CO(7--6) observations, while B and C are detected in CO(11--10). Remarkably B and C may be marginally detected in CO(16--15) due to the sensitivity of the deep Band-7 ALMA data presented in \citet{Hill20}, although strictly they are upper limits. All the new line channel maps and one-dimensional spectra are shown in 
Appendix~\ref{sec:appendixB}.
SLEDs are shown in Fig.~\ref{fig:sled} from the  available $J\,{=}$2, 4, 7, 11, and 16 transitions, and compared to AGN and starburst templates. The ATCA detection of
CO(2-–1) \citep{Miller} is shown as an upper limit as it is
an unresolved measurement over the core region including at
least B, C, and G. However for illustration, we show the division into B, C, and G of the integrated $J=2$ luminosity  assuming they scale with the $J=4$ fluxes.
 None of B, C, or G  appear to have high excitation SLEDs similar to AGN like Mrk231 (e.g. \citealt{vanderWerf}),  and are all similar to or less excited than M82 at high-$J$ \citep{kamenetzky2012}, with the caveat that G has only upper limits beyond $J\,{=}\,7$. 

Source B  has the highest excitation SLED confirmed of the three, lying near the M82 SLED. Source B also has a stronger cool/warm gas component similar to Mrk231. However, all three are reasonably characterized with a combination of cool and warm star-forming Photo Dissociation Regions (PDR) components, and without significant XDR contributions.
Detailed SLED modelling of \spt\ sources will appear in a future contribution.

\begin{figure}
    \centering
   \includegraphics[width=1\linewidth]{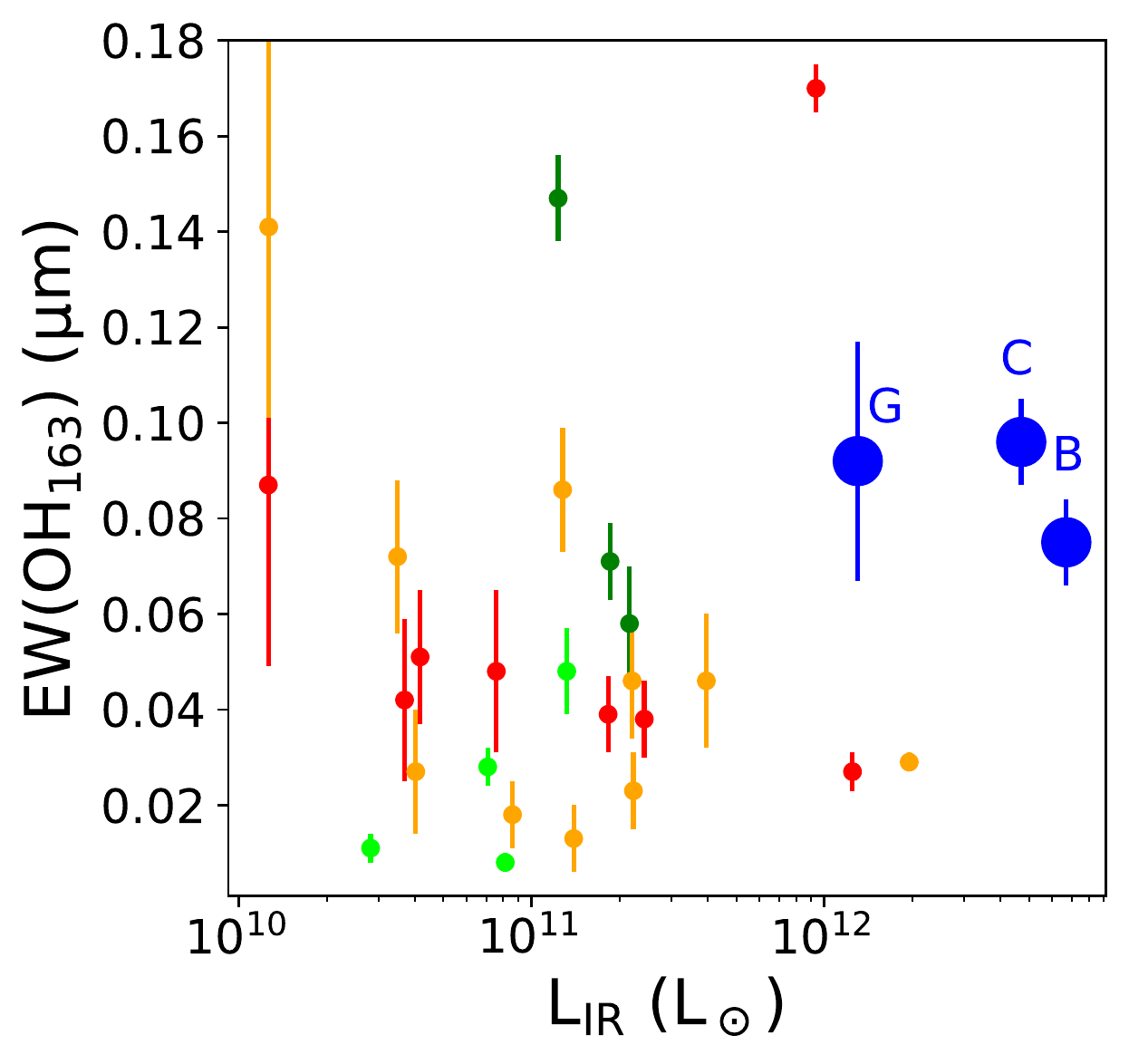}
    \caption{ 
  The equivalent width of the OH$_{\rm 163\mu m}$ doublet  versus IR luminosity. B, C, and G (blue circles) have similar values, although the error in G is large. 
  We compare to all local galaxies detected in OH$_{\rm 163\mu m}$ from the compilation in \cite{Runco2020}. Galaxies are colour coded by their classification as starbursts (lime), LINERs (green), Seyfert-1 (red), and Seyfert-2 and intermediate types (orange). 
  B, C, and G all appear well above local starbursts, although all three AGN types have a few examples this high.
}
    \label{fig:oh}
{ \ }\\
\end{figure}

Finally, the 163$\mu$m OH doublet in B, C, and G can be compared to 
\cite{Runco2020}, who studied 178 local galaxies in six of the 14 OH transitions in the FIR range. They found the highest frequency OH$_{\rm 163\mu m}$ (detected in 25 galaxies) is the only OH doublet which is always in emission, with most transitions often appearing in absorption. 
%
\cite{Runco2020} presented the correlations of the equivalent width, EW(OH), with various galaxy properties and line ratios, finding
 EW(163$\mu$m) is not well established as a direct AGN indicator. For example, while galaxies with lower X-ray luminosities exclusively have low EW(OH), the full range of EW is seen for the highest X-ray luminosities \citep{Runco2020}. However, a strong correlation is found for EW(OH) with the ratio of AGN activity to SFR, suggesting this is a better predictor of EW(OH) than the total AGN power. 
%
 In figure~\ref{fig:oh}, we compare the equivalent width, EW(163$\mu$m), to local starbursts, LINERs and Seyfert galaxies from \cite{Runco2020}.
The EW(163$\mu$m)$=$0.076$\mu$m and 0.095$\mu$m measured for B and C respectively are amongst the highest found locally. (G is similarly high, but is only marginally detected in OH). Their EW(163$\mu$m)  are more similar to values in local Seyfert galaxies than starburst galaxies, the latter having EW(163$\mu$m)$\sim$0.02--0.05$\mu$m.
%
It is not yet clear at $z>4$ what is ``normal'' for EW(163$\mu$m), since the line has only before been detected locally. However, these results may provide initial evidence 
 that B and C do in fact present as AGN through some submm-wave diagnostics.

%

\subsubsection{Optical and near-infrared properties}

Finally, we summarize optical and near-infrared spectra taken with the Gemini and VLT observatories. 
A VLT XSHOOTER spectrum ($\lambda_{\rm obs}=0.35$--$2.4\,\mu$m) was obtained which targeted B and C \citep{Rotermund21}, covering redshifted Ly$\alpha$ through [O{\sc II}]$_{3727}$. No lines were detected.  
A VLT-MUSE spectral cube was used to extract one-dimensional spectra at the locations of each of B, C, and G (Y.\ Apostolovski in prep.), but no lines are detected in any of these galaxies (nor any of the \spt\ SMGs).

Non-detections are not particularly surprising given the faintness of the galaxies at 6450\,\AA, the wavelength of redshifted Ly$\alpha$, where C has S$_{\rm 0.63\,\mu m}=0.061$\,$\mu$Jy, while B and G are undetected to S$_{\rm 0.63\,\mu m}<0.01$\,$\mu$Jy \citep{Rotermund21,hill2022}. The difficulty of spectroscopy in the near-infrared at 19,768\,\AA\ in the vicinity of the redshifted [O{\sc II}]$_{3727}$ means these limits on line equivalent widths are also not particularly constraining. 
Nonetheless, strong AGN often exhibit detectable high excitation lines in optically faint, obscured SMG hosts \citep{Chapman03N,Chapman2004,Chapman05,Danielson2017}, and it is surprising that this AGN in \spt\ eludes all optical and near-infrared spectral detection.


\subsubsection{Concluding remarks}

Thus while the radio observations do not have sufficient spatial resolution to uniquely identify one of the three galaxies as the AGN, the source properties themselves suggest source C could be a likely host, considering mainly its large stellar mass, along with narrow emission lines, and high EW(OH). Sources with similar radio luminosities in the local Universe are typically found in  massive hosts.  
However, the more FIR-luminous and much more dust-obscured source B might also be a possible host for the AGN, given the large dynamical mass from kinematic modelling and the higher excitation SLED.
Of course all three SMGs could have AGN components at the same time. 
The fact that they are likely strongly interacting dispels the typical duty-cycle arguments that would disfavor this scenario. 
To make progress, we will need deeper radio data with better resolution, 
and sensitive infrared spectroscopic observations now possible with the {\it James Webb Space Telescope}.






\section{Discussion} \label{sec:discussion}


\subsection{Inferring AGN properties from radio power}

\subsubsection{Jet power and energy input to the ICM}

Radio jets are thought to provide an important feedback mode in galaxy clusters by preventing the cooling of hot (X-ray) gas surrounding central galaxies (e.g., \cite{McNamara2012}).
This is named ``jet-mode'' feedback and is associated to radio sources characterised by radiatively-inefficient accretion. 
 However, radio jets can also drive massive gas outflows on galactic scales, another signature of AGN feedback. 

A theoretical relation between radio luminosity and radio jet power was determined by \cite{willott1999}, and can be used to estimate the kinetic energy output of AGN (e.g., \citealt{Hardcastle2007}).
The jet power can be estimated by assuming that the mechanical power of the jet can be approximated as the energy of the detected radio cavity averaged over some timescale (e.g., \citealt{Birzan2004}). 
The X-ray-detectable ``cavities'' that result from AGN jet activity \citep{osullivan2011}
allow us to quantify the heating experienced by the intra-cluster medium (ICM). The energy contained in these cavities comes from the product of the pressure and volume ($pV$) over the cavity. This
is the work done by the jet to create the cavity, and the internal energy of the radio lobes. 
Under the assumption that the cavity is dominated by relativistic plasma, this becomes $4pV$. 
Dividing the energy of the cavity by the cavity age gives the power, $P_{\rm cav}$.

Thus the most direct inference we can make from the radio properties of \spt\ adopts a relatively tight correlation observed between radio power and cavity power
\citep{cavagnolo2010,osullivan2011,panessa2015}, where a fitted relation follows:
\begin{equation}
\log P_{\rm cav} = (0.35 \pm0.07) \log L_{1.4} + (1.85 \pm0.10),   
\end{equation}
yielding $P_{\rm cav}$\,${=}$\,$(3.3\pm0.7)\times10^{38}$\,W.  
This is strictly a lower limit to the jet power, and therefore energy injection into the ICM. The true jet power depends on how the radio cavity is inflated (as described in \citealt{nusser2006}),  with some energy from the jet being carried away by shocks. 
The relation of L$_{1.4}$ to P$_{\rm cav}$ is still affected by uncertainties due to the assumption that the cavity is dominated by relativistic plasma and the detectability of cavities within the sample used in \cite{osullivan2011}, 
as discussed in their work.

This jet power is a sizeable amount of energy, given the potential well of the ${\sim}10^{13}$\,M$_\odot$ \spt\ halo (see below)
constrained from the central velocity dispersion and radial distribution of cluster members \citep{Miller,Hill20}.
This is also a significant addition to the already abundant energy injection from the 6600\,M$_\odot$\,yr$^{-1}$ of star formation being experienced by the core of \spt\ from summing the SFRs of all member galaxies found in these works \citep{Miller,Hill20,Rotermund21}.  We take  the
instantaneous injection of energy at $z=4.3$ 
as
\begin{equation}
\dot{E}_{\mathrm{kin}} = \frac{1}{2} \dot{M}_{\mathrm{out}} v^2,
\label{eq:einst}
\end{equation}
where $\dot{M}_{\mathrm{out}}$ is the total amount of gas ejected per unit time by galaxies and $v$
is the outflow velocity. While $v$ is not measured in \spt\
galaxies, the average outflow in high$-z$ SMGs and other starforming galaxies has been
constrained with increasingly large samples (e.g., \citealt{Banerji2011,forster2014}).
We adopt a typical 500\,\kms\ wind speed for the SN-driven
outflows in each \spt\ galaxy. A mass outflow rate can then be found 
by converting SFRs into
mass outflow rates $\dot{M}_{\mathrm{out}}$
by multiplying by a conservative mass loading factor $\eta\,{=}\,\dot{M}_{\mathrm{out}}/\mathrm{SFR}\,{=}\,1$. 
$\eta$ could even be greater than one 
based on observational (i.e., \citealt{Newman2012}) and theoretical work (i.e., \citealt{Hopkins2012}). 
However, the same amount of metals is found in stars and 
the ICM, which suggests equality,
$\dot{M}_{\mathrm{out}}\,{\approx}\,\mathrm{SFR}$, \citep[e.g.,][]{Renzini2014}. 
We therefore obtain 
$\dot{E}_{\mathrm{kin}}\,{=}\,(3.3\pm1.1)\,{\times}\,10^{38}$\,W, where the uncertainty reflects both the range in SFR estimates and the range of likely wind velocities. This energy injection
is remarkably similar to that found from the radio-loud AGN above from equation~6.

Estimating the total mechanical energy injected by the radio jets requires an estimate of the radio source lifetime.
\cite{Brienza2017} and \cite{Hardcastle2019} have suggested that remnant sources fade rapidly, with most of the observed remnant radio galaxies being relatively young, with  ages between $50$ to $100$\,Myr.

With 
$\tau\,{=}\,100$\,Myr for \spt, we find 
$E_{\rm mech}\,{=}\,P_{\rm cav}\,{\times}\,\tau\,{=}\,(1.0\pm0.2)\,{\times}\,10^{54}$\,J, 
assuming only the uncertainty in the $P_{\rm cav}$ scaling relation.

\subsubsection{Binding energy of the halo gas}

We then turn to estimating the binding energy of the gas in the \spt\ halo.
\cite{giodini2010} demonstrated that the mechanical energy from jets is comparable to the binding energy ($E_\mathrm{binding}$) in galaxy groups, while it is lower by a factor of 10$^2$--10$^3$ in clusters. Since the \spt\ halo mass is comparable to a large group today, and the
entire protocluster is expected to form a massive cluster by  $z=0$, it is thus of interest to investigate how $E_\mathrm{binding}$ compares to our estimate of $E_{\rm mech}$.

We define the binding energy as the total potential energy needed to push the ICM gas within $R_{500}$ (the radius where the mean dark matter halo density drops to 500 times the critical density) beyond $R_{200}$ (the radius where the mean dark matter halo density drops to 200 times the critical density, which we assume to be equal to the virial radius). \citet{hill2022} estimated $M_{200}$ (the mass contained within $R_{200}$) to be $(9\,{\pm}\,5)\,{\times}\,10^{12}\,$M$_{\odot}$, corresponding to $R_{200}\,{=}\,(120\,{\pm}\,70)$\,kpc at $z\,{=}\,4.3$, and $R_{500}$ can be computed 
if one assumes a density profile for the dark matter.

Following \citet{giodini2010}, the binding energy is computed as 
\begin{eqnarray}\label{ebin}
E_\mathrm{binding}&=&\int^{M_{\rm gas,500}}_{0}\left[\phi(r)-\phi(R_{200})\right]\,dM_{\rm gas} \nonumber \\
&=&4\,\pi \int^{R_{500}}_{0}\phi(r)\,\rho_{\mathrm{gas}}(r)\,r^2\,dr,
\end{eqnarray}
where the constant term $\phi(R_{200})$ is small compared to the potential within $R_{500}$ and can be ignored, and $\rho_{\mathrm{gas}}$ is the gas mass density.

Assuming the gas mass density follows the dark matter density but scaled by a single gas-mass fraction parameter, $f_{\rm gas}$, we can adopt an NFW dark matter profile to write the binding energy as (see \citealt{giodini2010} for details)
\begin{equation}
E_\mathrm{binding}=  f_{\mathrm{gas}}4\,\pi\,\rho_{\mathrm{crit}}\,\delta_{\rm c}\,A\,r_{\rm s}^3\int^{c_{500}}_{0}\frac{\ln(1+x)}{(1+x)^2}\,dx,
\end{equation}
where  $x=r/r_{s}$, with $r_{s}$ being the characteristic radius related to the halo concentration parameter by $c\,{=}\,R_{200}/r_s$, $\delta_c$ is a numerical factor that depends only on the halo concentration parameter $c$, $A$ scales with $M_{200}$ and also depends on $c$, and $\rho_{\mathrm{crit}}$ is the critical density at the redshift of interest (here 4.3). We compute the concentration parameter using the mass-dependent relation of \citet{maccio2007}; they find a linear trend between $\log c$ and $\log M_{\rm vir}$ (which we assume is equal to $M_{200}$), and we find $c\,{=}\,7.9$, corresponding to $r_s\,{=}\,15$\,kpc. We note that $c\,{=}\,5$ is typically adopted for massive clusters ${>}\,10^{14}$\,M$_\odot$. With the concentration parameter known, we calculate $R_{500}\,{=}\,80$\,kpc and $c_{500}\,{=}\,R_{500}/r_s\,{=}\,5.3$.


We cannot estimate the halo gas mass directly in \spt, beyond summing the measured cold gas masses in individual galaxies from the core region and inferring additional cool and warm gas components in the halo.
Summing the H$_2$ gas masses from the 23 SMGs within the cluster core from \cite{Hill20} yields $M_{\rm gas, cool}\,{=}\,3\,{\times}\,10^{11}$\,M$_\odot$.
The unseen gas components in the halo are more uncertain. 
A trend observed in groups and clusters is an increase of the fraction of hot gas with total system mass \citep{Connor2014}, approximately following 
$f_{\rm gas}\,{\propto}\,M^{0.1-0.2}$, where 10$^{13}$\,M$_\odot$ groups typically have $f_{\rm gas}$ of around 10\%.
%


The $L_{\rm X}$-$M$ relation has been shown to remain approximately self-similar out to $z\,{=}\,2$ \citep{mantz2018}, 
 including X-ray-detected clusters at $z\,{=}\,2$ \citep{gobat2011}. 
However, for low-mass systems the gas mass fractions may evolve with redshift \citep{Connor2014}. 
Regardless, this in itself does not constrain the ICM gas fraction, which requires  more detailed X-ray properties than $L_{X}$ to be detected.

Based on the above, we will assume for \spt\ a gas mass of 10\% of the halo mass, or $9\,{\times}\,10^{11}$\,M$_\odot$, which nominally requires that $M_{\rm gas, hot}\,{=}\,6\,{\times}\,10^{11}$\,M$_\odot$, unless there are substantial cold flows feeding the submm galaxies \citep{dekel2009}.
%
%
We estimate $E_\mathrm{binding}\,{=}\,(1.5_{-1.4}^{+0.7})\,{\times}\,10^{54}$\,J, where the uncertainty has been propagated from $M_{200}$ and the uncertainty in the $M_{200}$-$c$ scaling relation using a Markov chain Monte Carlo (MCMC) approach.

The radio feedback alone therefore conceivably provides all of the energy required to unbind the total gas in the cluster core. The stellar feedback has a comparable energy input, and could also be unbinding the cluster gas.
However, the total energy is a minimum condition; the energy must also couple efficiently to the ICM. An energetic jet may not couple to the bulk of the ICM gas \citep{babul2013,yang2016,cielo2018}.



Any hot ICM established at $z\,{>}\,4$ may not be in hydrostatic equilibrium 
since cold inflows likely dominate the flow of gas in  protocluster halos \citep{dekel2009}. The infalling gas only increases the energy required to inflate a bubble in the nascent ICM, acting as an additive term to $E_\mathrm{binding}$.  
While $L_{1.4}$ is fixed, the work done on an inflowing medium will be higher than for an ambient static medium. Therefore the $P_\mathrm{cav}$  (${\propto}\,4pV$) to $L_{1.4}$ relationship might not hold when inflows dominate the halo. 
\cite{Yajima2022} and \cite{Trebitsch2021} have begun to explore some of these issues in hydrodynamical simulations of protoclusters, aiming to better understand AGN feedback and the impact of massive starburst galaxies in forming clusters.
We leave more detailed calculations to future work (D.\ Rennehan, in prep.).

%

\subsubsection{Inferred X-ray luminosity and accretion rate}

A correlation also exists between radio power and X-ray luminosity ($L_X$) for radio-loud AGN \citep{ballo2012}, although there is substantial scatter in this relation. While the correlation appears to be similar over a large range (nine orders of magnitude) in X-ray luminosity, there is a range of over 100 in $L_X$ for a given radio luminosity in well populated areas of the correlation.
The relation plotted in \cite{ballo2012}  is characterized at 5\,GHz rest frame, which we measure almost directly (through the ASKAP detection). Using the correlation, we find that $L_5$\,${=}$\,$7\,{\times}\,10^{25}$\,W\,Hz$^{-1}$ in the radio corresponds to $L_{X}\,{=}\,10^{38}$\,W (where the X-ray luminosity is between 2 and 10\,keV). 
%
We conclude that the X-ray emission from the central AGN  in \spt\ can be easily detected by XMM-{\it Newton} or {\it Chandra} under the full range of possible L$_{X}\,{=}\,10^{37-39}$\,W suggested by this correlation.

Finally, taking source C as the most likely host, we can infer the SMBH  mass from the stellar mass that has been well characterized for C \citep{Rotermund21,hill2022}. For M$^*\,{=}\,4\,{\times}\,10^{11}$\,M$_\odot$, the SMBH mass is $7\,{\times}\,10^{8}$\,M$_\odot$ (e.g. \citealt{ding2020}).
From this, we can infer the range of Eddington luminosities with respect to the range in X-ray luminosity constrained by the radio power. In other words, how close to the maximal rate of accretion is the \spt\ AGN if its SMBH is close to that implied by the stellar mass of source C.  In particular, following \cite{ballo2012}, our measurements of L$_{\rm 5\,GHz}$/M$_{\rm BH}$ constrain $L_{X}/L_{\rm Edd}$ to the range of roughly 0.005 to 0.05, based on their distribution shown (their figure~10).
Directly measuring the X-ray properties of \spt\ will allow substantial progress in characterizing the system and its environment.

\subsection{Implications of the steep spectrum}\label{sec:results:4.1.2}

For an optically-thin synchrotron source, the spectrum will steepen
 in spectral index from low to high frequencies by $\Delta\alpha\,{=}\,-0.5$ 
if the source lifetime is greater than the timescale for energy-loss from the radiating electrons. This leads to a concave spectral shape with a characteristic bend frequency, $\nu_{\rm b}$ \citep{Kellermann1969}.  Thus the age of the electron population within radio jets contributes to the steepness of the spectrum.
Three effects will then decrease $\nu_{\rm b}$ as the source redshift increases \citep{krolik1991}: (1) for a fixed  bend frequency $\nu_*$ in the rest frame, the observed bend $\nu_{\rm b}\,{=}\,\nu_*\,{/}\,(1+z)$; (2) losses due to inverse Compton scattering off the microwave background rise with redshift as $(1+z)^4$, so that for a fixed time electrons spend in the radiating region, 
the lowest energy electron that can cool has a frequency (or energy), which decreases 
with increasing redshift; and (3) 
flux-limited samples result in a selection effect that favors low $\nu_*$ at high-$z$.  
Sources must have higher emissivity at higher redshift to be included in the sample. They also must have stronger implied magnetic fields, and therefore more rapid synchrotron losses. 

A combination of these effects has been used to explain the observed trend that higher-redshift radio galaxies have steeper spectral indices \citep{Carilli13,vanbreukelen2009}.
%
The ultra-steep spectral indices of HzRGs (up to the $\alpha\,{=}\,1.6$ we find in \spt) is a main selection criterion for identifying these powerful radio sources in the distant Universe
 \citep{debreuck2000,broderick2007}.
All three HzRGs shown in Fig.~\ref{fig:sed} in fact have $\alpha$ very close to 1.6. 
It is of note that \spt\ would have been discovered by these HzRG surveys over one to two decades ago had the radio source been  10--100 times more radio luminous, and even cursory submm followup would then have revealed the extended $S_{850}=110$\,mJy source that belies its nature as a submm-luminous protocluster.



A steep spectrum generally argues for self-absorbed synchrotron, and a lack of electron injection (e.g., \citealt{radcliffe2021}).
Thus the steep $\alpha$ in 
\spt\ could represent a dying radio source.
In this case, the ATCA flux should be extended over the same area as the ASKAP data. 
Thus if there is no physical offset between ATCA and ASKAP, and this is just a measurement uncertainty, \spt\ could be a young and completely unresolved compact radio source. On the other hand, this might be a ``contained'' or ``frustrated'' radio source inside a dense medium, sometimes referred to as a compact steep spectrum source, or CSS \citep{padovani2017},
but an issue with this interpretation is that the luminosity of the source is low relative to these typical GHz-peaked sources.
%
If self-absorbed synchrotron is contributing to the steep spectrum, the observational constraints would mean that the break frequency is well below about 5\,GHz in the rest frame. 
In principle this break frequency can provide a constraint on the age of the radio source, but since we do not constrain this break with the current data, we do not pursue this further here.
However if the radio emission is due to a CSS then it would have to be older than 500\,Myr to have a break frequency below 0.9\,GHz (4.8\,GHz rest) \citep{padovani2017}. 

\begin{figure}
    \centering
 \includegraphics[width=0.99\linewidth]{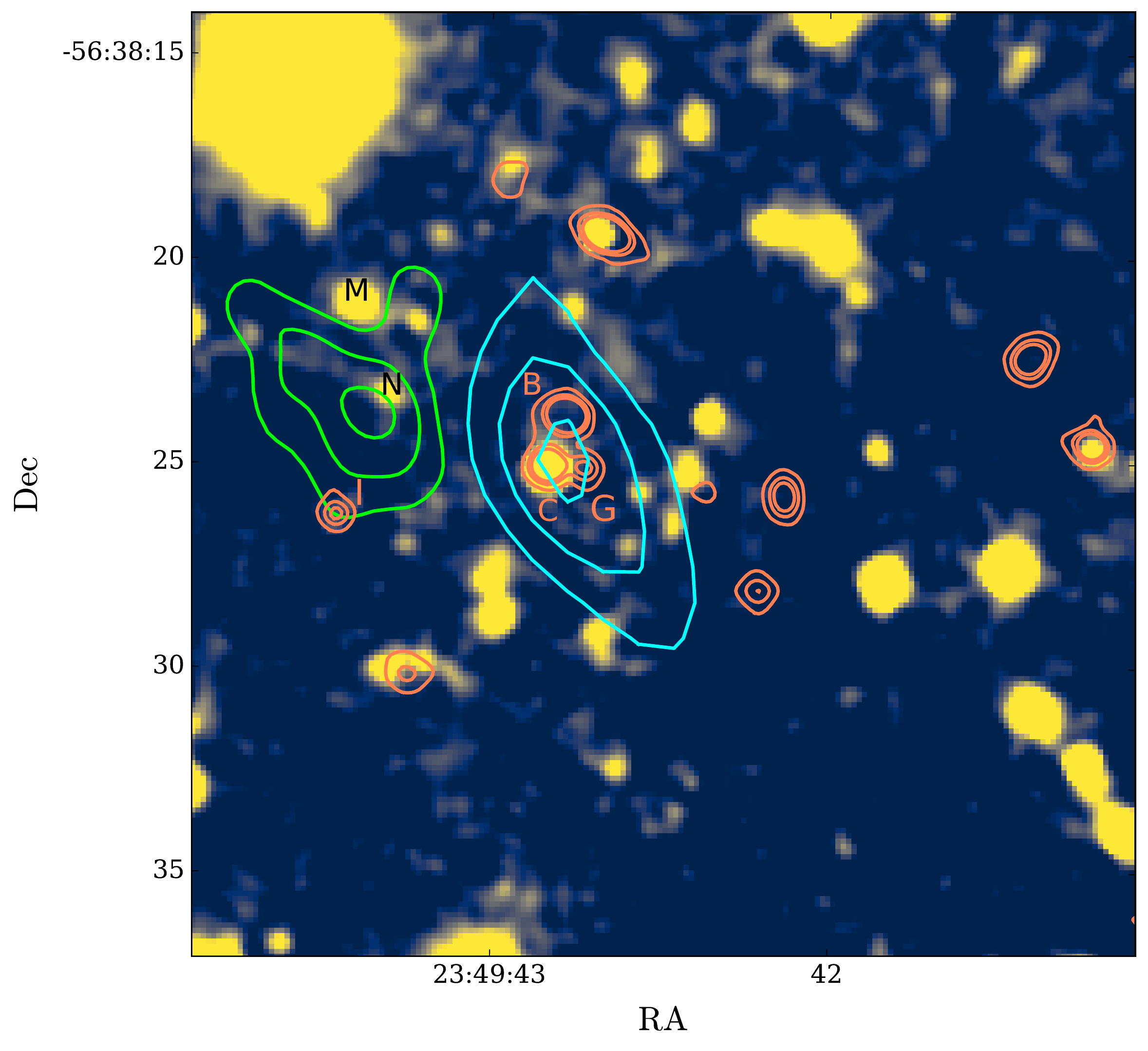}
 \caption{{\it Does the radio-loud AGN power the LAB?}
 The background  shows HST F160W imaging with ATCA 2.2\,GHz contours (cyan) and the Y.\ Apostolovski et al.\ (in prep.) MUSE Ly$\alpha$ contours (lime). The center of the LAB lies 4.5$''$ (31\,kpc) from the radio source centroid. 
 ALMA 850$\mu$m contours are shown (coral), but sources M and N are too weak to see in this representation. The LAB is centered near the SMG, N, which was originally identified through its \CII\ emission and undetected in continuum \citep{Miller}. 
 N has S$_{\rm 1.1mm}= 0.18$\,mJy and an L$_{\rm IR}=4\times10^{11}$\,L$_\odot$, with an implied SFR=35\,M$_\odot$\,yr$^{-1}$.
 }  
    \label{fig:muse}
\end{figure}

\subsection{Connection to the LAB}\label{sec:results:4.2}

The powering sources of Ly-$\alpha$ blobs (LABs)
have often been  identified broadly with the photoionizing emission from a close ionizing source (e.g., a QSO,  \citealt{Geach2009,overzier2013}), 
shocks (e.g., \citealt{Taniguchi2000}), 
or ``cooling radiation'' 
during  gravitational collapse of the gas (e.g., \citealt{haiman2000}).
The \spt\ LAB (shown in figure~\ref{fig:muse}) was originally hypothesized to be heated by some combination of the three ALMA sources that reside near or within it (Y.\ Apostolovski et al.\ in prep.). 
However, given that the LAB center is only 4.5$^{\prime\prime}$ (31\,kpc in projection) offset from ALMA source C,  it could instead be heated by the radio-loud AGN.
The LAB is centered on the weak SMG, N, which was originally identified through its \CII\ emission \citep{Miller}. 
N is a luminous infrared galaxy (LIRG) with $S_{850}\,{=}\,0.27\pm0.04$\,mJy, $L_{\rm FIR}\,{=}\,4\,{\times}\,10^{11}$\,L$_\odot$, and a substantial $M^*\,{=}\,3\,{\times}\,10^{10}$\,M$_\odot$. It is a plausible, but somewhat unlikely power source for the luminous LAB (whose total luminosity is $3\,{\times}\,10^{42}$\,erg\,s$^{-1}$, or $3\,{\times}\,10^{35}$\,W); source N fails to provide the necessary UV ionizing photons by at least a factor of ten, scaling from its meager $R$-band flux density of 0.37\,$\mu$Jy (similar to the analysis in Y.\ Apostolovski et al.\ in prep.). 
We can directly estimate the AGN X-ray emission expected for powering the Ly-$\alpha$ blob following \cite{overzier2013},
assuming  that the fraction of ionizing photons that will cascade to Ly$\alpha$ is 68\% (case B recombination). This number likely exceeds the actual amount of ionizing radiation available due to the absorption by dust by a factor of around 10, which we account for here. 
We then assume a radio-quiet QSO spectrum given by \cite{Richards2006}. The predicted observed frame (0.2--12\,keV) X-ray luminosity would be 2$\,{\times}\,$10$^{37}$\,W, which is comparable to the low end of the expected range of 
$L_{\rm X}$ from the \spt\ radio source, as discussed above. 
The radio AGN may therefore be at least as plausible a heating source as N.

Regarding the Ly$\alpha$ blob being spatially offset from the AGN position, we note that in the  radio source B3~J2330 at $z\,{=}\,$3.1 \citep{matsuda2009},  the peak of the Ly$\alpha$ emission was also found to be similarly offset from the HzRG itself.
Even in the $z\,{=}\,4$ Distant Red Core (DRC) LAB, there is a roughly 3$^{\prime\prime}$ (21\,kpc) offset from the X-ray-emitting AGN that is proposed as the LAB's power source \citep{vito2020}.
However, these are rare cases.  \cite{venemans2007}
showed that generally the AGN is very near the center of the Ly$\alpha$ halo, which grants some geometrical credence to the idea that the Ly$\alpha$ halo is ionized by the central AGN's photons. In \spt, this is harder to argue, but  the Ly$\alpha$ could be  completely absorbed by the copious amounts of dust in the core. The SPT pre-selection (as with the {\it Herschel} selection of the DRC) may favor finding sources with such offsets.

%


\subsection{AGN fractions in protoclusters}\label{sec:results:4.3}

As described in Section \ref{sec:results:3.1}, with 27\,$\mu$Jy RMS at 2.2\,GHz, we are sensitive to moderately-luminous and heavily-obscured $z\,{=}\,$4.3 AGN among the 30 sub-millimeter galaxies identified in the \spt\ structure. They need to lie approximately 5 times above the radio-FIR relation to be significantly (5$\sigma$) detected by ATCA. In GOODS-N (Fig.~\ref{fig:distribution}), there are seven radio sources (all lacking submm detection) that satisfy this threshold, all of which lie at at redshifts less than 2. Another seven such radio sources lie 2--3.5 times above the relation, extending to a redshift of about 4, which would not be detected by our observations.
The fact that all submm-detected sources in GOODS-N, and 74 of 76 SMGs in ALESS, are consistent with the radio-FIR relation does signify that radio-loud AGN are not common amongst the submm-luminous population.
No significant radio emission 
is found from any other (non-SMG) cluster members or candidates.  
 With our current radio depth,  the radio-AGN content among SMGs in this protocluster is constrained to be less than 10\% (three of 30 members), and most likely 3\% (assuming C is the host of the ATCA radio source). However, the radio-loud AGN are only about 10\% of the total AGN population in the field \citep{barger2007,radcliffe2021}. The X-ray AGN fraction remains unconstrained, and given that many X-ray AGN are not radio emitters \citep{barger2007},  our AGN fraction estimates in \spt\ are lower limits.

 
 In the $z\,{=}\,$4.0 DRC protocluster \citep{Oteo18}, a central galaxy is radio-undetected, but is a Compton-thick X-ray AGN. Only one of the three X-ray-identified AGN is detected in the radio -- DRC6 ($S_{5.5}\,{=}\,128\,\mu$Jy, $S_{9}\,{=}\,120\,\mu$Jy), indicating a flat-spectrum source. In this case, the radio-AGN in the DRC lies towards the edge of the projected distribution of SMGs (offset from the core of the cluster).  
 Thus without X-ray data, we cannot tell if the total AGN fraction of \spt\ is different from that in the DRC (23\%, \citealt{vito2020}).
 As another example, in the core of the $z\,{=}\,3.09$ SSA22 protocluster, the SMGs have a 50\% X-ray AGN fraction, with four of eight SMGs detected by {\it Chandra} \citep{Umehata19}, significantly larger than the DRC.

\subsection{Radio sources and cluster evolution}\label{sec:results:4.4}

Given that \spt\ is conceivably the most massive and active halo we know of at $z\,{>}\,4$, an open question concerns the feedback or radio mode that this AGN is operating in, and how it is shaping the early core evolution of the cluster.
With the current data, having only the two photometric points characterizing the radio emission, and not even localizing it uniquely to one galaxy, we cannot definitively address these issues.
Most radio-loud
AGN appear to be hosted in recent or ongoing mergers (e.g., \citealt{ramos2012,chiaberge2015}).
In this light it may not be too surprising to find a radio-loud AGN in the core of \spt.
Given that the radio luminosity of \spt\ is modest for an HzRG, we may be seeing a radio-loud AGN fueled via radiatively-inefficient flows with low accretion rates \citep{best2012}. In this picture, the gas supplying the radio galaxy is frequently associated with hot X-ray halos surrounding massive galaxies, groups and clusters, as part of a radio-AGN feedback loop. 
This contrasts with more luminous radio sources (e.g. TN\,J1338) thought to be fuelled at higher rates through radiatively efficient standard accretion disks by cold gas \citep{best2012}.
 These more luminous  radio sources are hypothesized to have fuel brought in through mergers and interactions, which are in fact abundant in \spt. 
The debate thus remains open as to whether we are seeing a decaying radio source, or a radio source quickly building in luminosity.
By better specifying the radio emission and its origin, we could learn about the build-up and state of the ICM that may already be present at $z\,{=}\,$4.3.


\section{Conclusions} \label{sec:conclusion}

We have presented ATCA radio observations of \spt, a starbursting and gas-rich protocluster, consisting of over 30 SMGs at $z\,{=}\,$4.3. We placed \spt\ in context with $\mu$Jy radio sources in the GOODS-N and ALESS fields, and with the other 22 gravitationally-lensed SPT SMGs also observed with ATCA in our program. We also studied in detail the central galaxies identified by ALMA in \spt\ near this strong radio detection. 

– We detected a single source at 2.2\,GHz in \spt, spatially coincident with the central three luminous members of the protocluster, denoted B, C, and G in \citet{Miller}.
While the ATCA radio centroid lies close to source C, which has the largest stellar mass in the  protocluster, we cannot rule out that the radio emission is coming from B or G, or even a combination of the galaxies.

- Under any of the possibilities above, the 214\,$\mu$Jy flux density at 2.2\,GHz translates to more than 20 times the radio luminosity expected from the FIR-radio correlation defined by star-forming galaxies, and suggests that an AGN 
is driving the radio emission.

- The radio source has a steep spectrum, with an index of $\alpha\,{=}\,-1.58\pm0.31$, constrained by the ASKAP 888MHz detection, and the non-detections at 5.5 and 9\,GHz, consistent with 
an AGN. 

- No other clear signs of AGN activity have yet been detected in this protocluster using any other diagnostics available to us (CO SLEDs; EW(OH$_{\rm 163\mu m}$), 
\CII/FIR ratios; 
optical spectra), highlighting the radio continuum as a powerful probe of obscured AGN in high-$z$ protoclusters.

- The three SMGs likely associated to the radio source have amongst the highest gas and dynamical mass  of the protocluster members \citep{Rotermund21}. Moreover, high resolution ALMA imaging resolves this system into multiple interacting, star-forming clumps, with a surrounding arc of \CII\ emission (\citealt{Hill20}; Sulzanauer et al.\ in prep.). This is consistent with the idea that the availability of large amounts of gas and galaxy interactions, both of which are enhanced in gas-rich overdensities at high redshift, can trigger fast and obscured SMBH accretion.

– No significant radio emission (nor any other robust AGN signature) is found from any other cluster member,
constraining the radio-loud AGN content among SMGs in this protocluster to no more than 10\% (three of 30 members), and likely just 3\%.
A radio stacking analysis on the remaining ten brightest \spt\ SMGs finds (11$\pm$10)\,$\mu$Jy, which is  consistent with the average 2.2\,GHz emission from star formation via the FIR-radio correlation. 
We thus  find no evidence that nuclear accretion powering radio emission exists below our detection threshold in other SMG members of \spt. However, radio-loud AGN represent only 10\% of all AGN, and X-ray observations and JWST infrared spectroscopy would be the next key steps to constrain AGN in this system and compare to  AGN fractions found in other protoclusters.

- The \spt\ radio-loud AGN  has a luminosity density of L$_{2.2}\,{=}\,4.4\,{\times}\,10^{25}$\,W\,Hz$^{-1}$, extrapolating to L$_{1.4,\,{\rm rest}}\,{=}\,(2.4\pm$0.3)$\,{\times}\,10^{26}$\,W\,Hz$^{-1}$ with the measured $\alpha=-1.6$, which is still over two orders of magnitude less luminous than the powerful radio galaxies normally studied at these redshifts. 
Many such HzRGs have rich protocluster environments, 
however it remains unclear if the opposite is true, that all massive $z\,{>}\,4$ protoclusters have a central radio galaxy. 

- The fact that the radio AGN is detected in the hypothesized central seed of a growing BCG galaxy with significant stellar mass already in place 
makes  this discovery an important new ingredient in understanding the formation and evolution of the cluster.

- The radio luminosity was used to infer a radio jet power of $P_{\rm cav}$\,${=}$\,$(3.3\pm0.7)\times10^{38}$\,W, sufficiently large as to provide a dominant feedback on the cooling gas in the 10$^{13}$\,M$_\odot$ halo.
The radio luminosity also suggests a strong X-ray source with $L_{X}\,{=}\,10^{38}$\,W (integrated between 2 and 10\,keV), easily detectable by {\it Chandra} or XMM-{\it Newton}.
\spt\ therefore has a high luminosity AGN, even if in the form of a highly obscured quasar, and {\it JWST} will be a powerful tool to uncover its properties through high ionization infrared emission lines.


\section*{acknowledgements}

The Australia Telescope Compact Array is part of the Australia Telescope National Facility (https://ror.org/05qajvd42), which is funded by the Australian Government for operation as a National Facility managed by CSIRO.
The Australian SKA Pathfinder is part of the Australia Telescope National Facility (https://ror.org/05qajvd42) which is managed by CSIRO. Operation of ASKAP is funded by the Australian Government with support from the National Collaborative Research Infrastructure Strategy. ASKAP uses the resources of the Pawsey Supercomputing Centre. Establishment of ASKAP, the Murchison Radio-astronomy Observatory and the Pawsey Supercomputing Center are initiatives of the Australian Government, with support from the Government of Western Australia and the Science and Industry Endowment Fund. We acknowledge the Wajarri Yamatji people as the traditional owners of the Observatory site.
%
The National Radio Astronomy Observatory is a facility of the National Science Foundation operated under cooperative agreement by Associated Universities, Inc.
%
This paper makes use of the following ALMA data: ADS/JAO.ALMA\#2015.1.01543.T, ADS/JAO.ALMA\#2018.1.00058.S, and\\ ADS/JAO.ALMA\#2021.1.01010.P. 
ALMA is a partnership of ESO (representing its member states), NSF (USA) and NINS (Japan), together with NRC (Canada), MOST and ASIAA (Taiwan), and KASI (Republic of Korea), in cooperation with the Republic of Chile. The Joint ALMA Observatory is operated by ESO, AUI/NRAO and NAOJ.
S.C., A.B., and D.S.\ gratefully acknowledge support for this research from NSERC.
Manuel A.\ acknowledges support from FONDECYT grant 1211951, CONICYT + PCI + INSTITUTO MAX PLANCK DE ASTRONOMIA MPG190030. M.A.\ and M.S.\ acknowledge support from CONICYT + PCI + REDES 190194 and ANID BASAL project FB210003.
K.A.P., Melanie A. are supported by the Center for AstroPhysical Surveys at the National Center for Supercomputing Applications as an Illinois Survey Science Graduate Fellow.


\appendix

\section{Appendix A} \label{sec:appendixA}

\begin{figure*}
    \centering
 \includegraphics[width=0.99\linewidth]{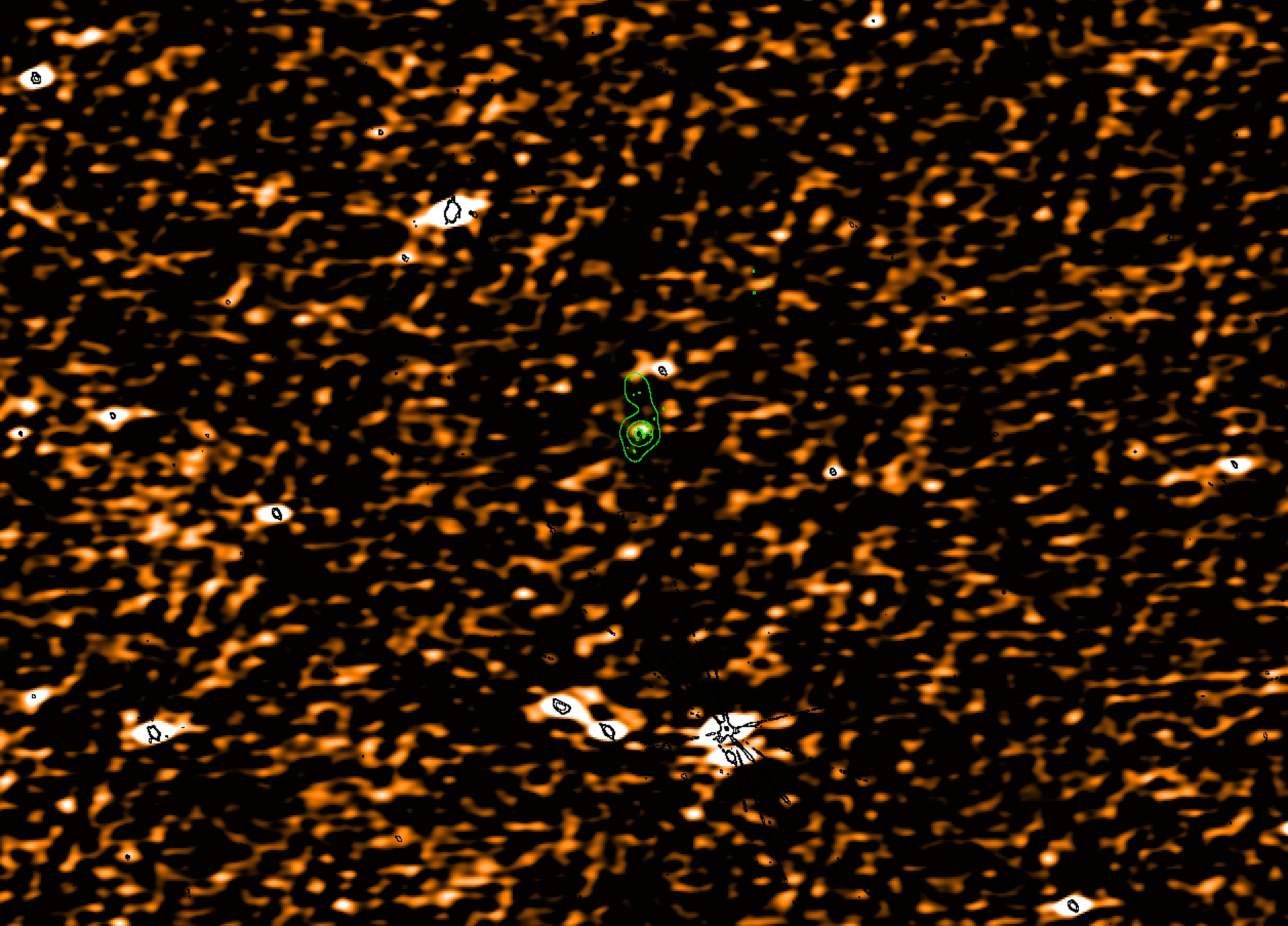}    
 \includegraphics[width=0.49\linewidth]{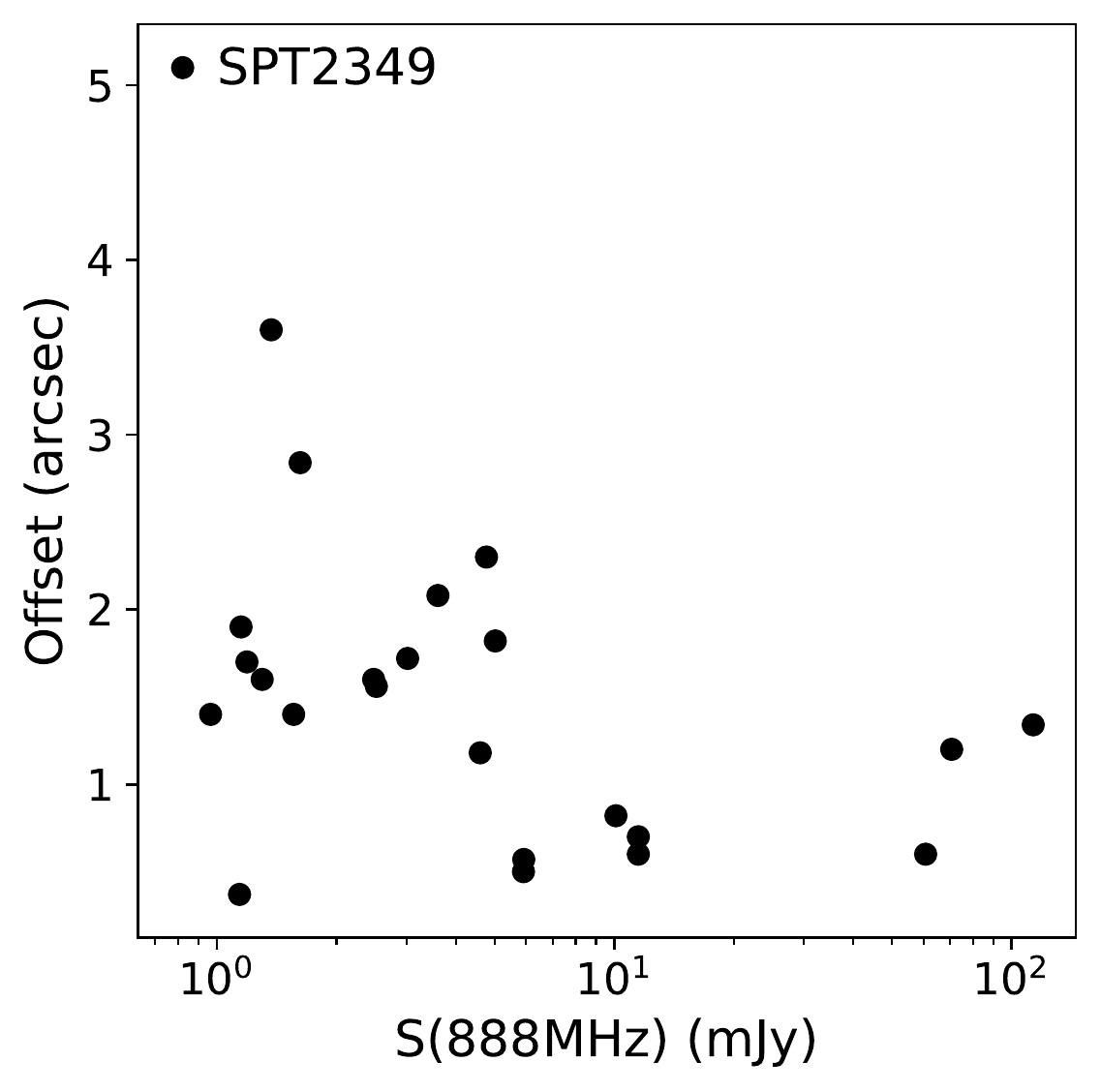}
 \caption{Assessing the offsets between ASKAP and ATCA sources. {\bf Top:} The 888\,MHz ASKAP image surrounding \spt\ (green LABOCA contours, showing ALMA sources as green dots) with ATCA 2\,GHz contours overlaid ($26^{\prime}\,{\times}\,19^{\prime}$ field shown). 
 {\bf Bottom:} The radial offsets between the centroids  for  sources common in both images.
 \spt\ shows the largest offset, which might therefore require a physical interpretation.
 It has a 4.6$\sigma$ deviation from the median (excluding \spt) offset of 1.4$^{\prime\prime}$ offset.
 }  
    \label{fig:offset}
\end{figure*}

In this appendix, we further assess the offsets between source centroids in the 888\,MHz ASKAP image and the ATCA 2\,GHz image that were discussed in section~\ref{sec:results:3.1} (Fig~\ref{fig:offset}). We measured peak fluxes in both images for all sources within a 13$^{\prime}$ radius of \spt,  measured their centroids, and calculated radial offsets for each. The offsets appear random in orientation, with the mean $x$ and $y$ offset being close to zero (0.3$^{\prime\prime}$, $-$0.2$^{\prime\prime}$).
In Fig.~\ref{fig:offset} we plot the radial offsets versus ASKAP flux.
\spt\ shows the largest offset, which could indicate that its origin may be physical. It has a 4.6$\sigma$ deviation from the median (excluding \spt) offset of 1.4$^{\prime\prime}$.
 Even restricting the analysis to those sources with comparable flux densitites and SNRs ($S_{888}\,{<}\,$2\,mJy) only increases the median offset to 1.6$^{\prime\prime}$.

\section{Appendix B} \label{sec:appendixB}

\begin{figure*}
    \centering
 \includegraphics[width=0.48\linewidth]{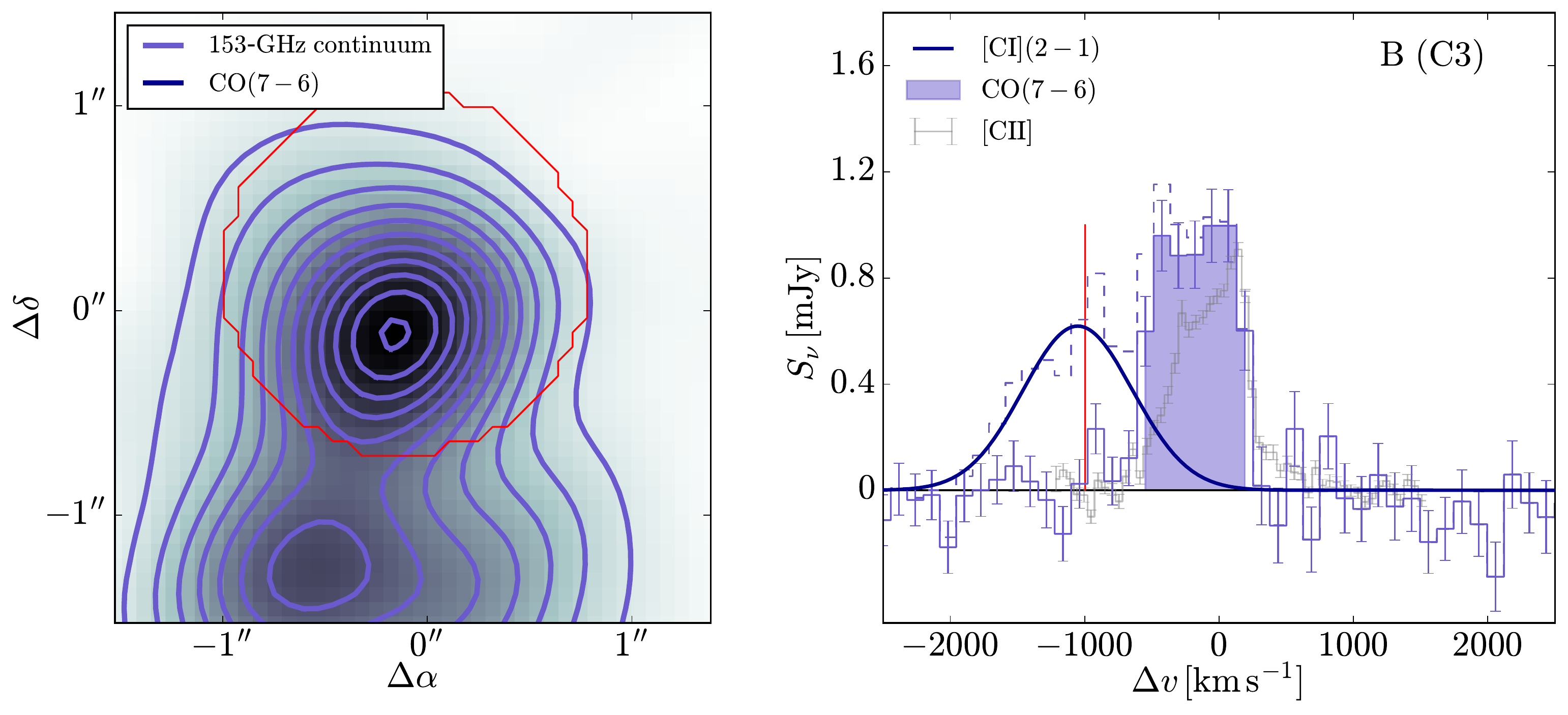}
 \includegraphics[width=0.48\linewidth]{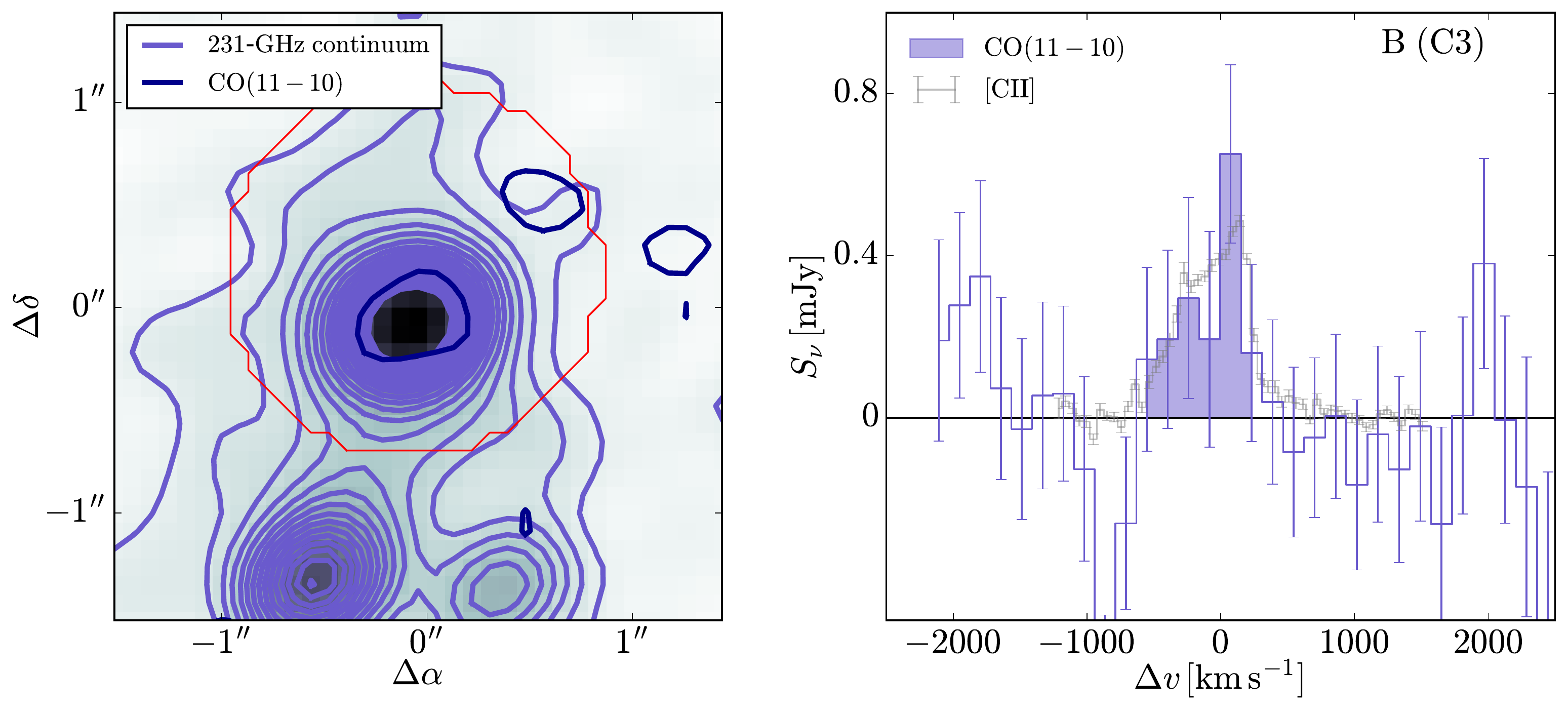}
 \includegraphics[width=0.48\linewidth]{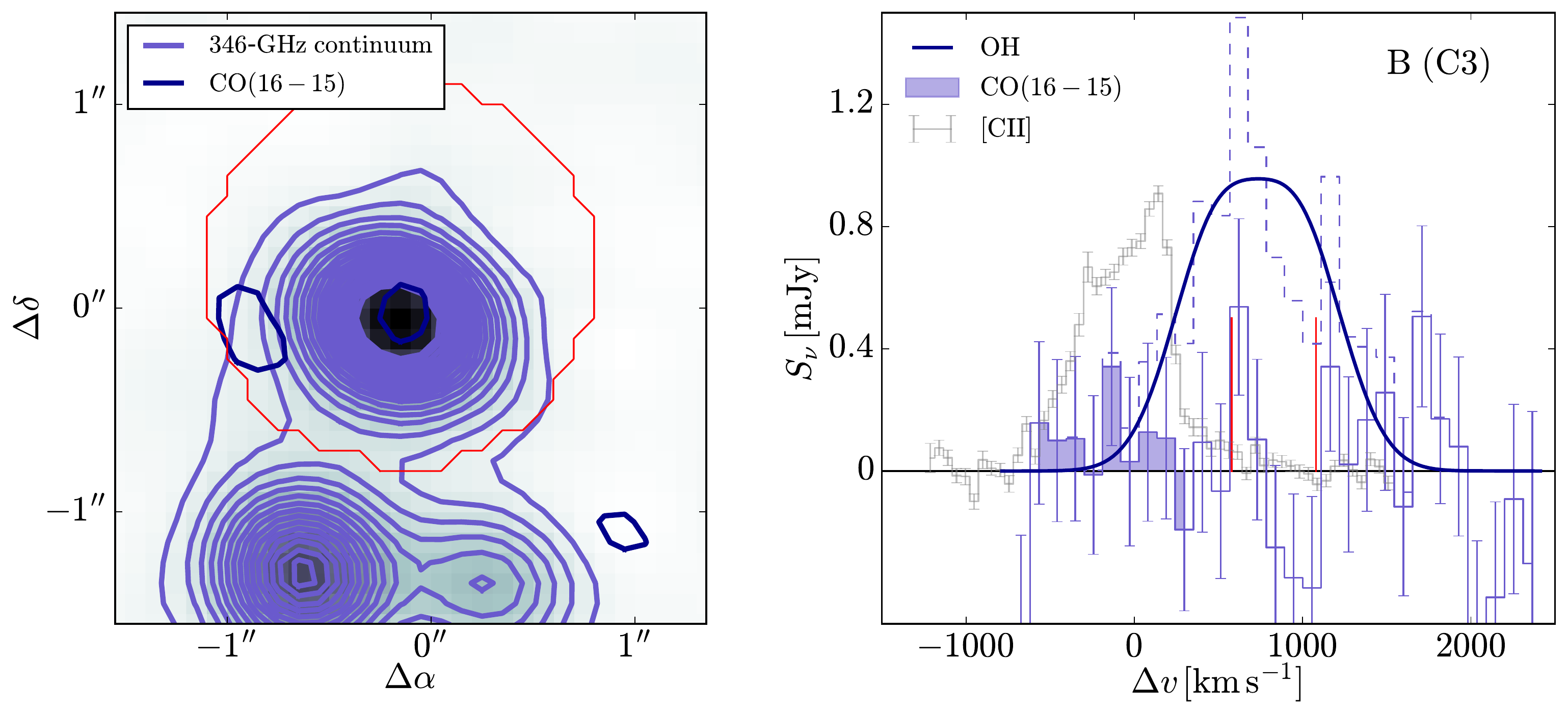}
 \includegraphics[width=0.48\linewidth]{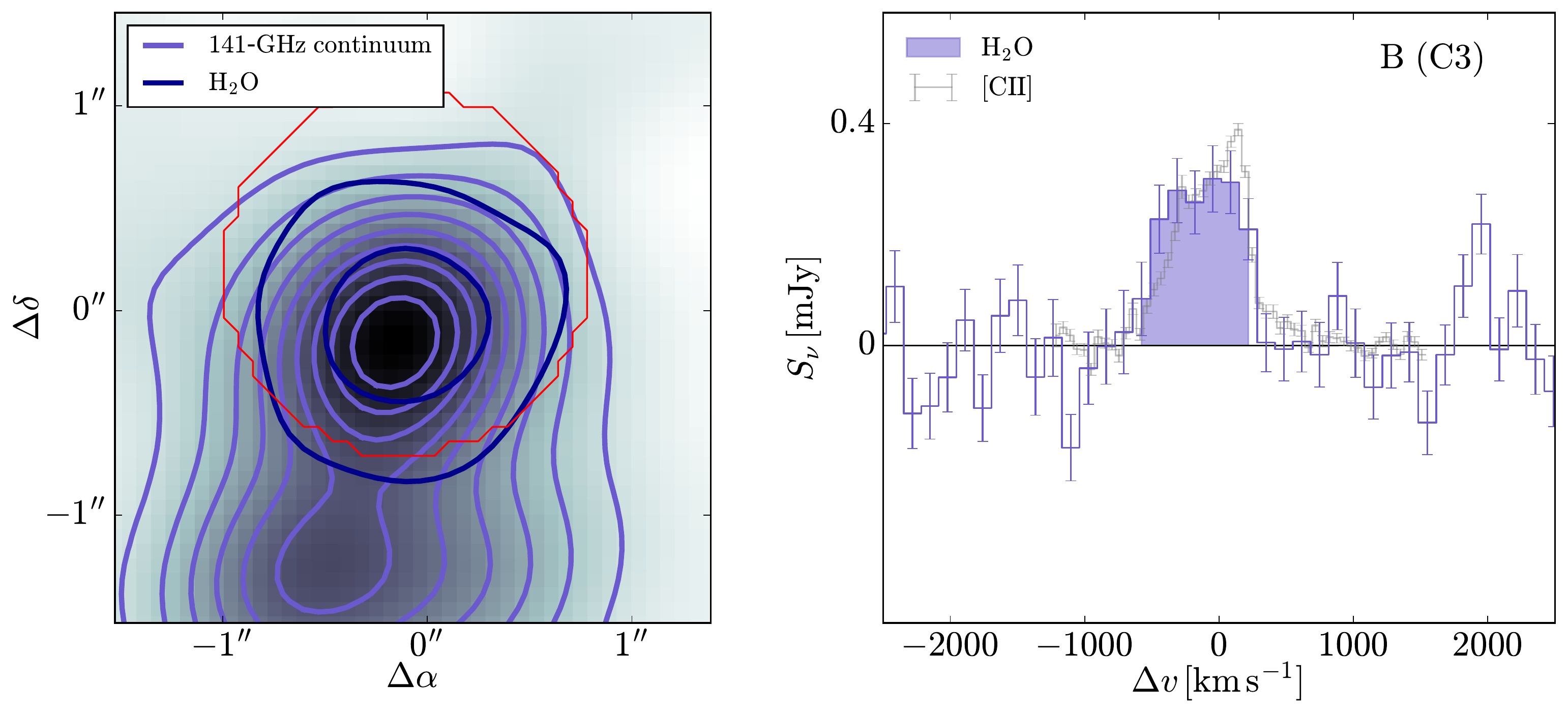}
\caption{Cutouts and spectra of CO(7--6), CO(11--10),  CO(16--15), and H$_2$O line emission for galaxy B. The cutouts in each panel show continuum emission (obtained by averaging over all line-free channels) and line emission (obtained by averaging over all channels where the line is expected -- see Section \ref{sec:data} for details), with contours starting at 2$\sigma$ and increasing in steps of 3$\sigma$. Apertures are shown as red circles, and used to obtain the spectra shown in the right panels. In each spectrum plot, we show the \CII\ profiles from \citet{Hill20}, scaled to the expected frequency of the given line, and arbitrarily normalized. The shaded regions show the integration ranges (set to be $\pm$2$\sigma$ about the \CII\ line -- see Section \ref{sec:data}) used to obtain line strengths. The CO(7--6) line is blended with the [CI](2--1) line, and the expected central frequency (or for G, two central frequencies as the \CII\ profile has two components) of the [CI](2--1) is marked in red. The [CI](2--1) line is fit by a Gaussian profile and subtracted, and the original spectra are shown by the dashed lines. Similarly, the CO(16--15) line is blended with the OH doublet, and we mark the mean frequency of each OH line in red (corresponsing to two frequencies for B and C, and four frequencies for G). The OH doublet is fit by a scaled \CII\ profile (see Section \ref{sec:data}) and subtracted, and the original spectra are shown by the dashed lines.
 }  
    \label{fig:line_cutouts1}
\end{figure*}

\begin{figure*}
    \centering
 \includegraphics[width=0.48\linewidth]{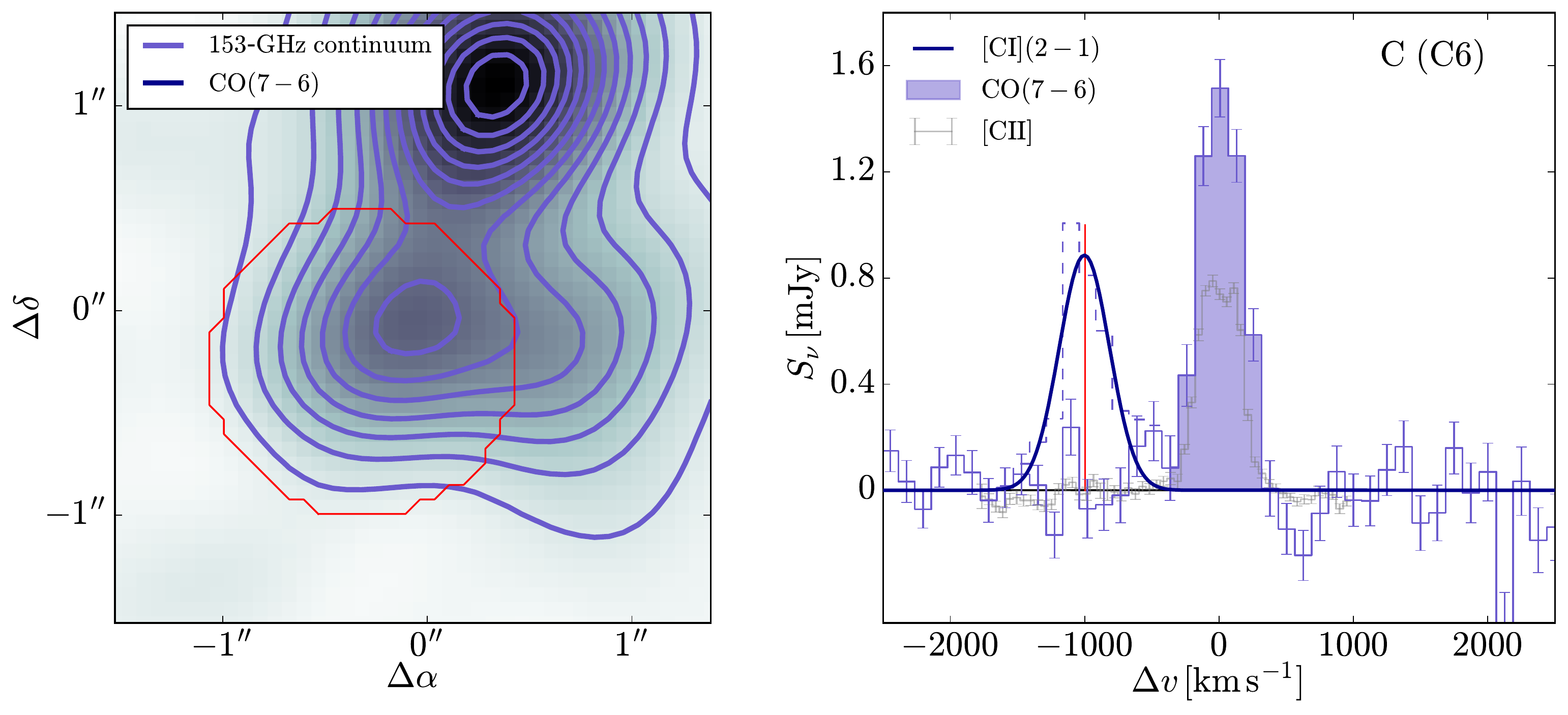}
 \includegraphics[width=0.48\linewidth]{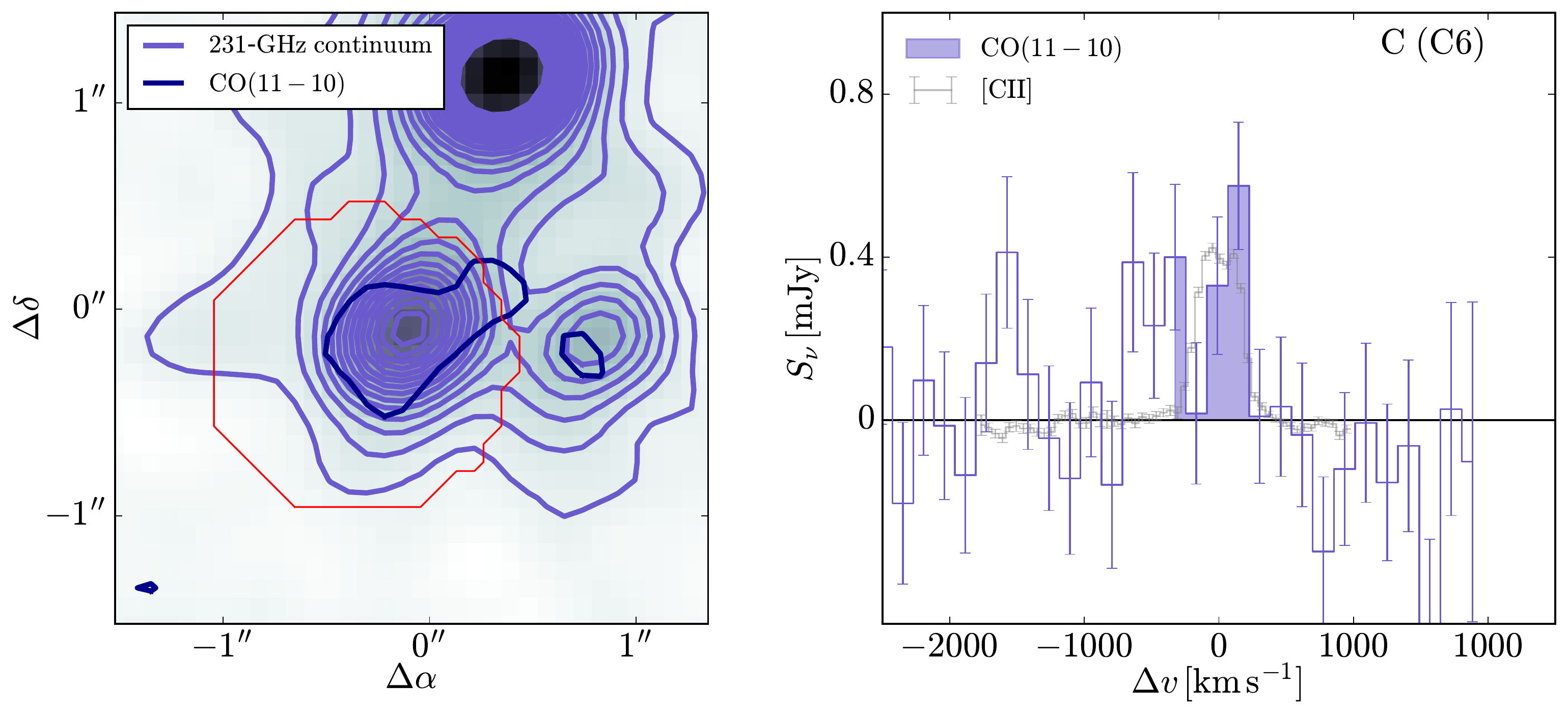}
 \includegraphics[width=0.48\linewidth]{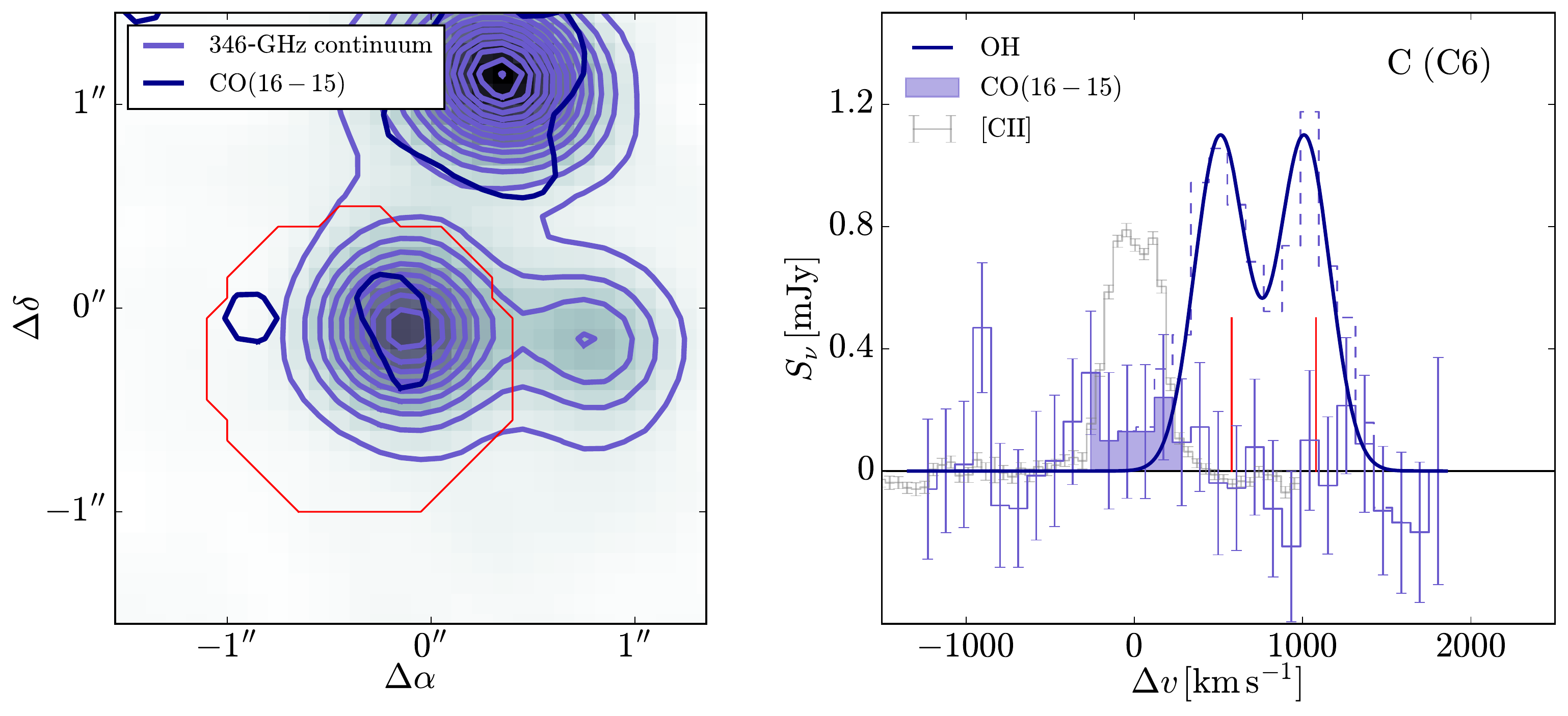}
\includegraphics[width=0.48\linewidth]{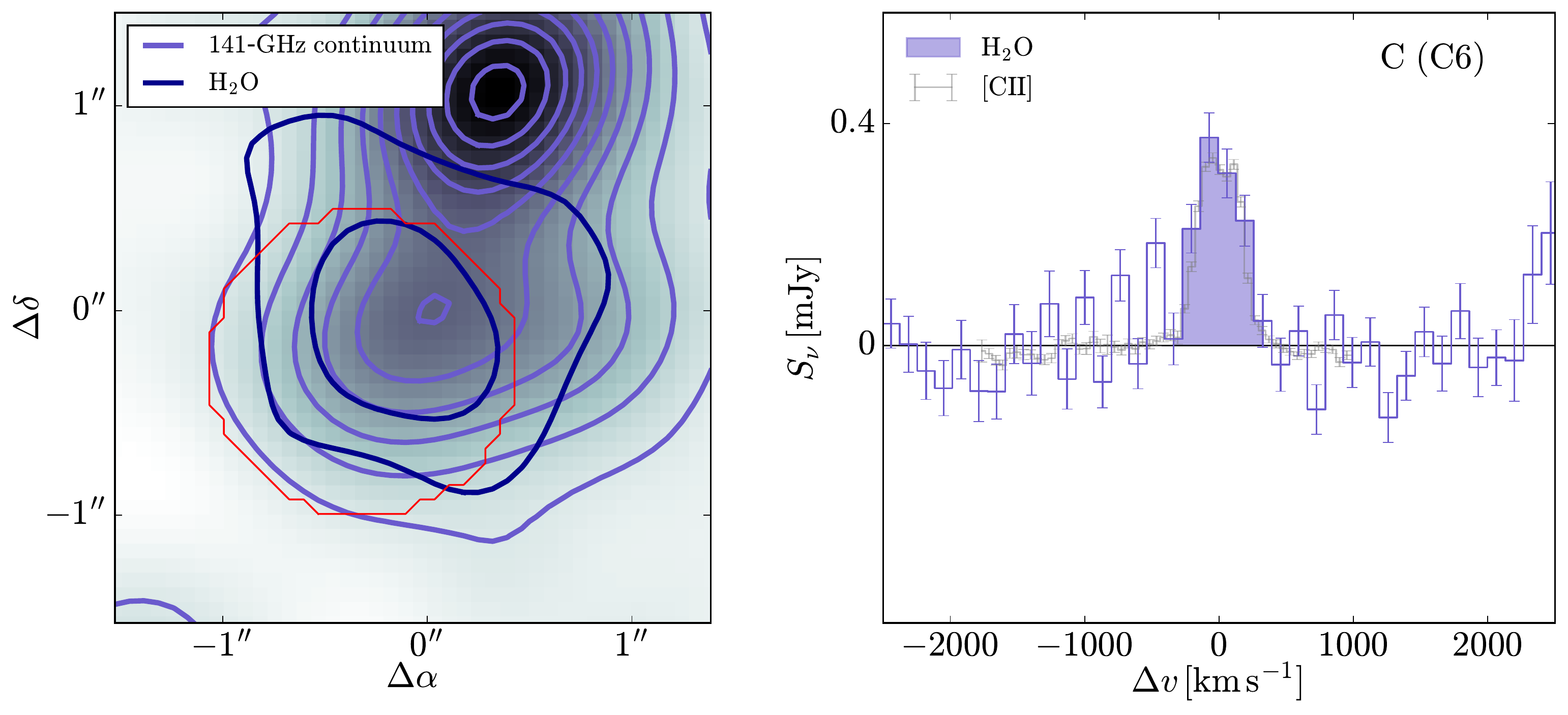}
\caption{Same as Fig.~\ref{fig:line_cutouts1} but for galaxy C.
 }  
    \label{fig:line_cutouts2}
\end{figure*}

\begin{figure*}
    \centering
 \includegraphics[width=0.48\linewidth]{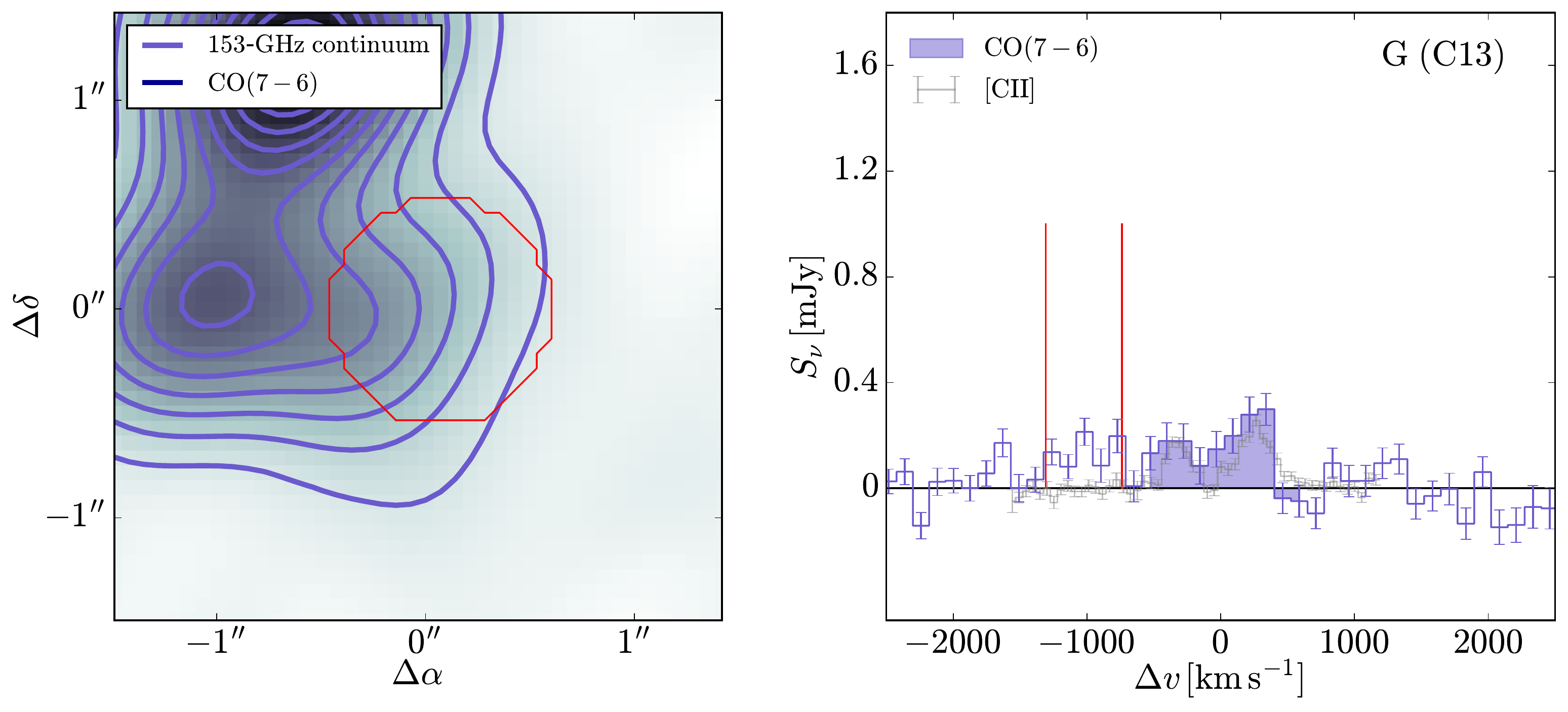}
 \includegraphics[width=0.48\linewidth]{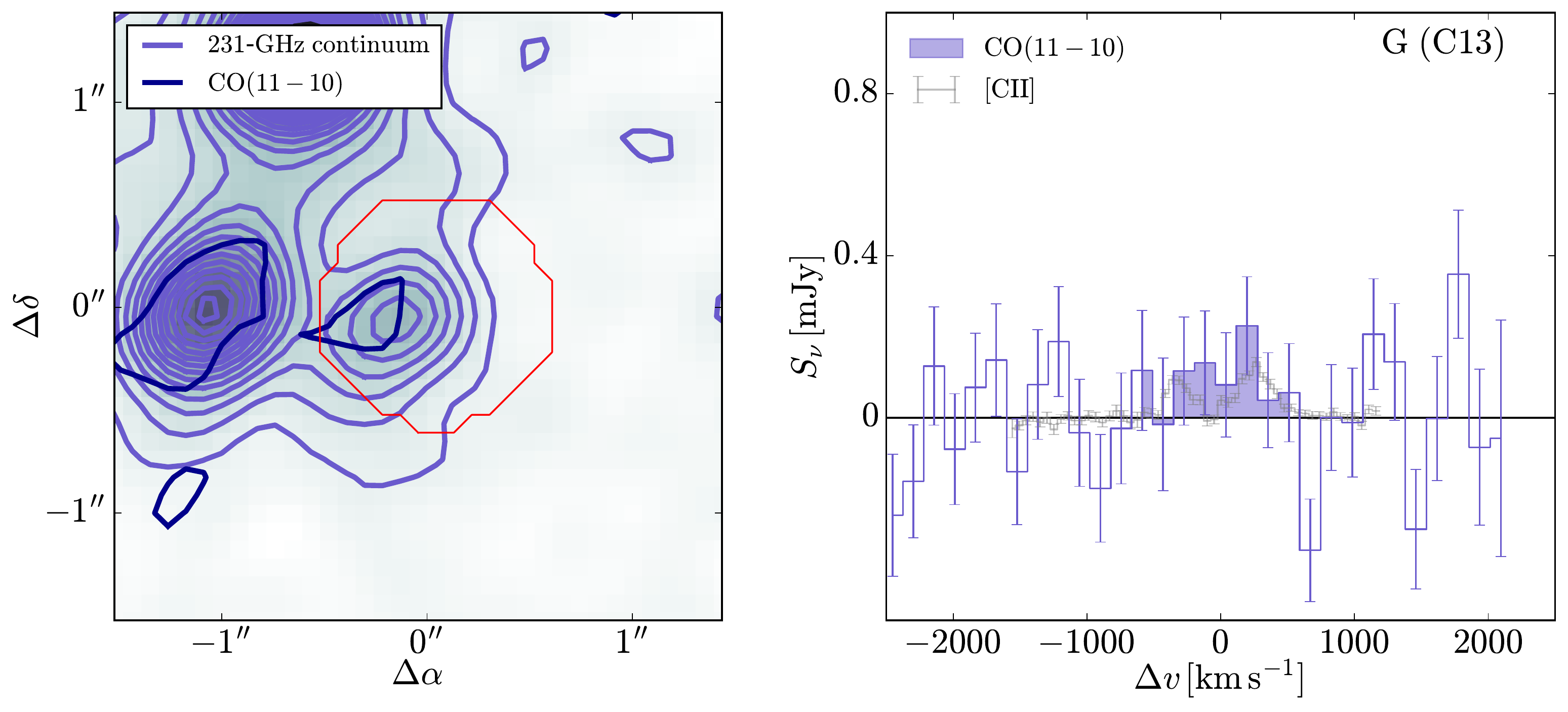}
 \includegraphics[width=0.48\linewidth]{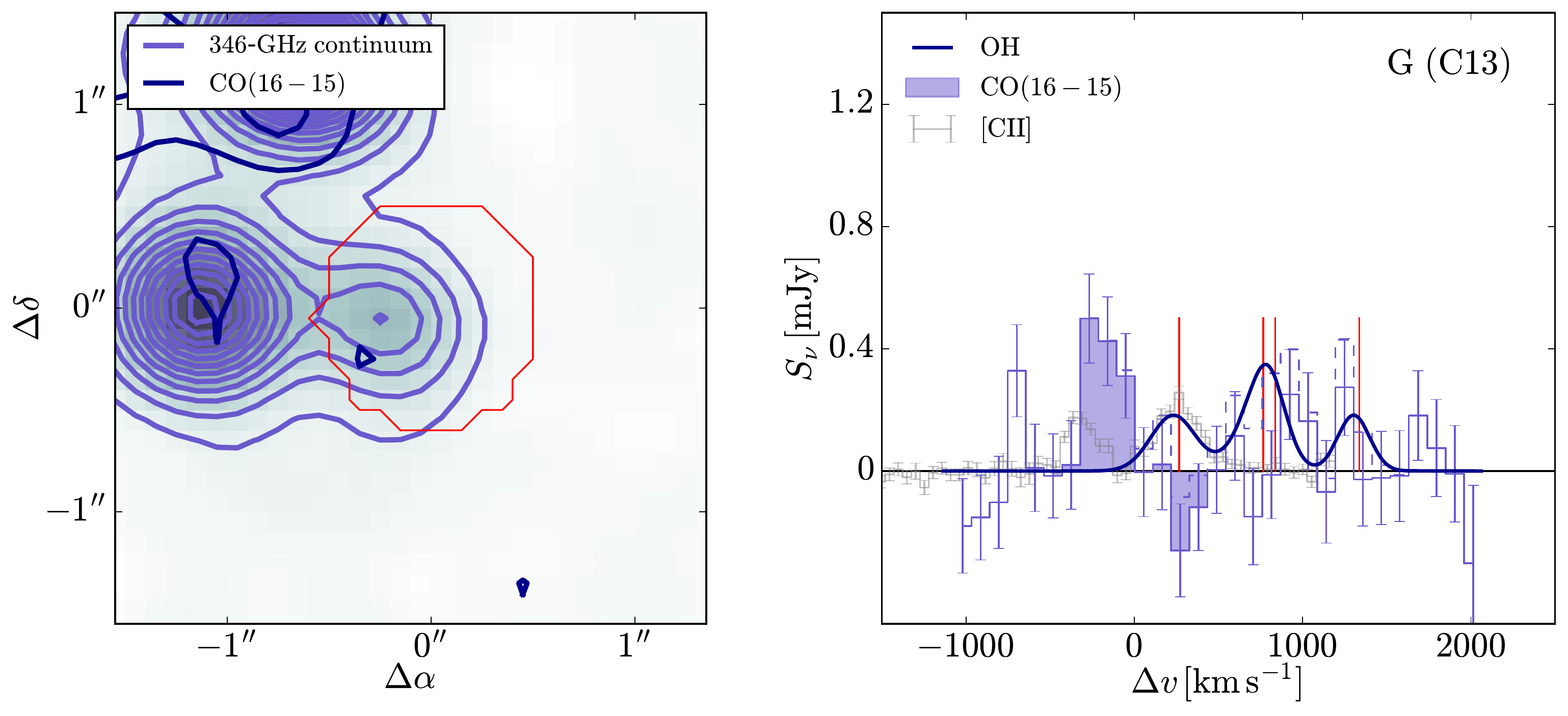}
 \includegraphics[width=0.48\linewidth]{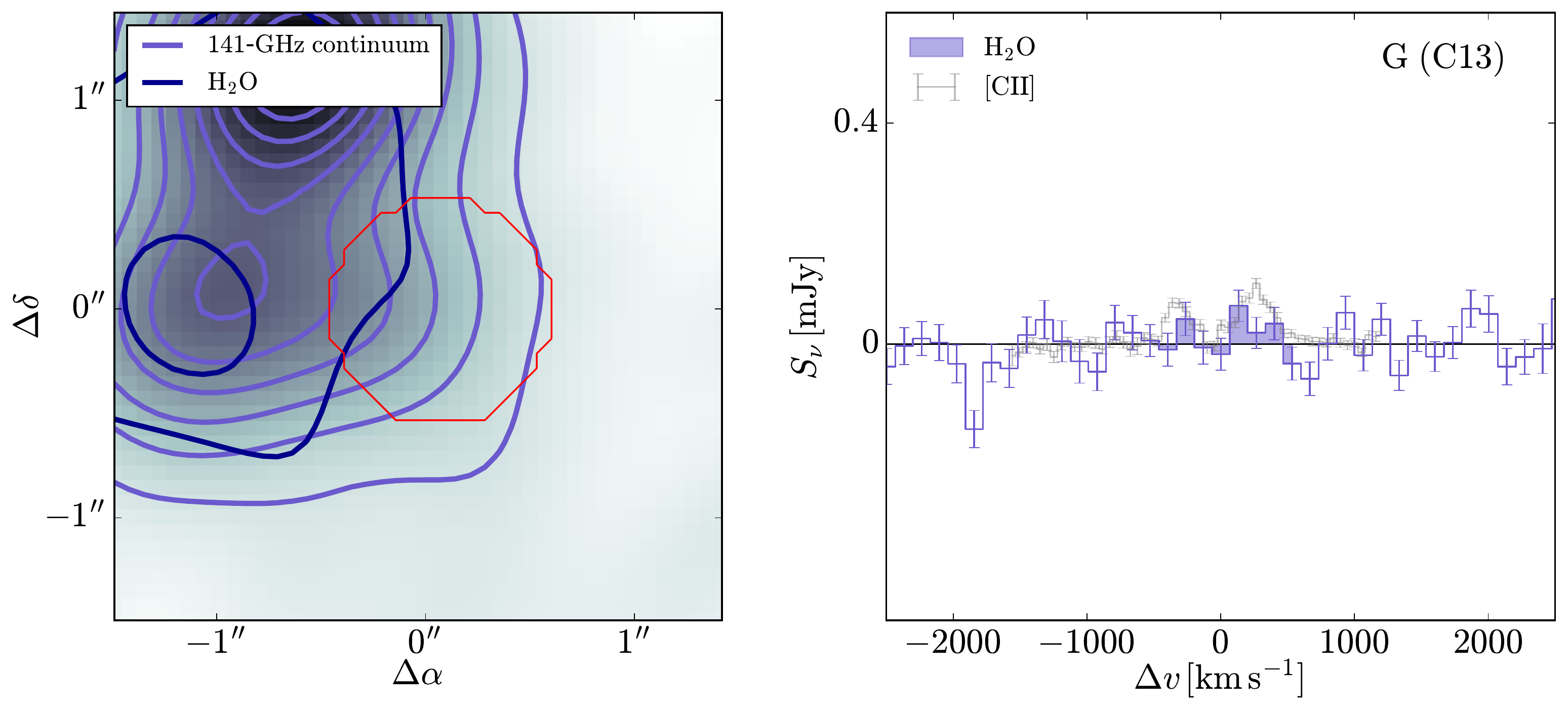}
\caption{Same as Fig.~\ref{fig:line_cutouts1} but for galaxy G.
 }  
    \label{fig:line_cutouts3}
\end{figure*}

In this appendix and Figs.~\ref{fig:line_cutouts1} -- \ref{fig:line_cutouts3}, we show the CO $J\,{=}\,$7, 11, and 16 lines for the central B, C, and G sources, whose line strengths are plotted in the SLED diagram (Fig.~\ref{fig:sled}). We also show the H$_2$O lines that are used to compare with the FIR luminosity estimates from \citet{Hill20}. 

In order to measure line strengths, the bright and well-detected \CII\ lines provided in \citet{Hill20} were used as a template. These \CII\ lines were fit by single and double Gaussian profiles, and we selected the integration range by scaling the \CII\ profile to the rest frequency of the line of interest and then summing channels between $-2\sigma$ and $2\sigma$ (where $\sigma$ is the standard deviation of the best-fitting linewidth), or for cases where two Gaussians were a better fit, from $-2\sigma_{\rm L}$ to $+2\sigma_{\rm R}$, where $\sigma_{\rm L}$ and $\sigma_{\rm R}$ are from the left and right Gaussian fits, respectively. 

The CO(7--6) line is blended with the \CI(2--1) line, and the CO(16--15) line is blended with the OH doublet, so these had to be fit and subtracted before integrating over the CO lines. For the former case, where both CO(7--6) and \CI(2--1) are both well-detected in B and C, we simultaneously fit single Gaussian profiles at the locations of the two lines, then subtract the best-fit \CI(2--1) model from the spectrum, and sum over the relevant channels as described above (then vice-versa to obtain \CI(2--1) line strengths). For source G, we do not see any strong line features around the expected \CI(2--1) frequency, so we simply sum over the CO(7--6) and \CI(2--1) channels in the raw spectrum. The CO(16--15) line is not well-detected for any sources but the OH doublet is, so we fit a Gaussian to these OH lines and subtract the models before summing over the CO(16--15) channels. In the fit we force the amplitude of each doublet component to be equal, and we fix the width of each doublet component to be equal to the width of the \CII\ line (described in Sec.~\ref{sec:results:3.3}; see Table \ref{table:properties}). Since the profile for G is two Gaussians, we include an additional OH doublet component of equal amplitude and fixed frequency separation/width to match the \CII\ profile. This leaves two free parameters in all fits: the frequency of the first doublet, and the amplitude of the all the components. In Fig.~\ref{fig:line_cutouts1} we can see that for source B the two OH doublet components are blended with each other due to the large FWHM of the system, and for G the four components blend into three peaks.

Lastly, Band 4 and Band 6 continuum flux densities were estimated by averaging over all line-free channels in the original (non continuum-subtracted) data cubes (again using the \CII\ line as a template). We combined channels from the lower and upper sideband of these observations, meaning they are at observed frequencies of 147 and 231\,GHz, respectively. Band 7 continuum flux densities (around the CO(16--15) and OH lines) are already provided in \citet{Hill20}.


\section{Appendix C} \label{sec:appendixC}

\begin{figure*}
    \centering
 \includegraphics[width=0.24\linewidth]{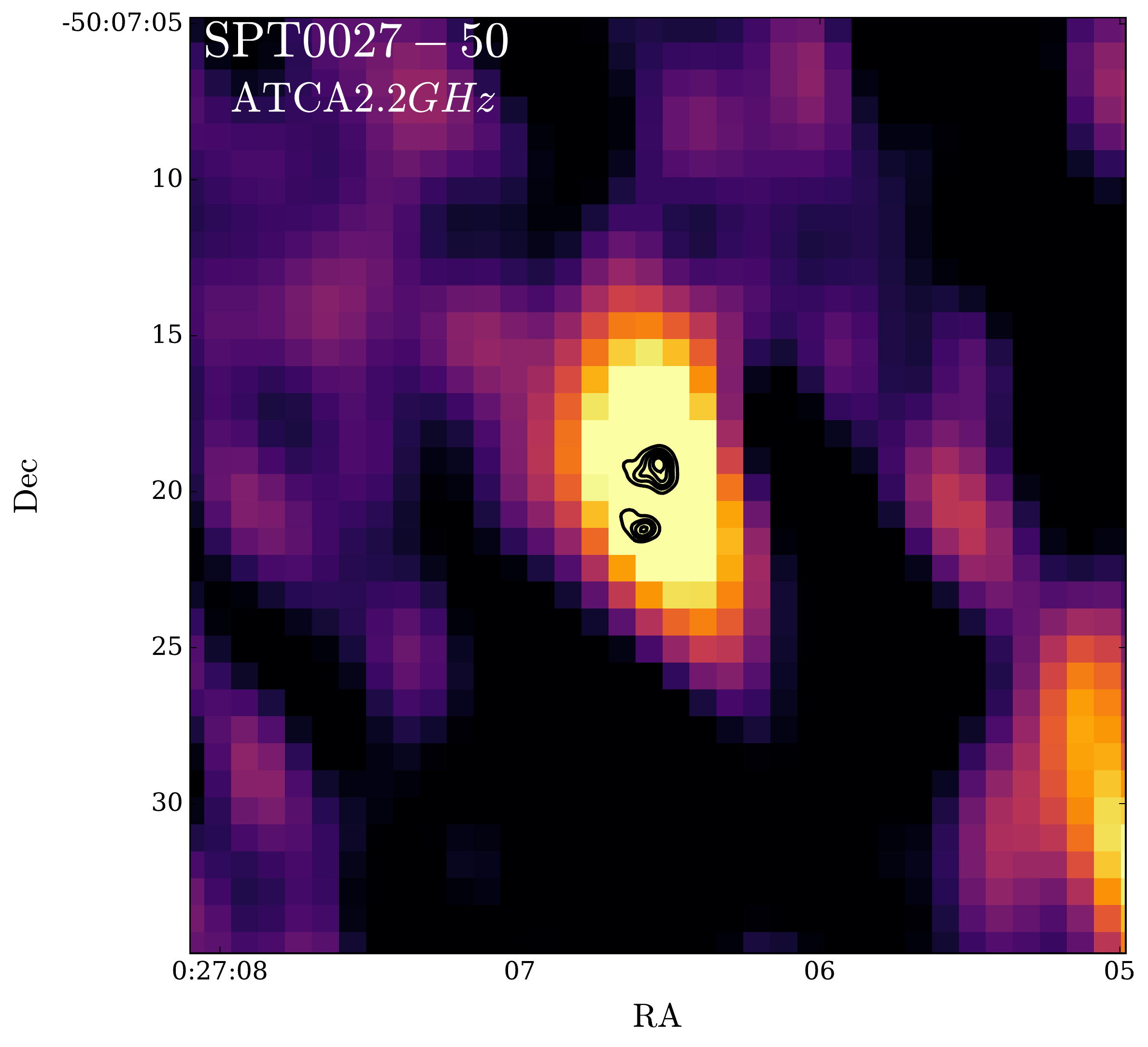}    
 \includegraphics[width=0.24\linewidth]{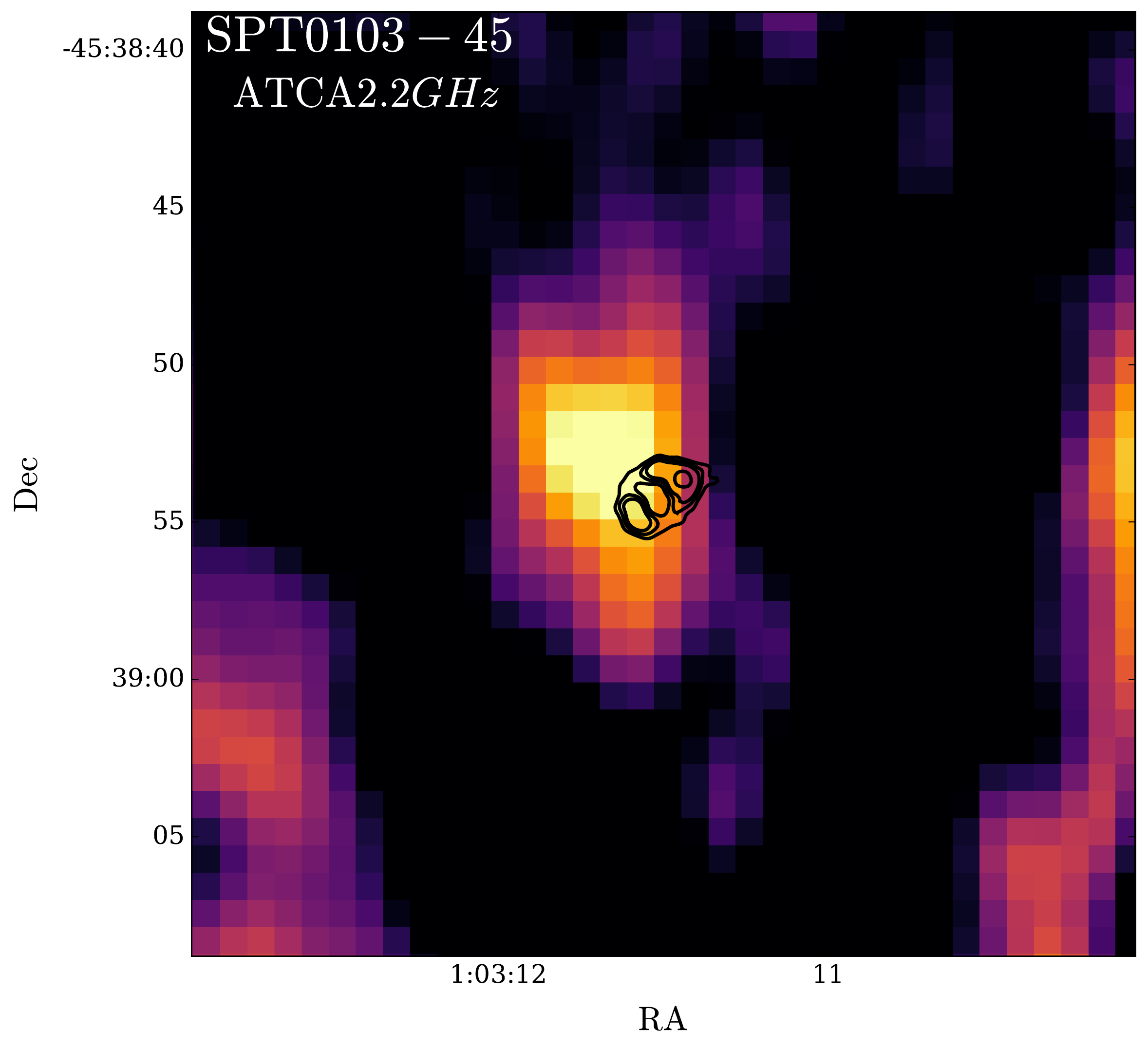}    
 \includegraphics[width=0.24\linewidth]{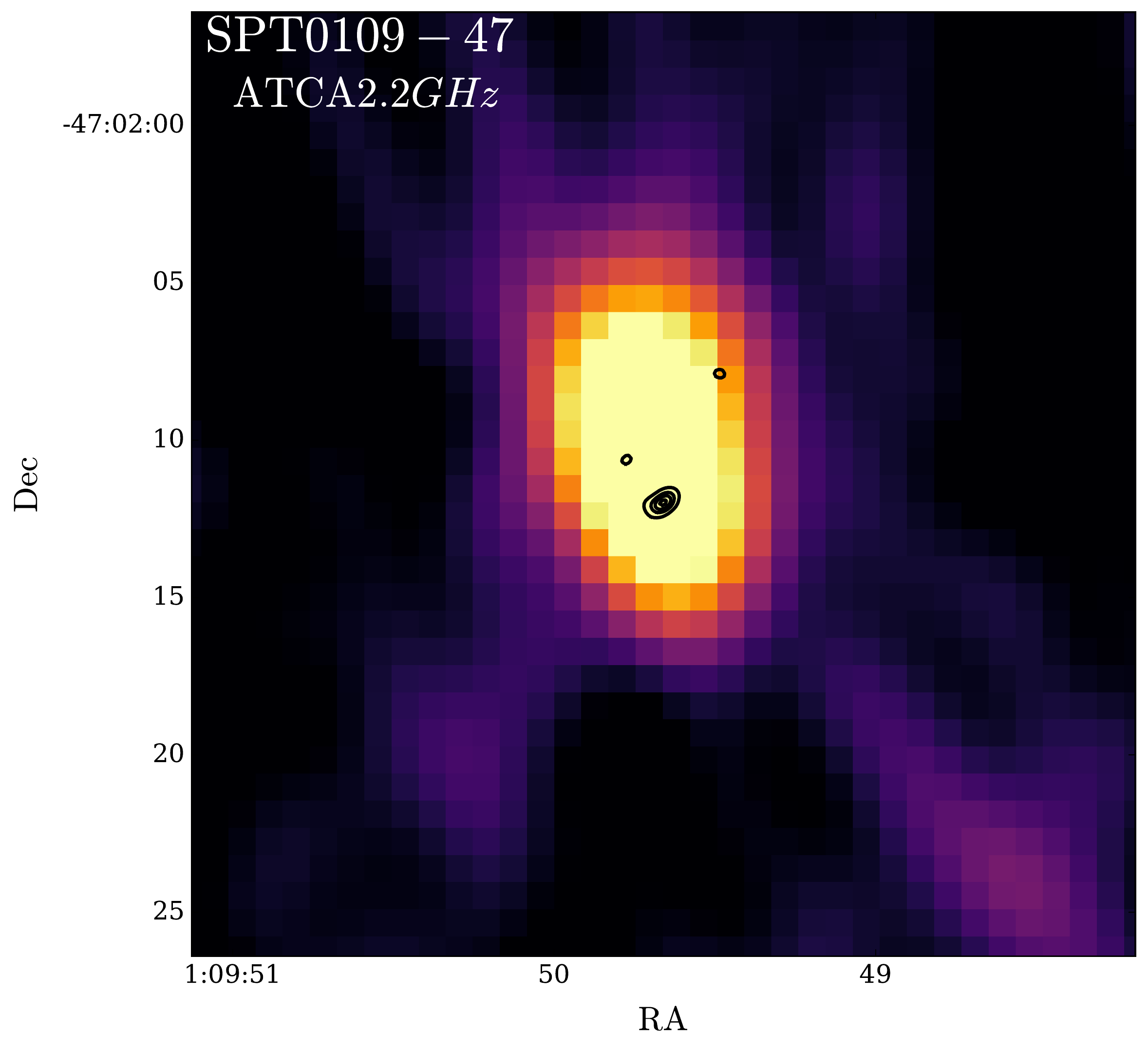}
 \includegraphics[width=0.24\linewidth]{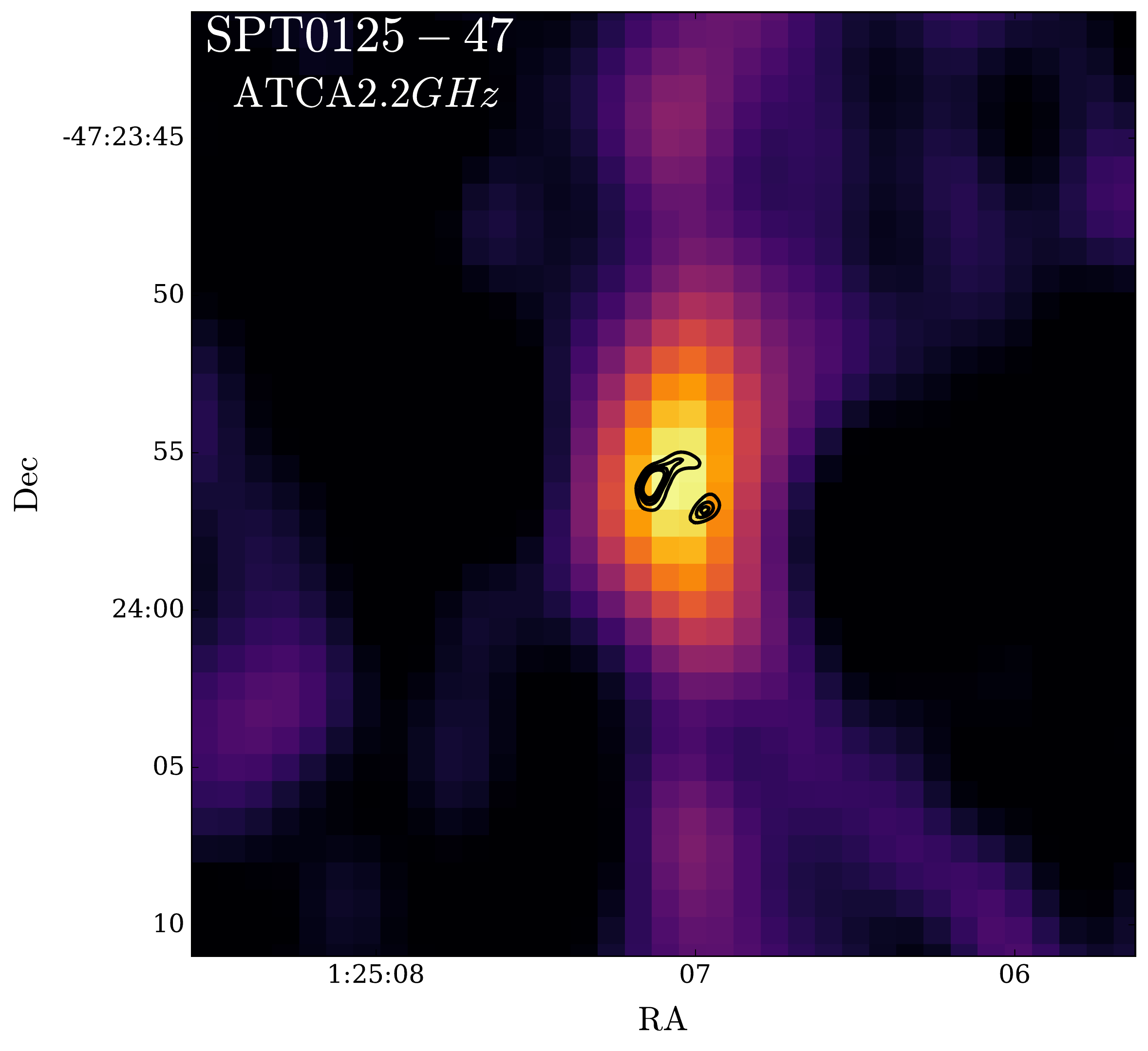}
 \includegraphics[width=0.24\linewidth]{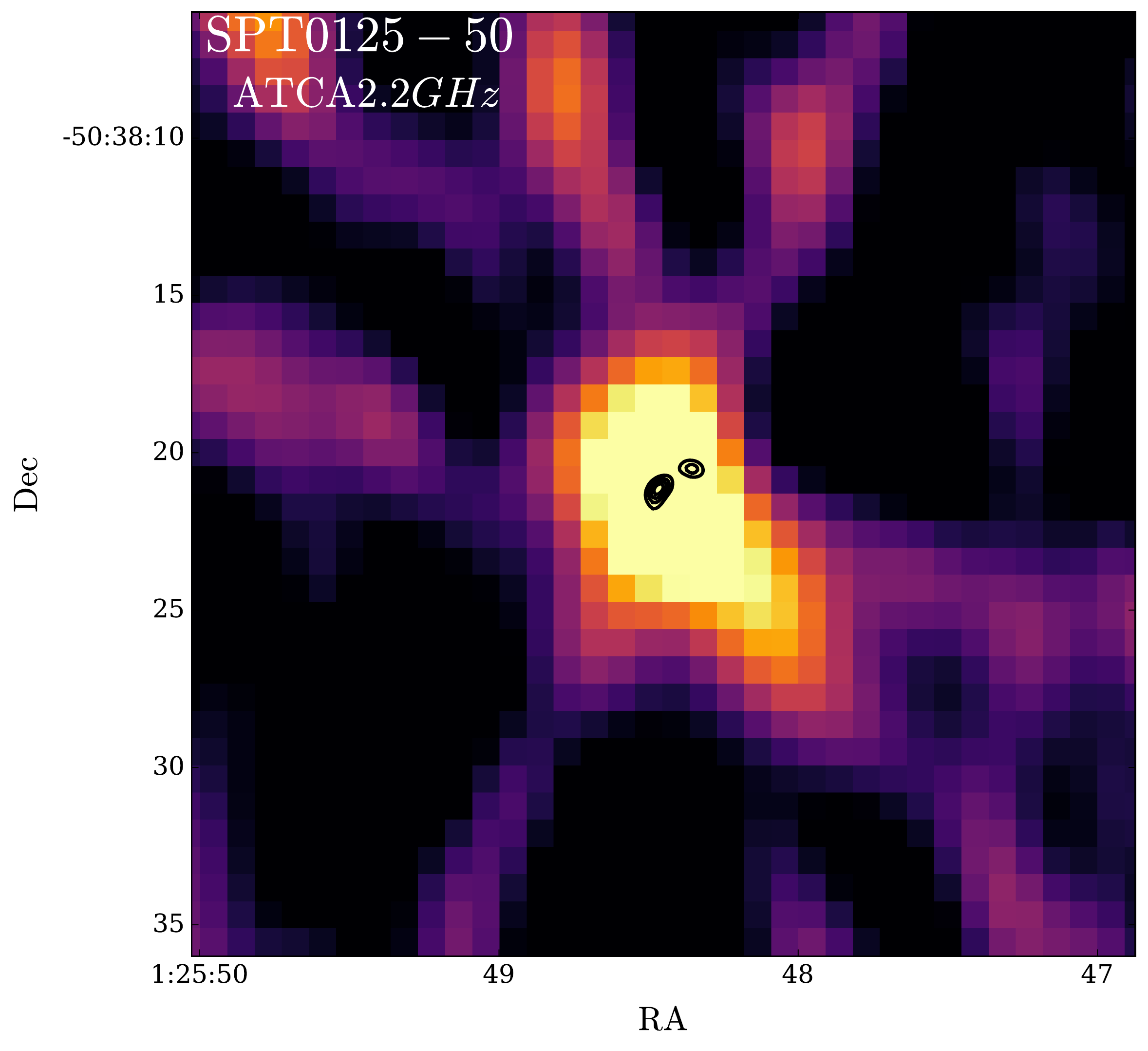}    
 \includegraphics[width=0.24\linewidth]{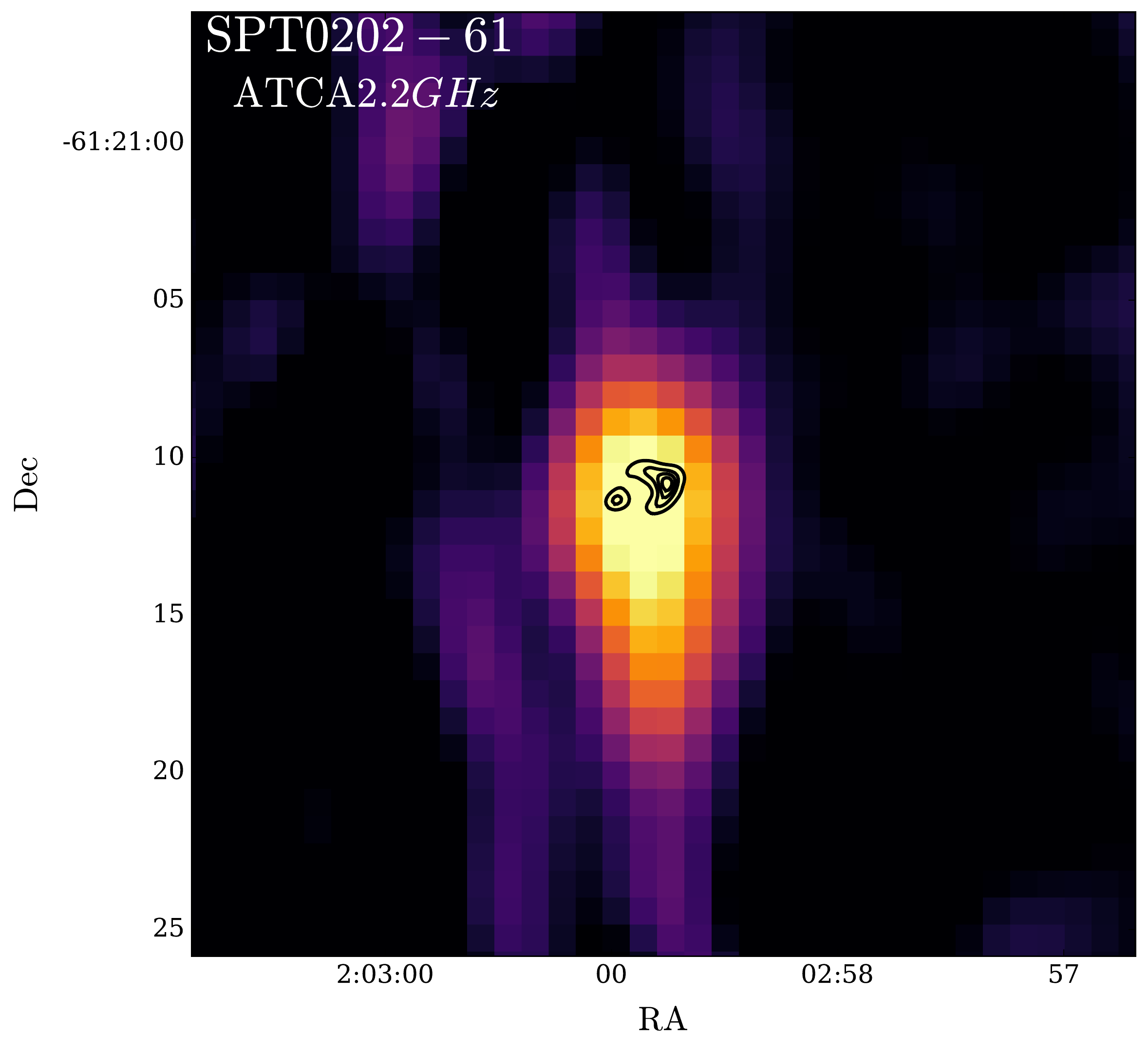}    
 \includegraphics[width=0.24\linewidth]{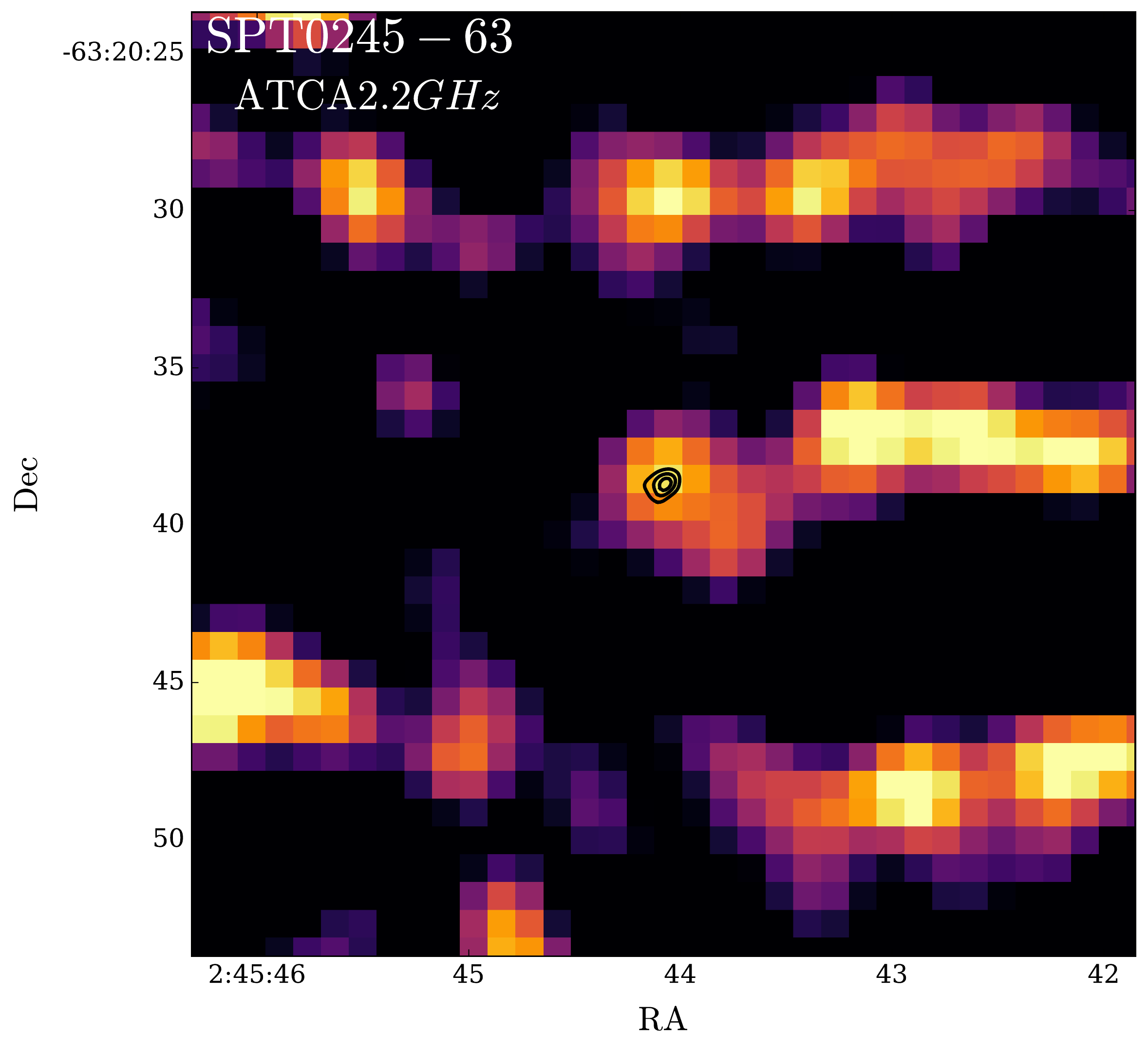}    
 \includegraphics[width=0.24\linewidth]{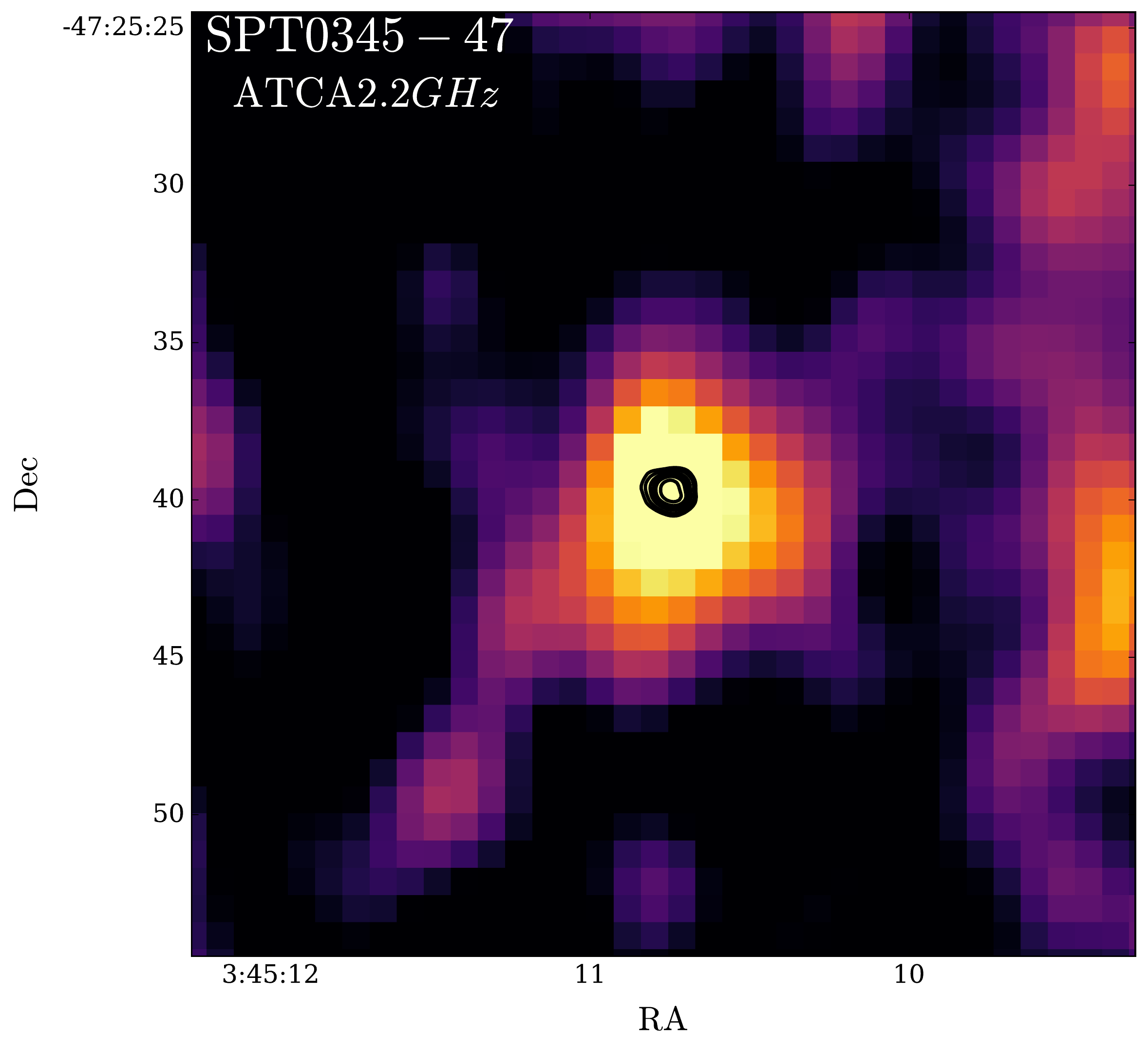}    
 \includegraphics[width=0.24\linewidth]{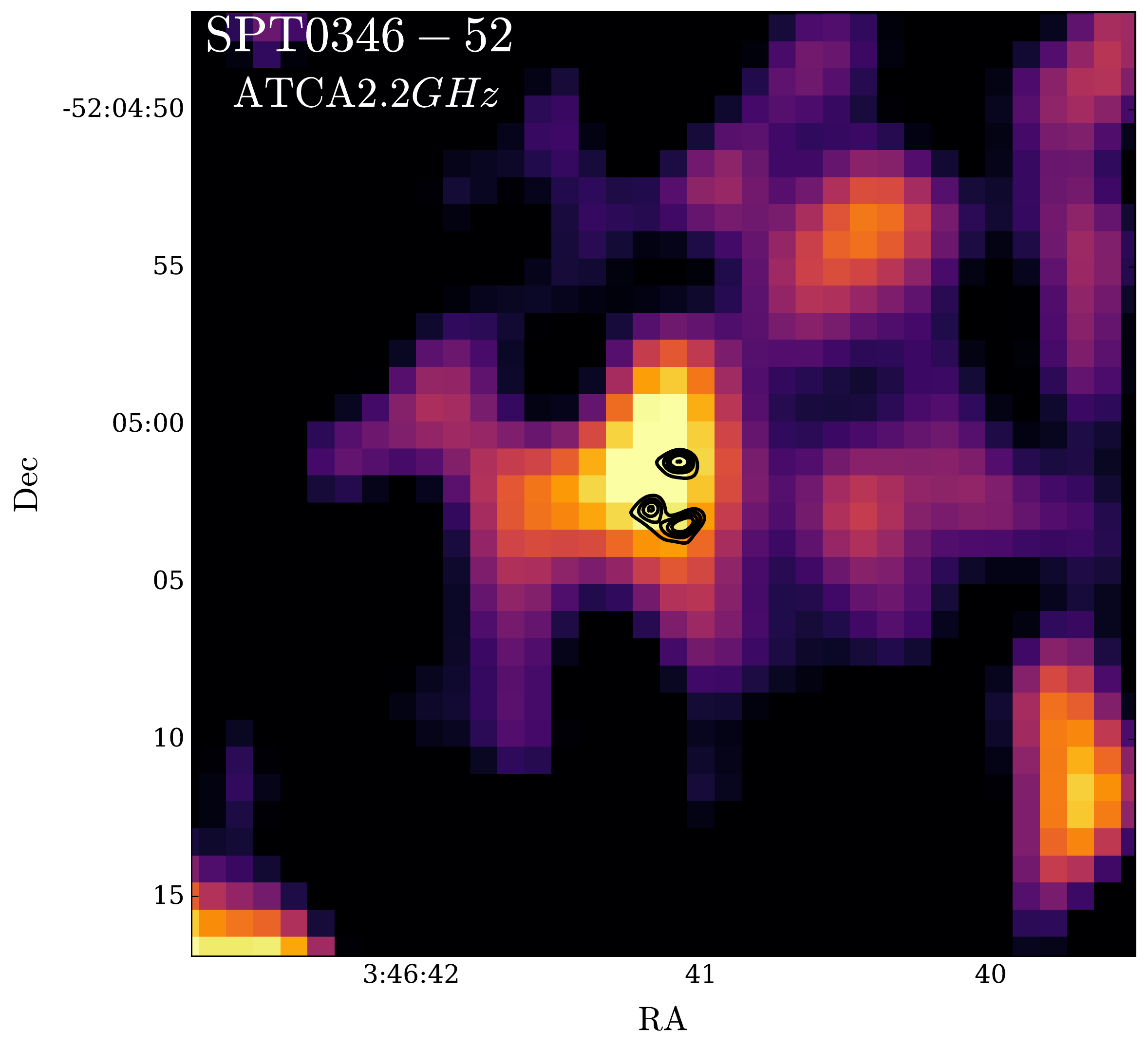}    
 \includegraphics[width=0.24\linewidth]{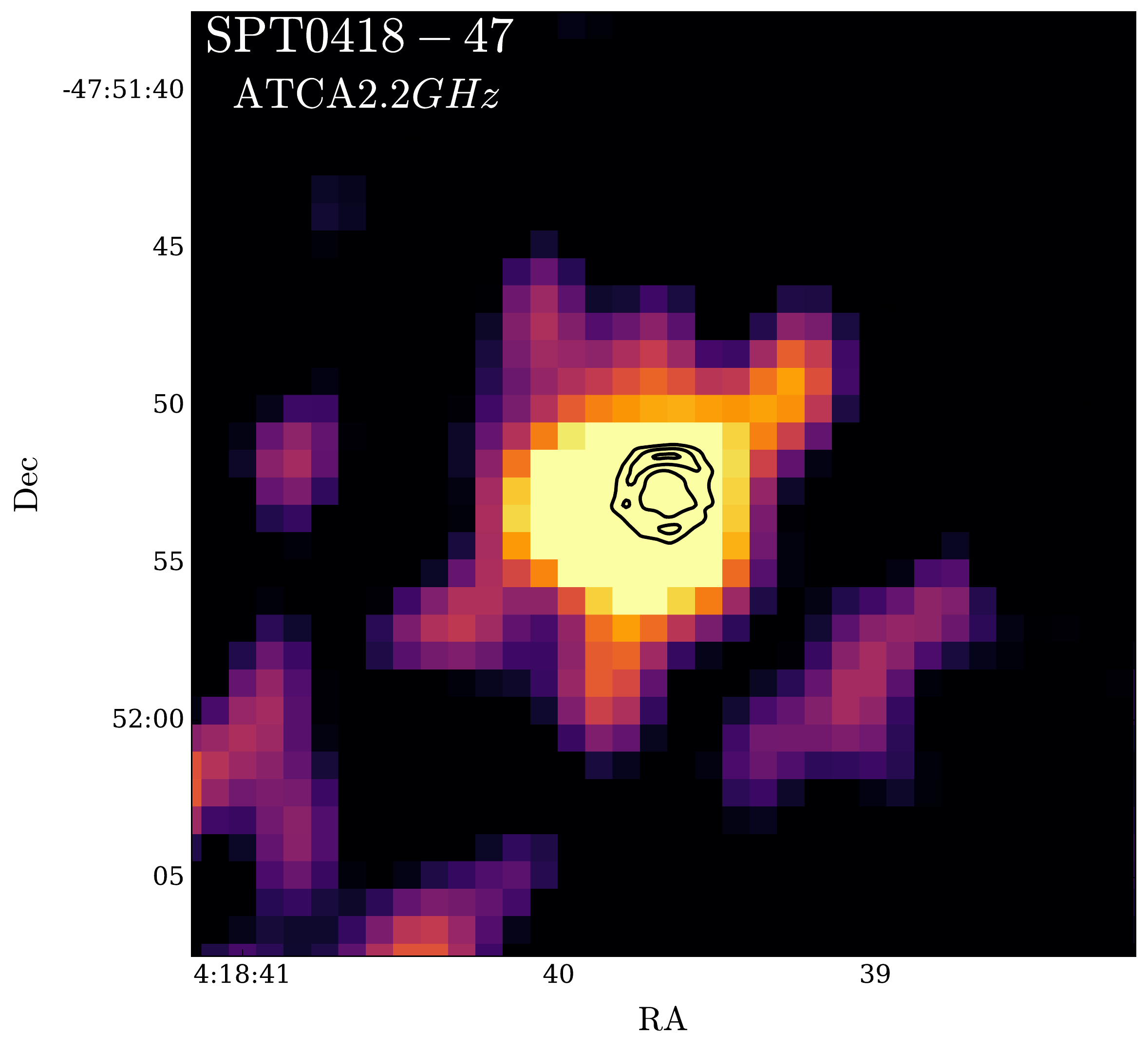}    
 \includegraphics[width=0.24\linewidth]{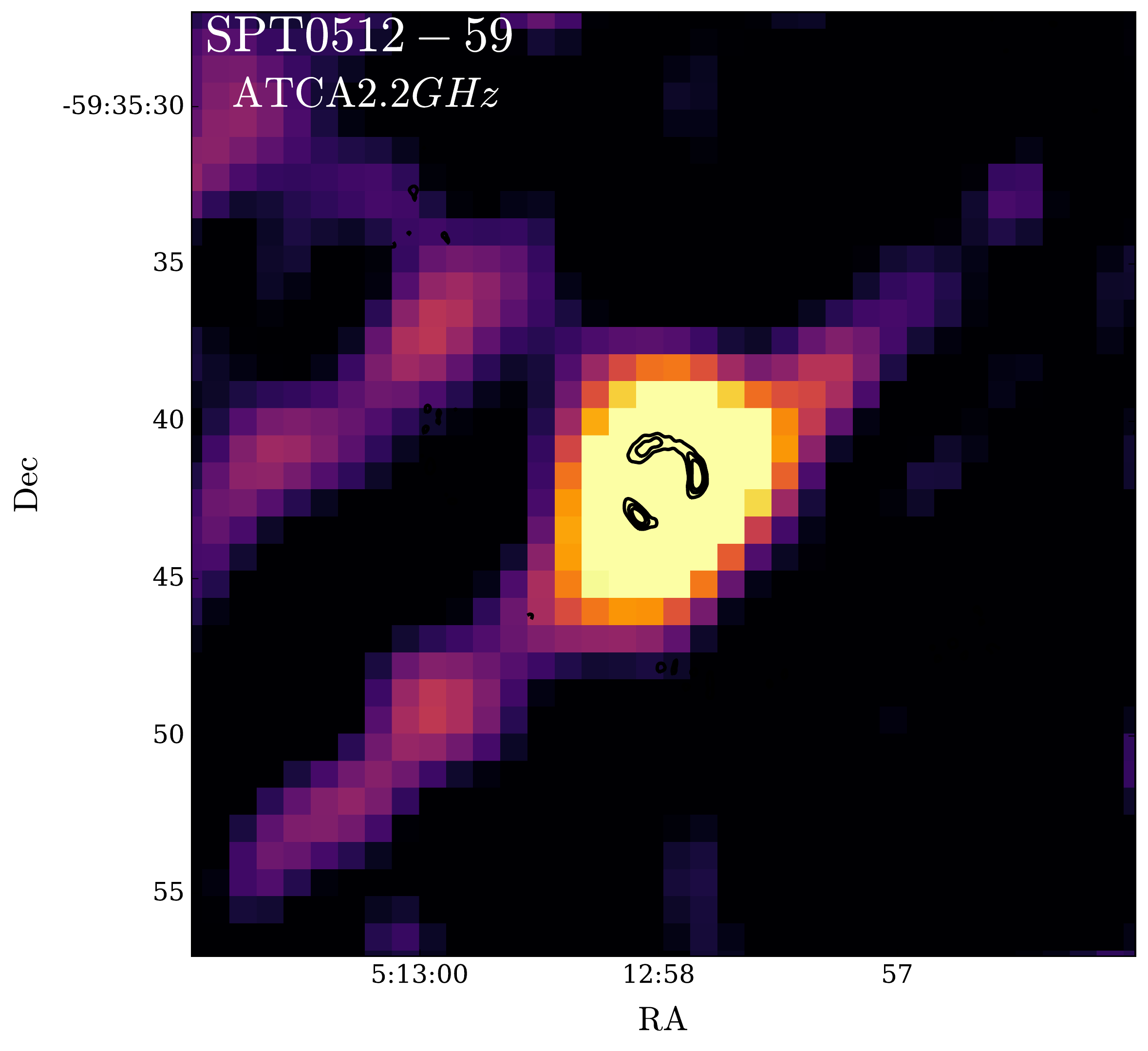}    
 \includegraphics[width=0.24\linewidth]{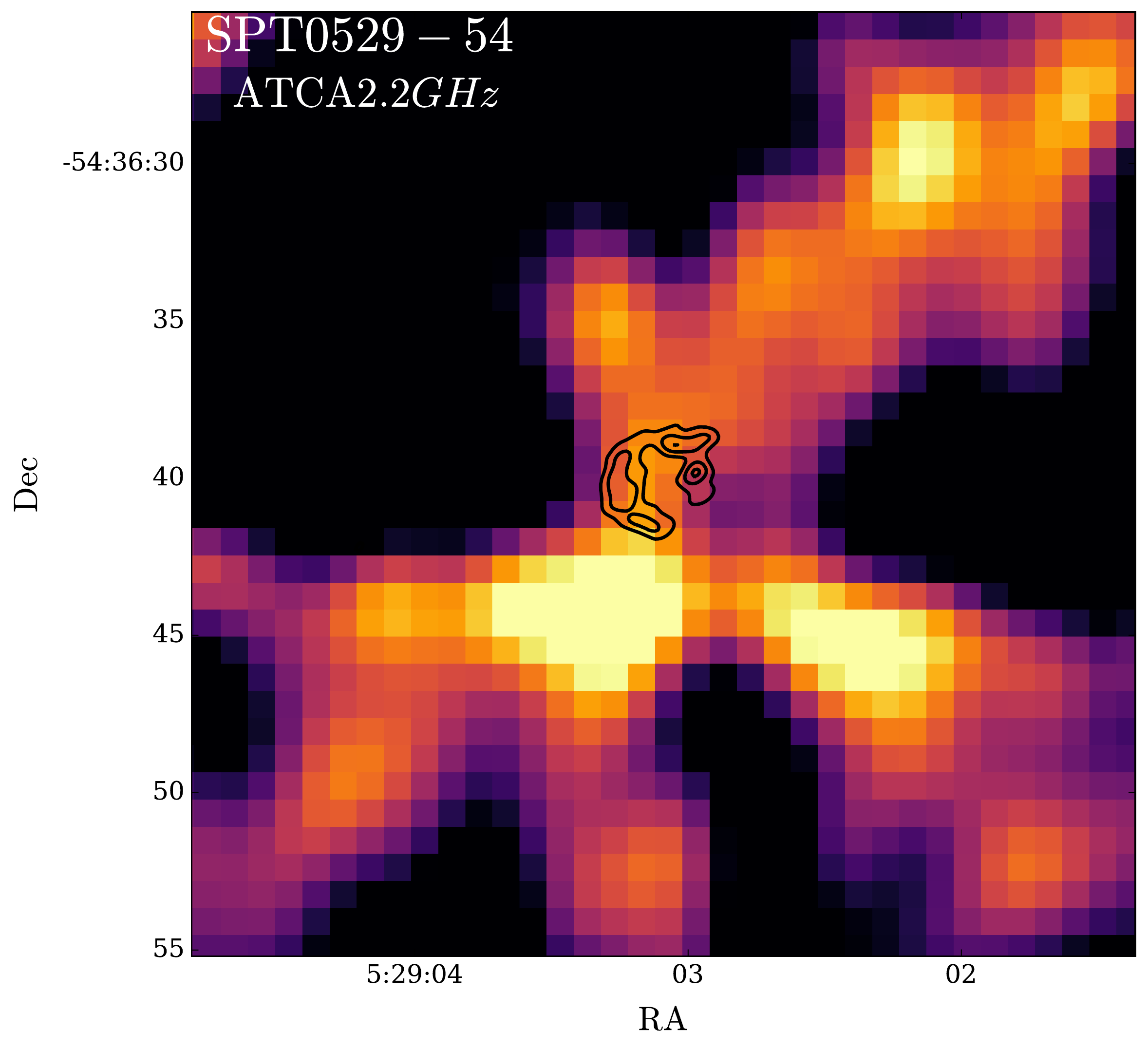}    
 \includegraphics[width=0.24\linewidth]{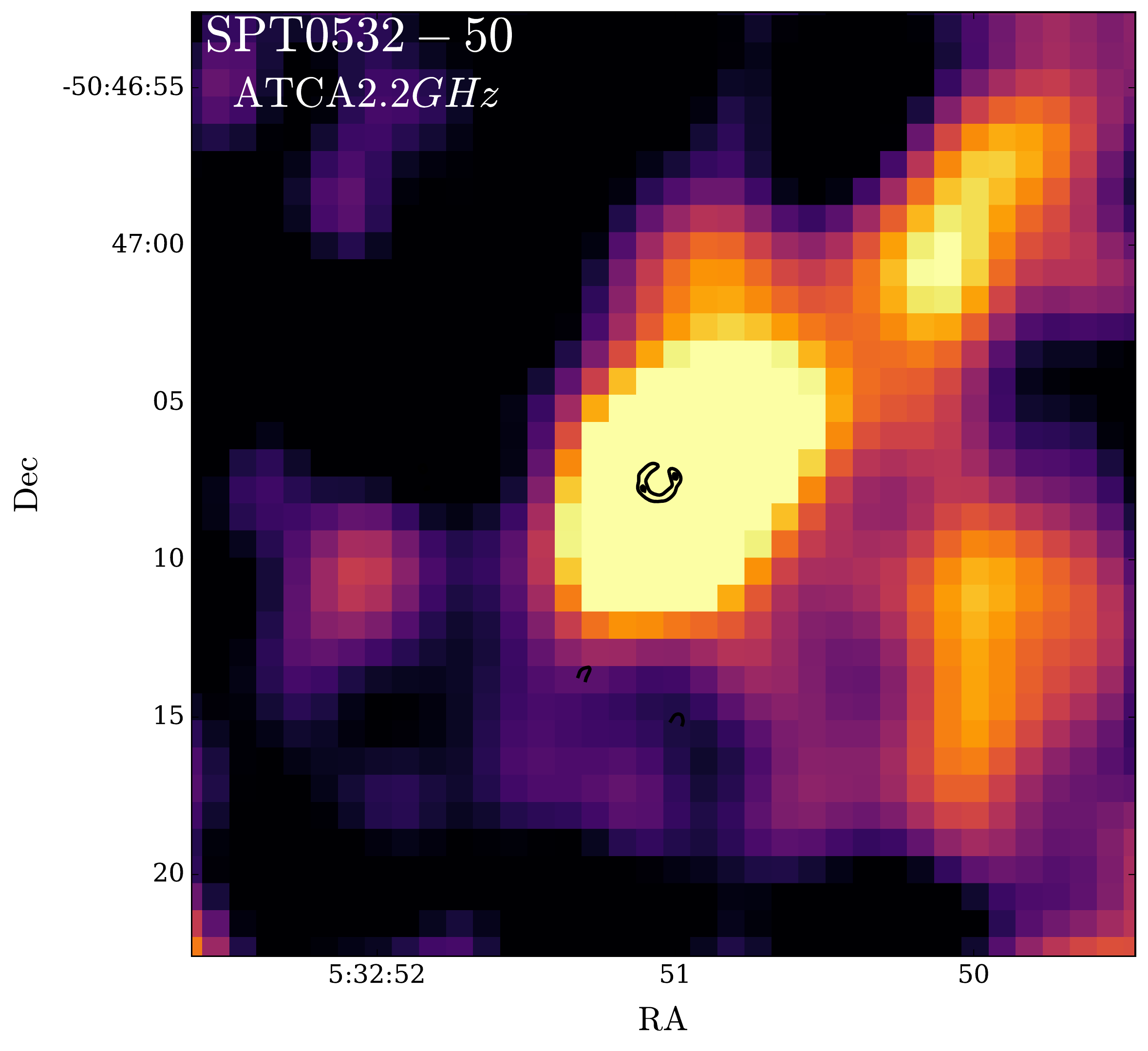}    
 \includegraphics[width=0.24\linewidth]{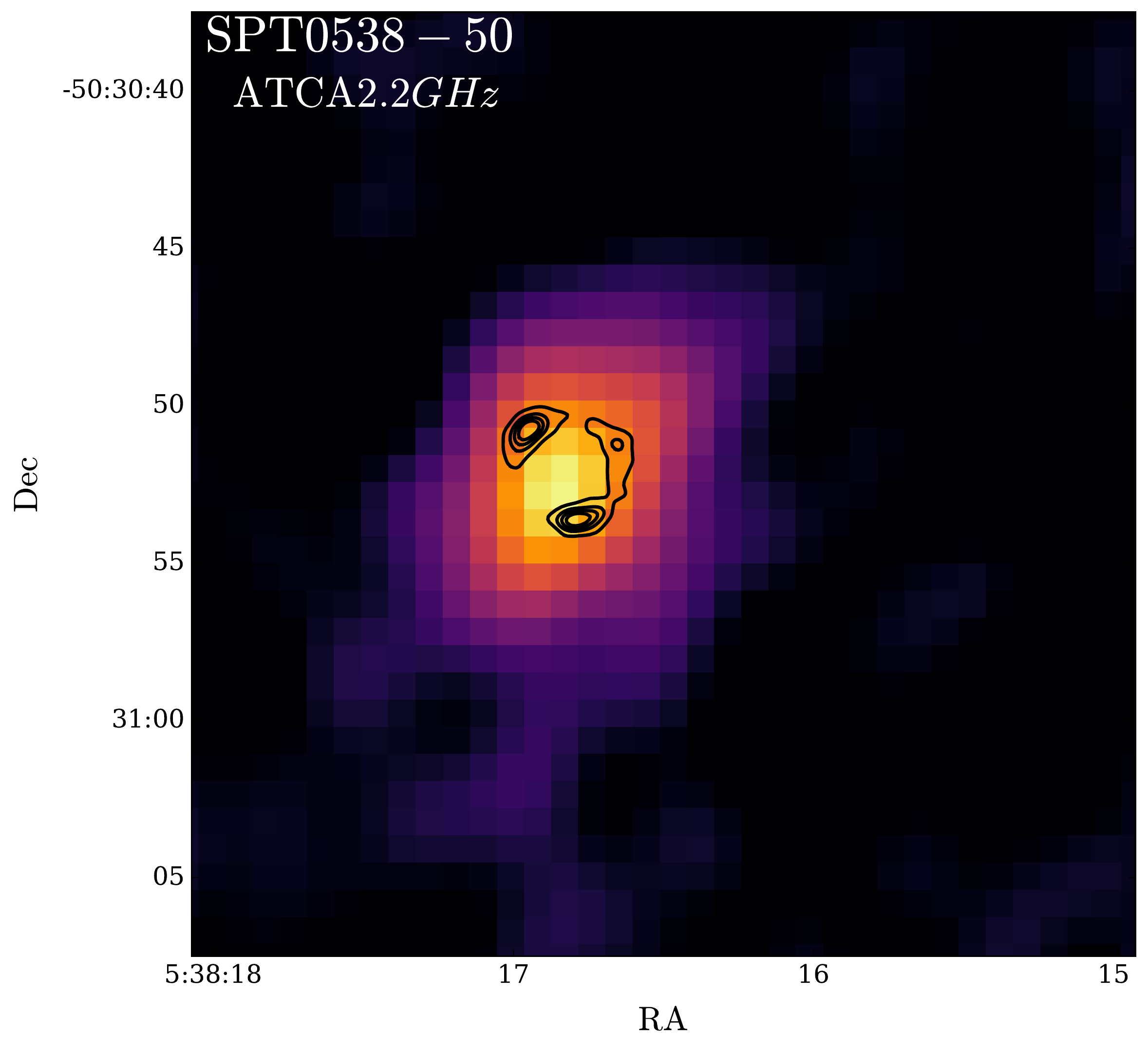}    
 \includegraphics[width=0.24\linewidth]{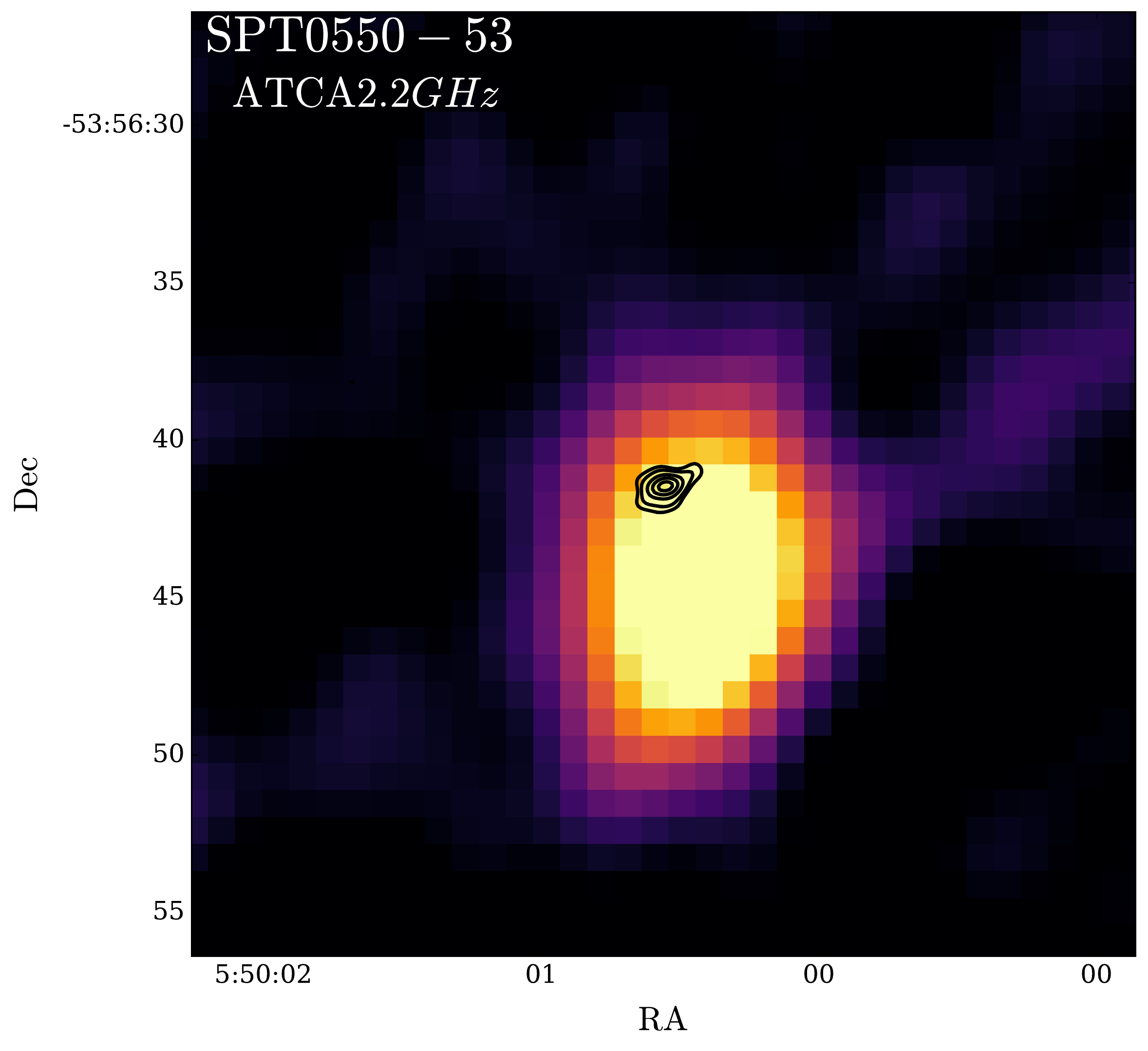}
  \includegraphics[width=0.24\linewidth]{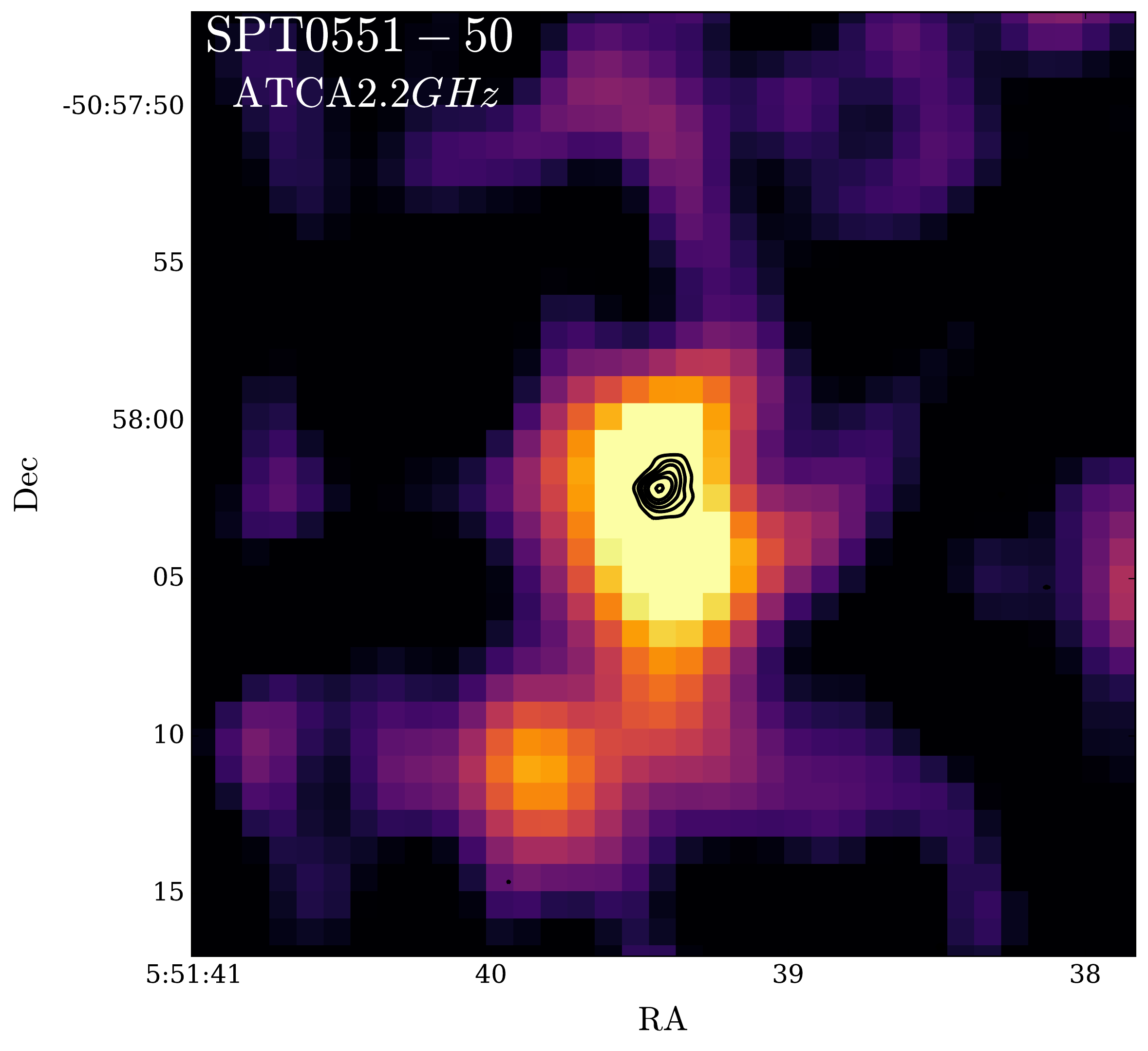}
  \includegraphics[width=0.24\linewidth]{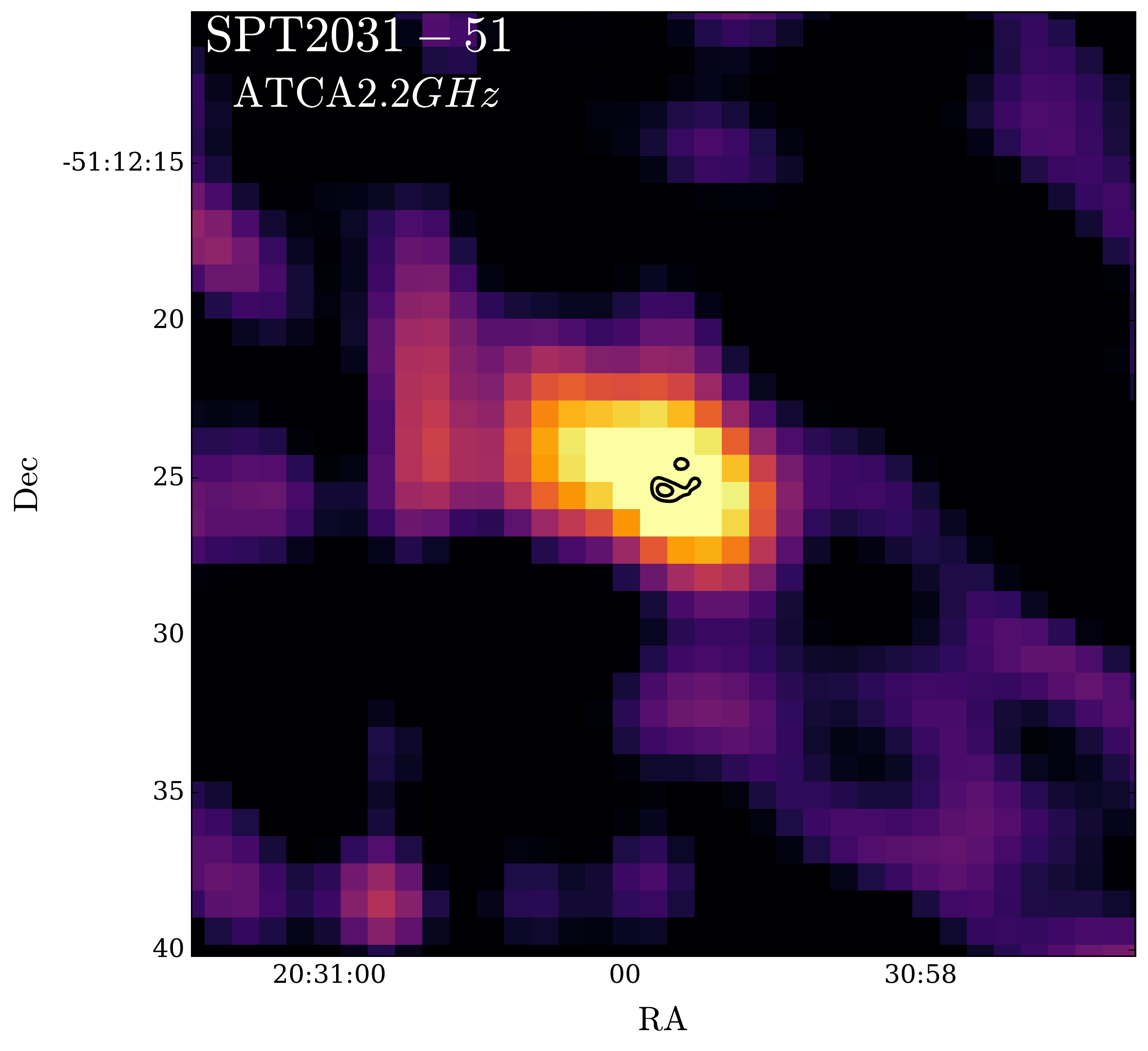}    
 \includegraphics[width=0.24\linewidth]{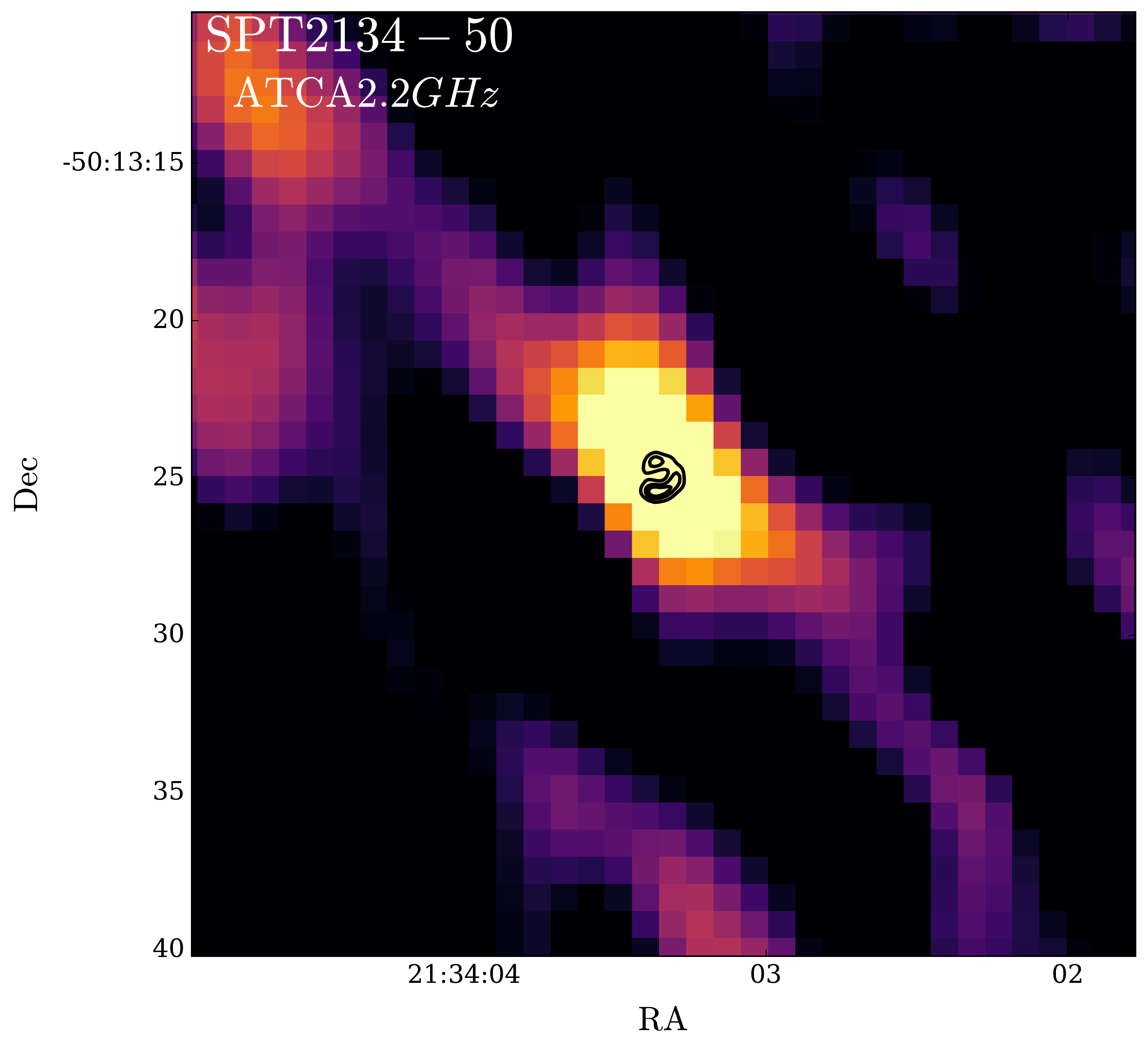}
 \includegraphics[width=0.24\linewidth]{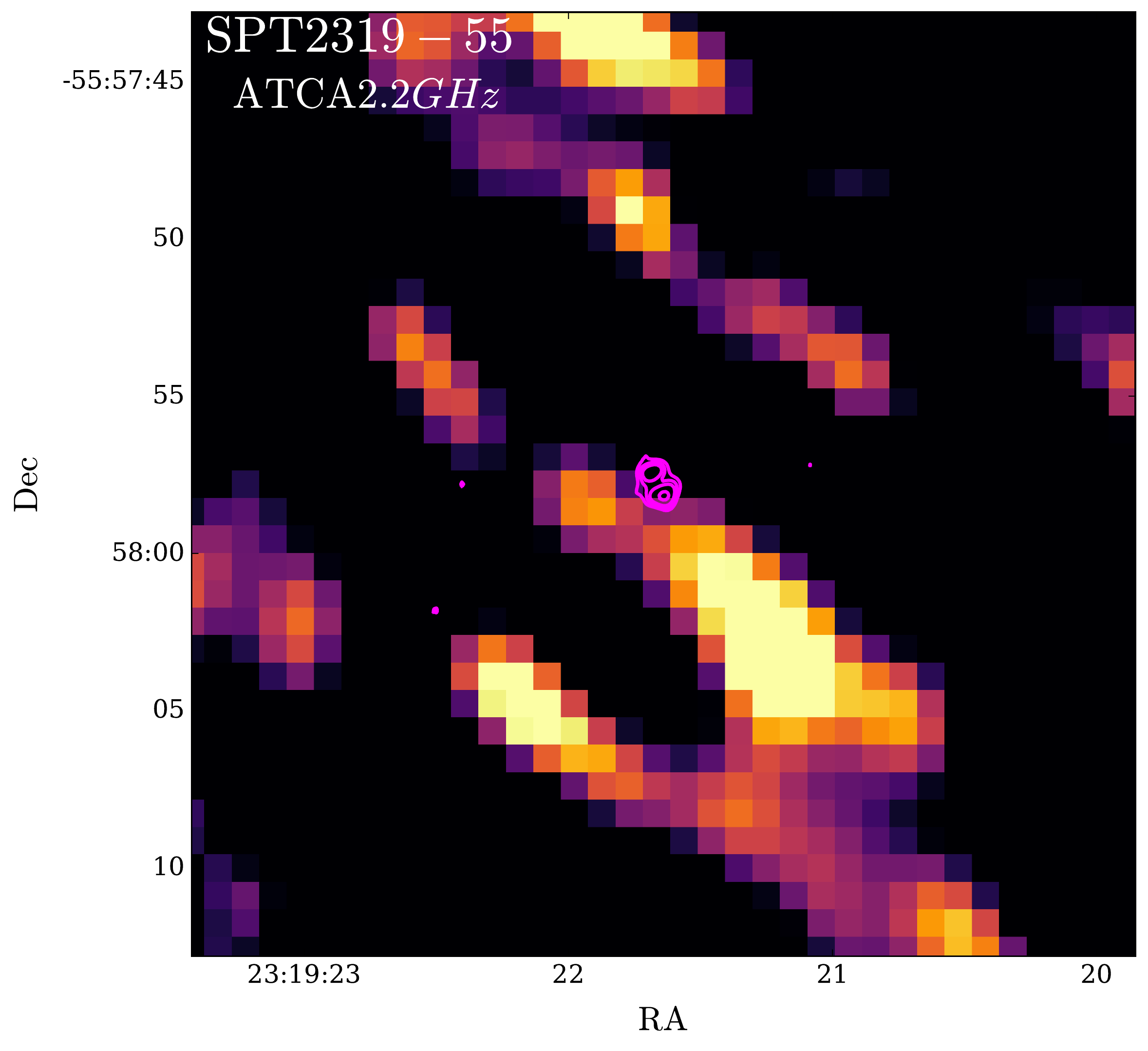} 
 \includegraphics[width=0.24\linewidth]{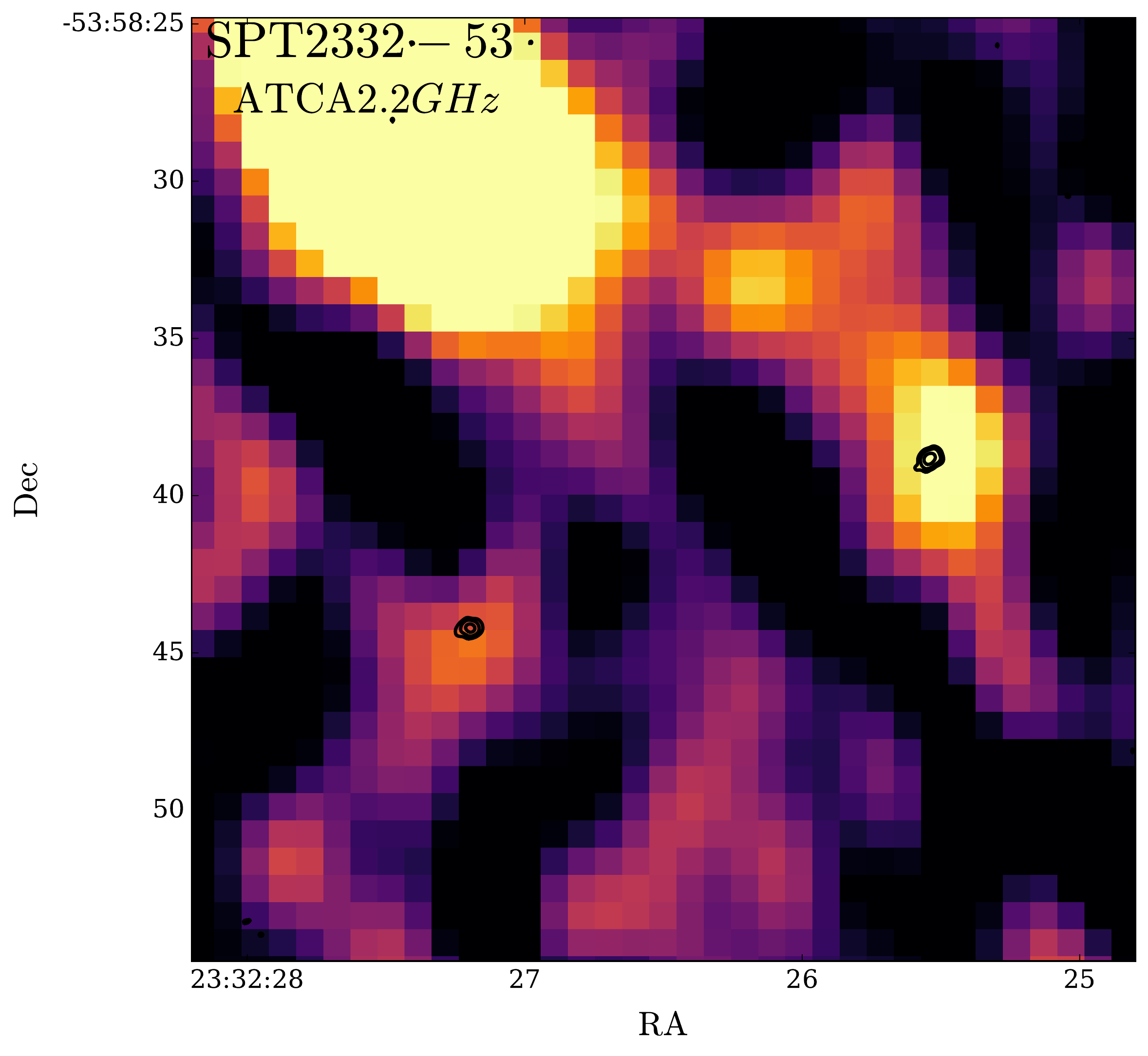}     
 \includegraphics[width=0.24\linewidth]{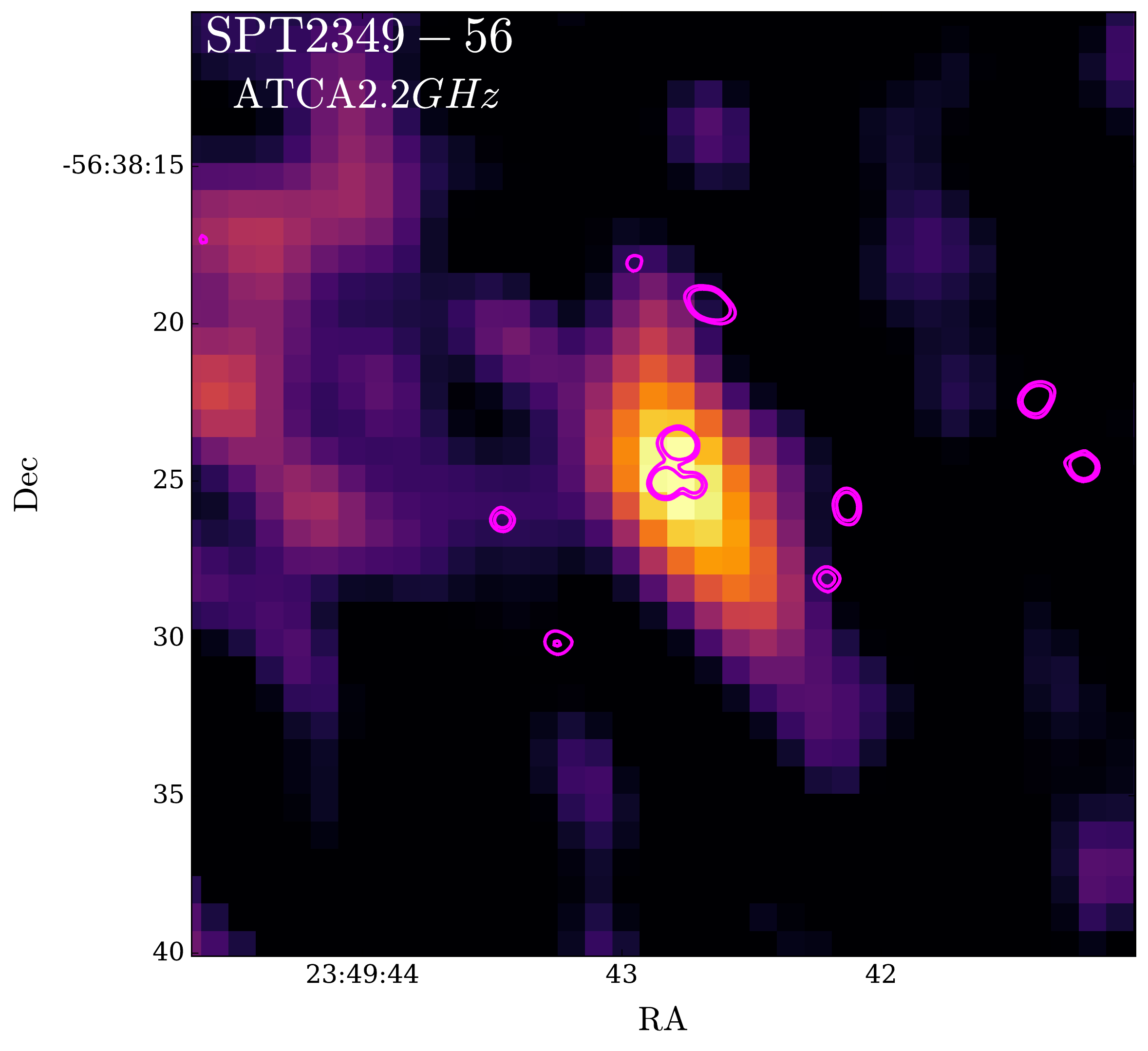}    
 \includegraphics[width=0.24\linewidth]{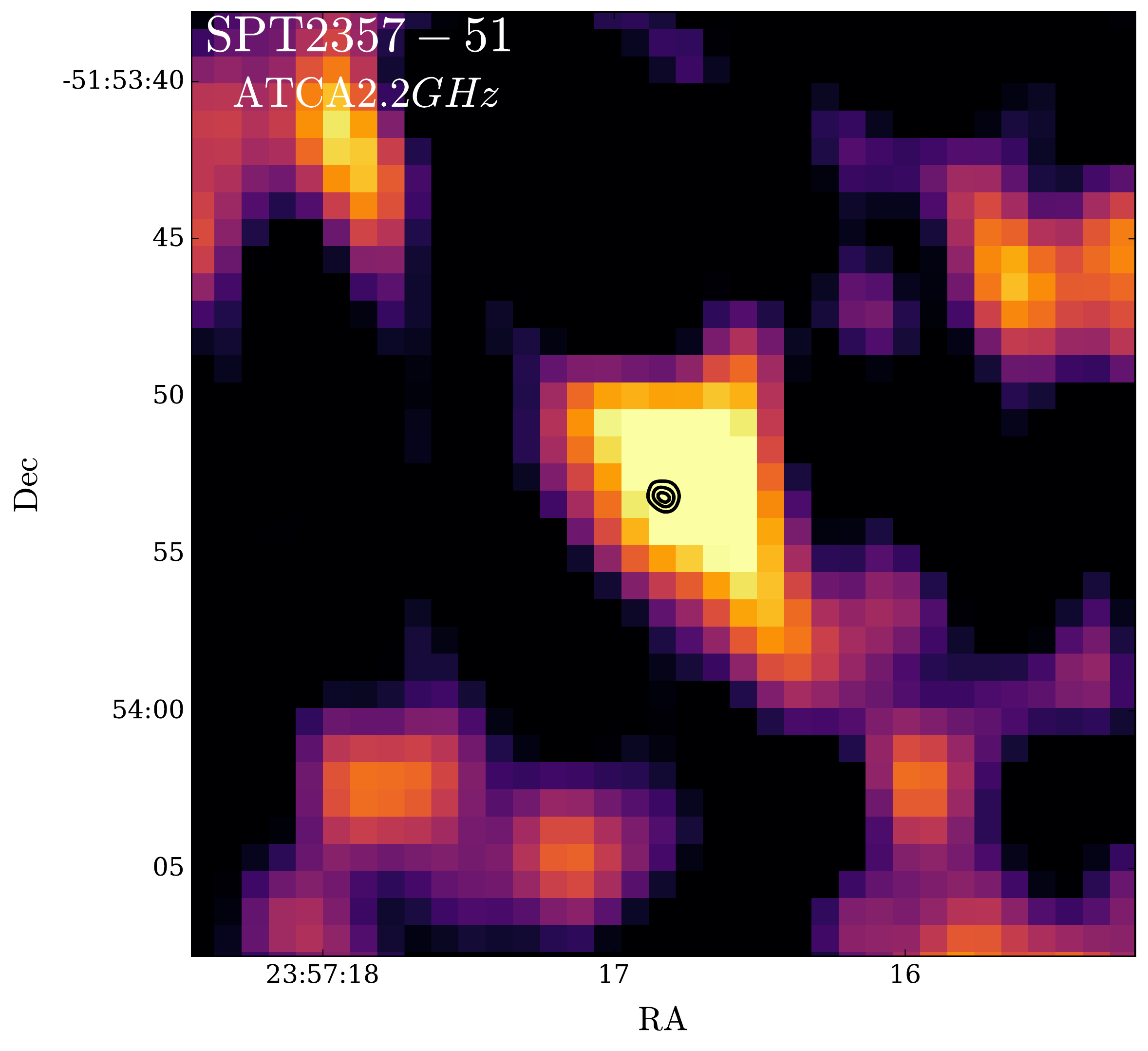}
 \includegraphics[width=0.24\linewidth]{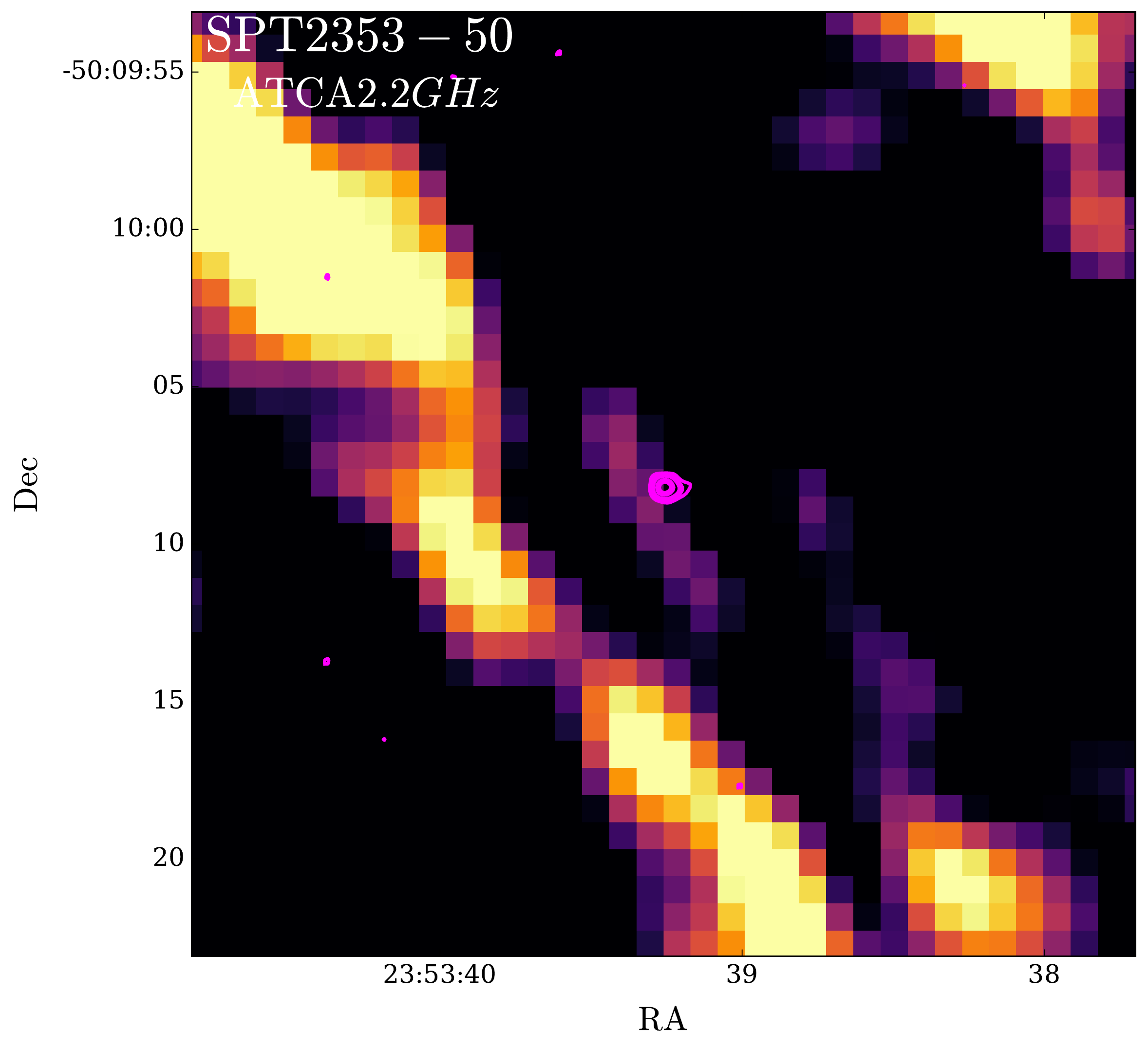}    
 \caption{ATCA 2.2\,GHz images ($30^{\prime\prime}\,{\times}\,30^{\prime\prime}$) of the 23 SPT-SMGs observed, with ALMA 850\,$\mu$m contours overlaid.
 }  
    \label{fig:radio_cutouts}
\end{figure*}

Here we describe the ATCA observations of the full sample of 23 SPT-SMGs observed in the survey program, shown in Fig.~\ref{fig:distribution}. 
These SPT-SMGs were drawn from the complete sample of 81 sources \citep{Reuter20}, selecting those that had the best redshift constraints at the time of observations. 
All but three of the 23 SPT-SMGs are detected at ${>}\,4\sigma$ significance at 2.2\,GHz.
The 2.2\,GHz flux densities are measured at peak pixels (Table~\ref{table:radio_full}), as in all cases the sources are unresolved in the $8^{\prime\prime}\,{\times}\,5^{\prime\prime}$ beam.  The restored beam sizes and position angles are also listed in Table~\ref{table:radio_full}.
ALMA 850\,$\mu$m overlays are shown in Fig.~\ref{fig:radio_cutouts} (data from \citealt{Spilker,Reuter20}).
LABOCA 850\,$\mu$m fluxes and ALMA-derived redshifts from \cite{Reuter20} are also listed in Table~\ref{table:radio_full} for completeness. 

\begin{figure*}
    \centering
 \includegraphics[width=0.25\linewidth]{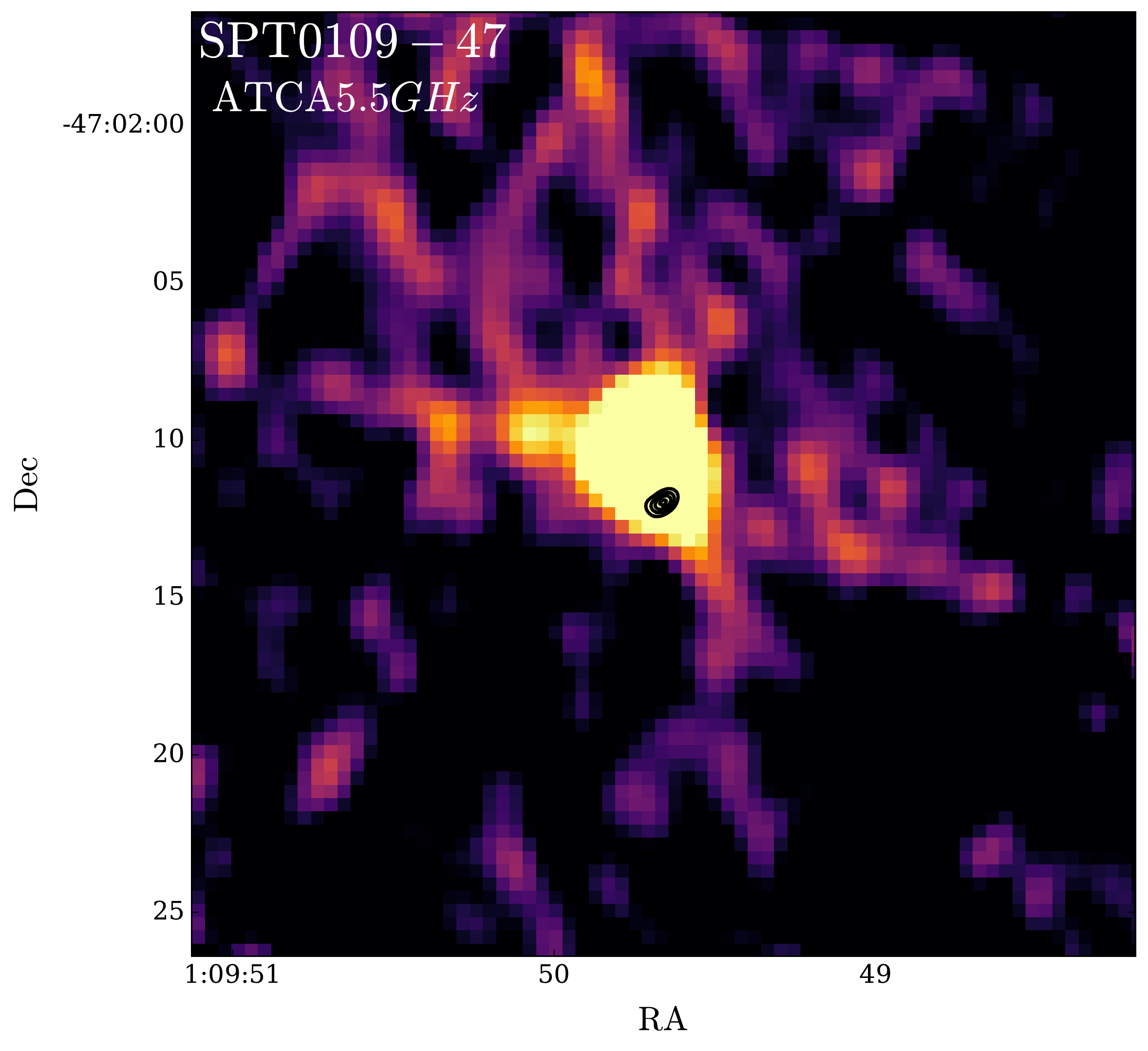}
 \includegraphics[width=0.25\linewidth]{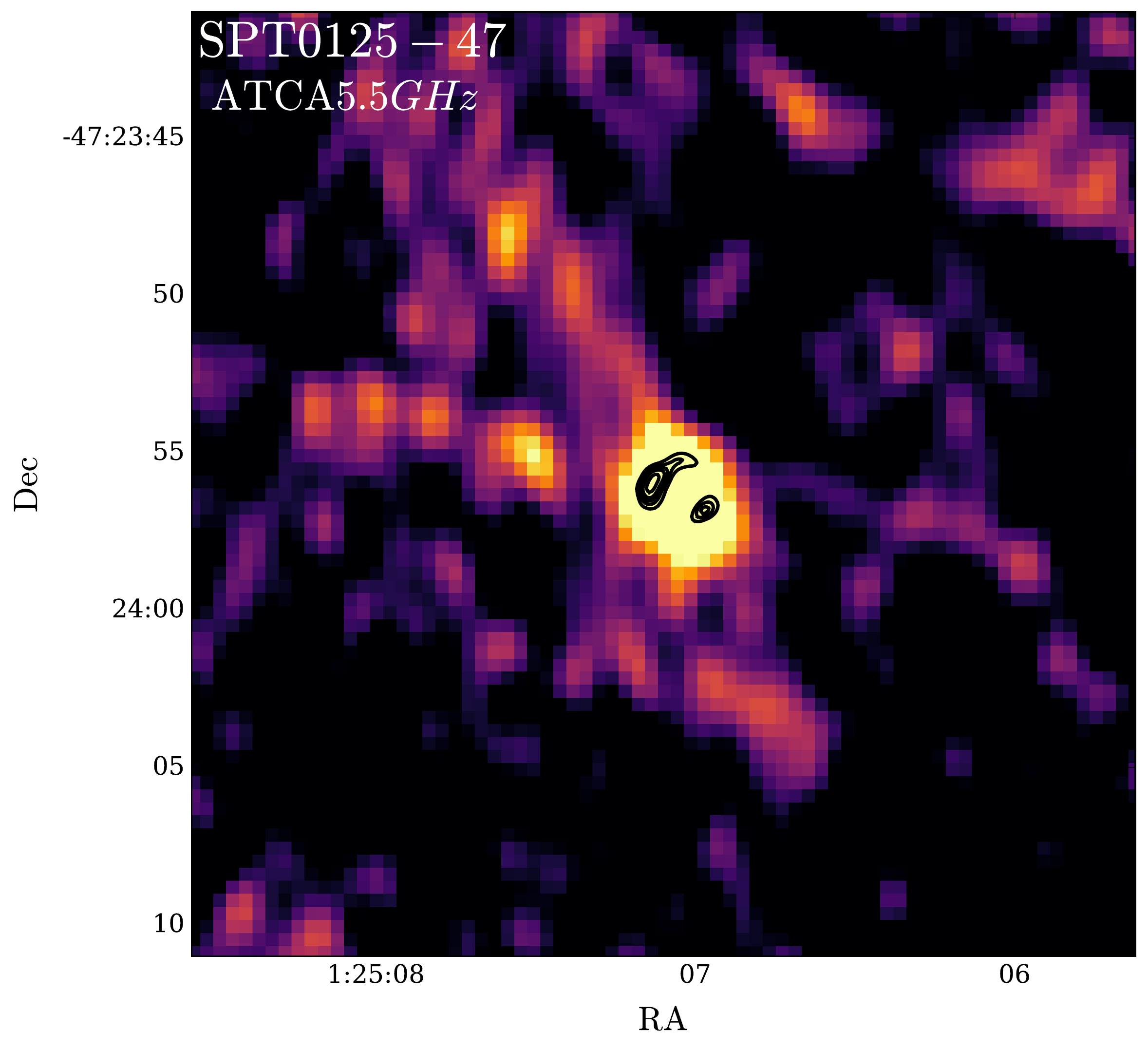}
 \includegraphics[width=0.25\linewidth]{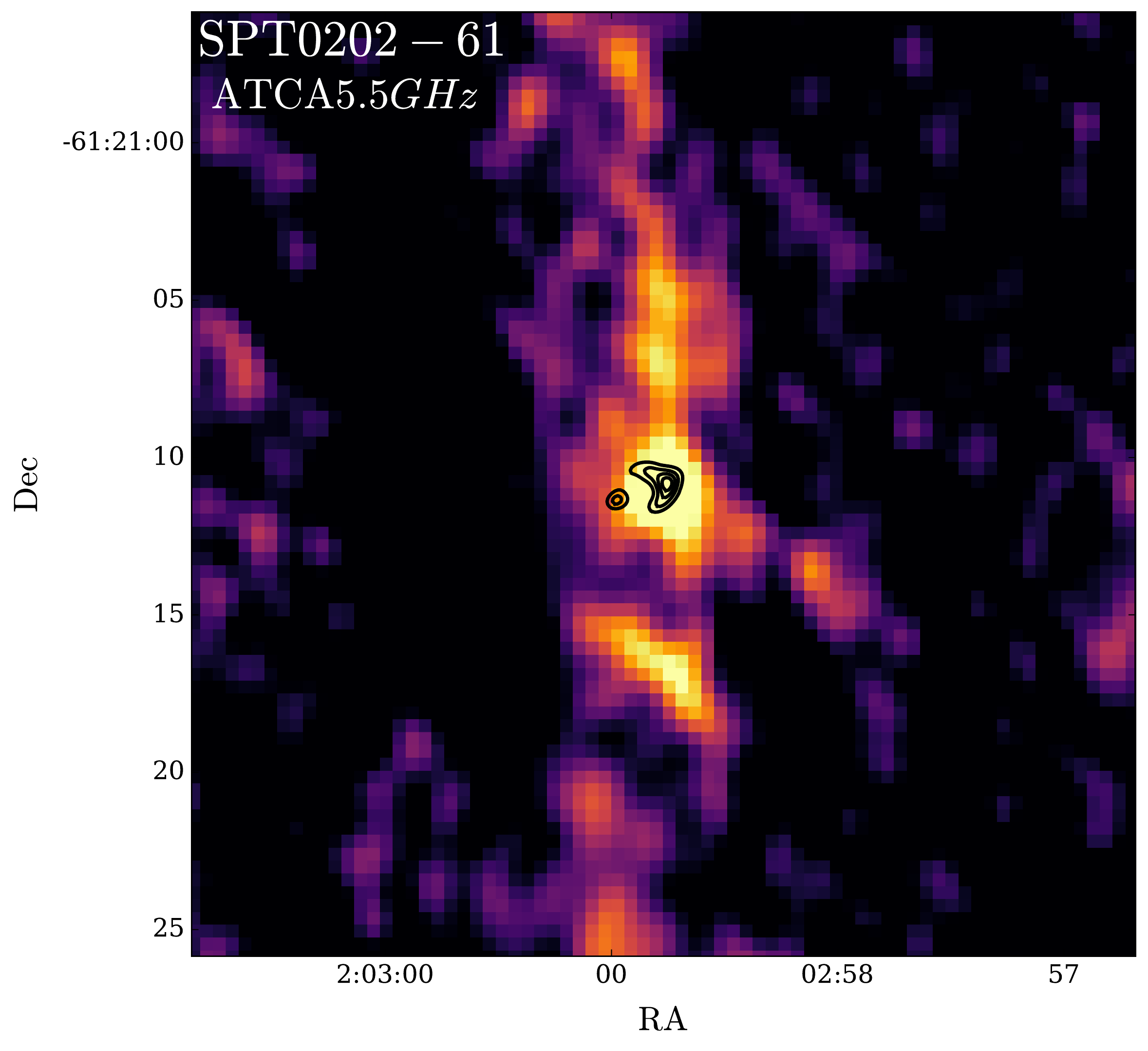}    
 \includegraphics[width=0.25\linewidth]{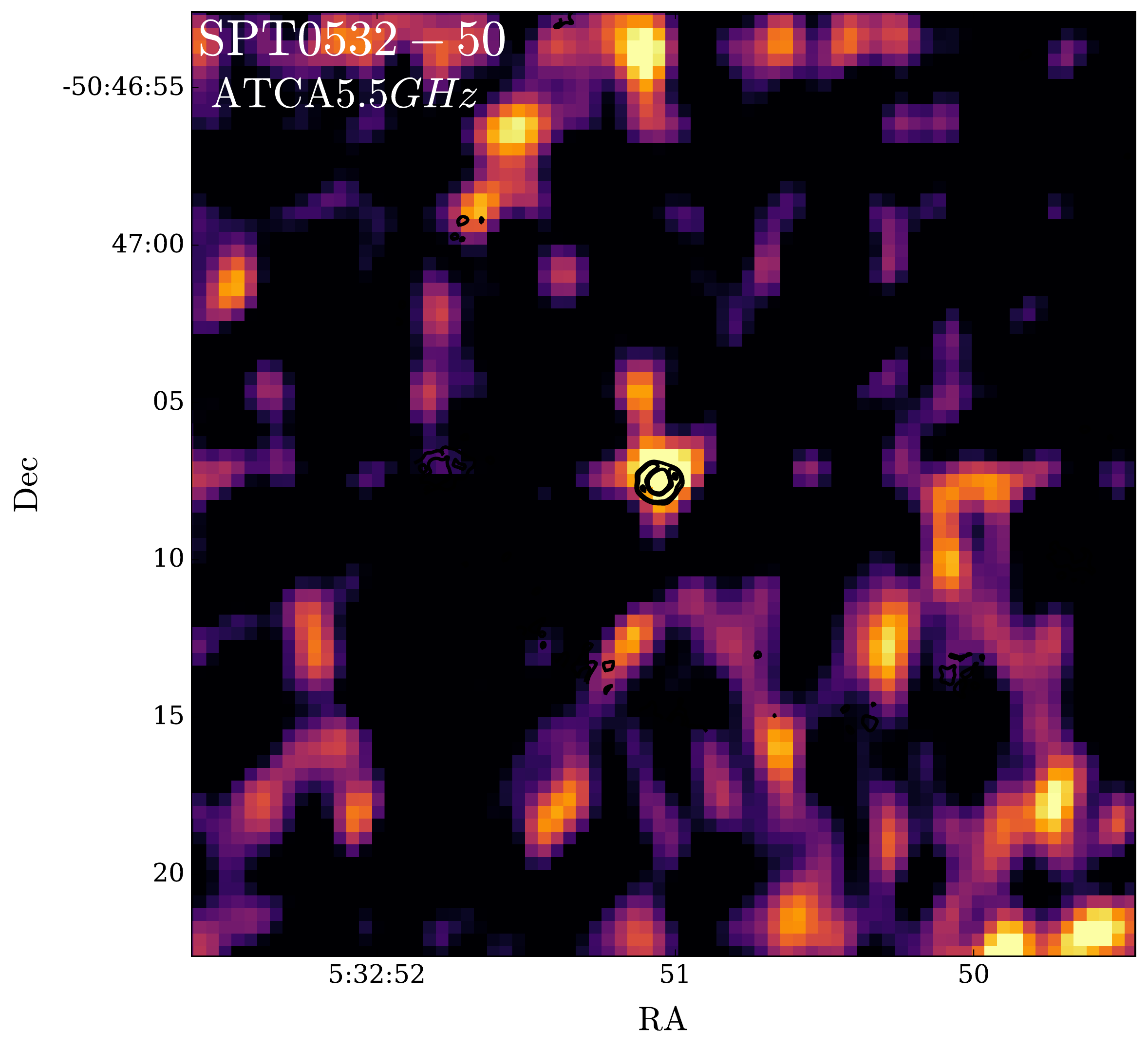}    
 \includegraphics[width=0.25\linewidth]{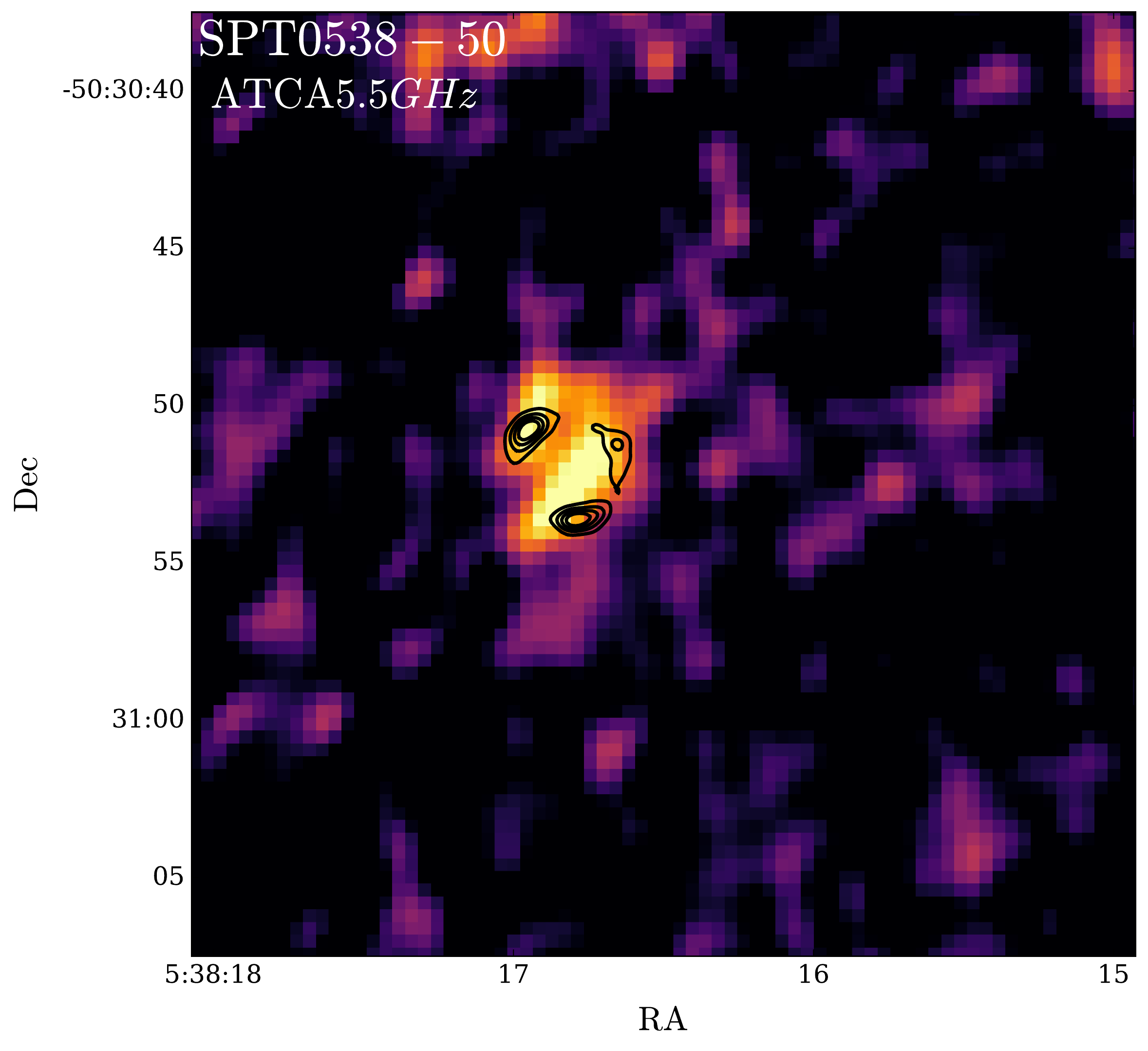}    
 \includegraphics[width=0.25\linewidth]{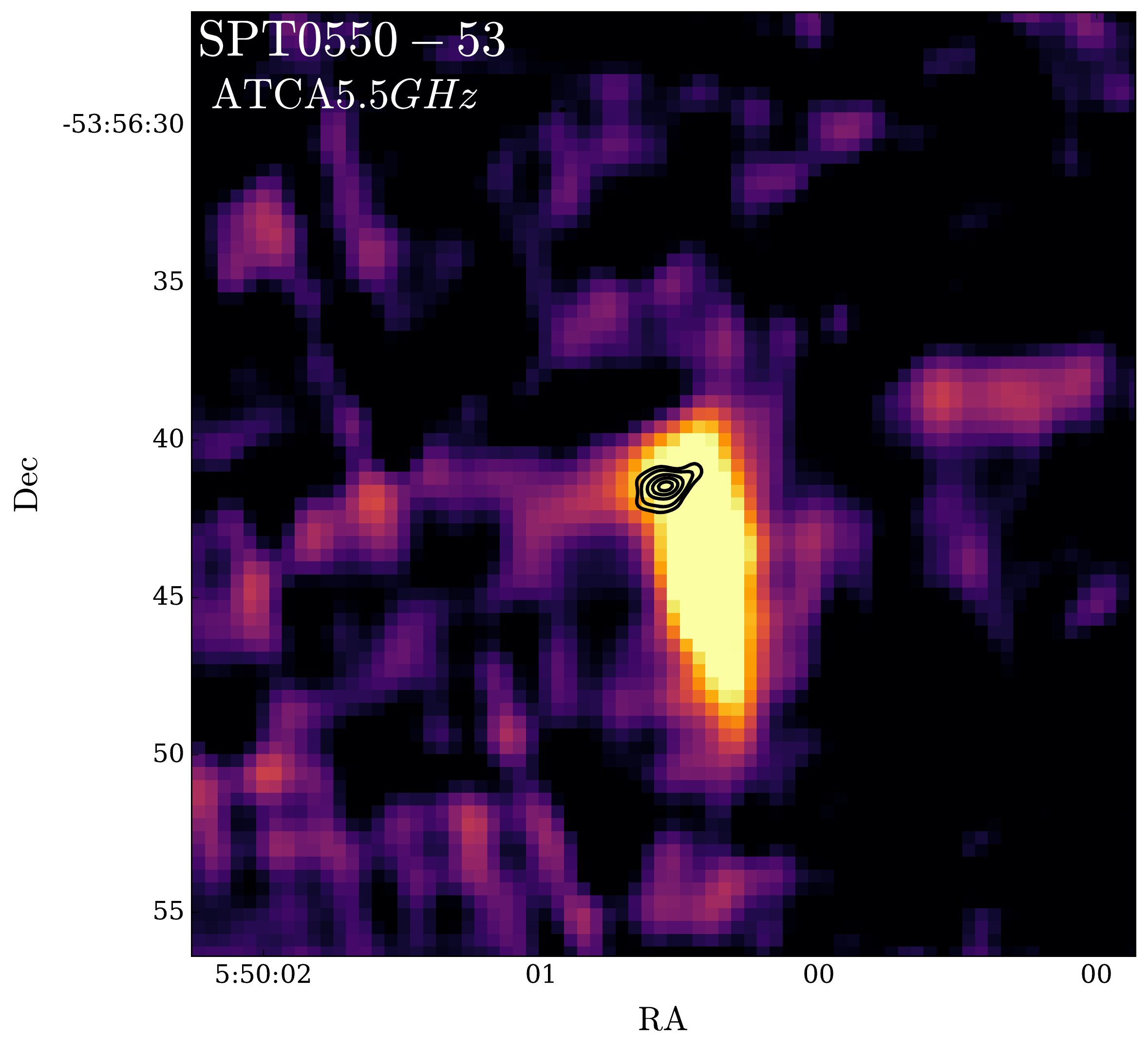}
  \includegraphics[width=0.25\linewidth]{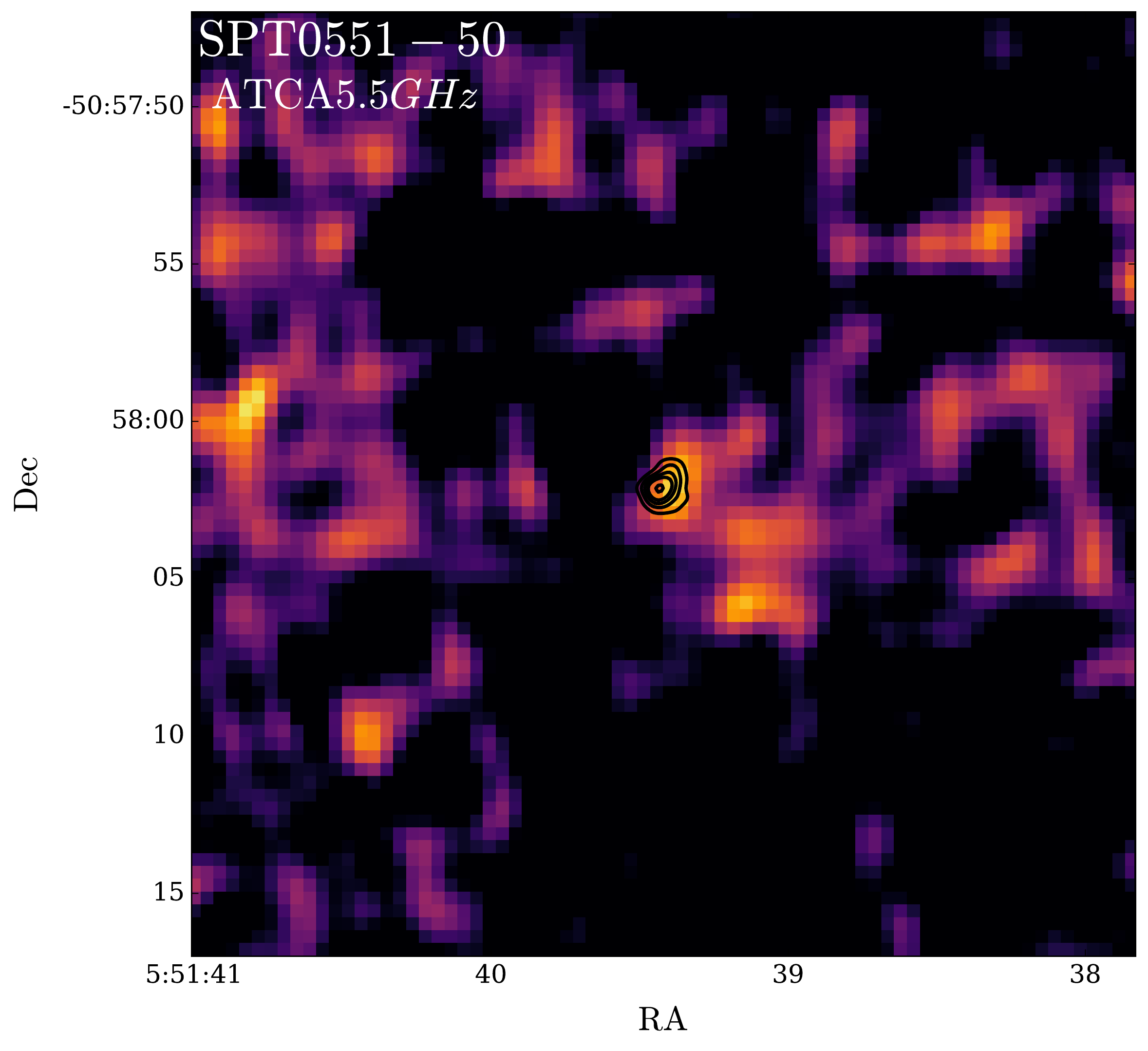}
 \includegraphics[width=0.25\linewidth]{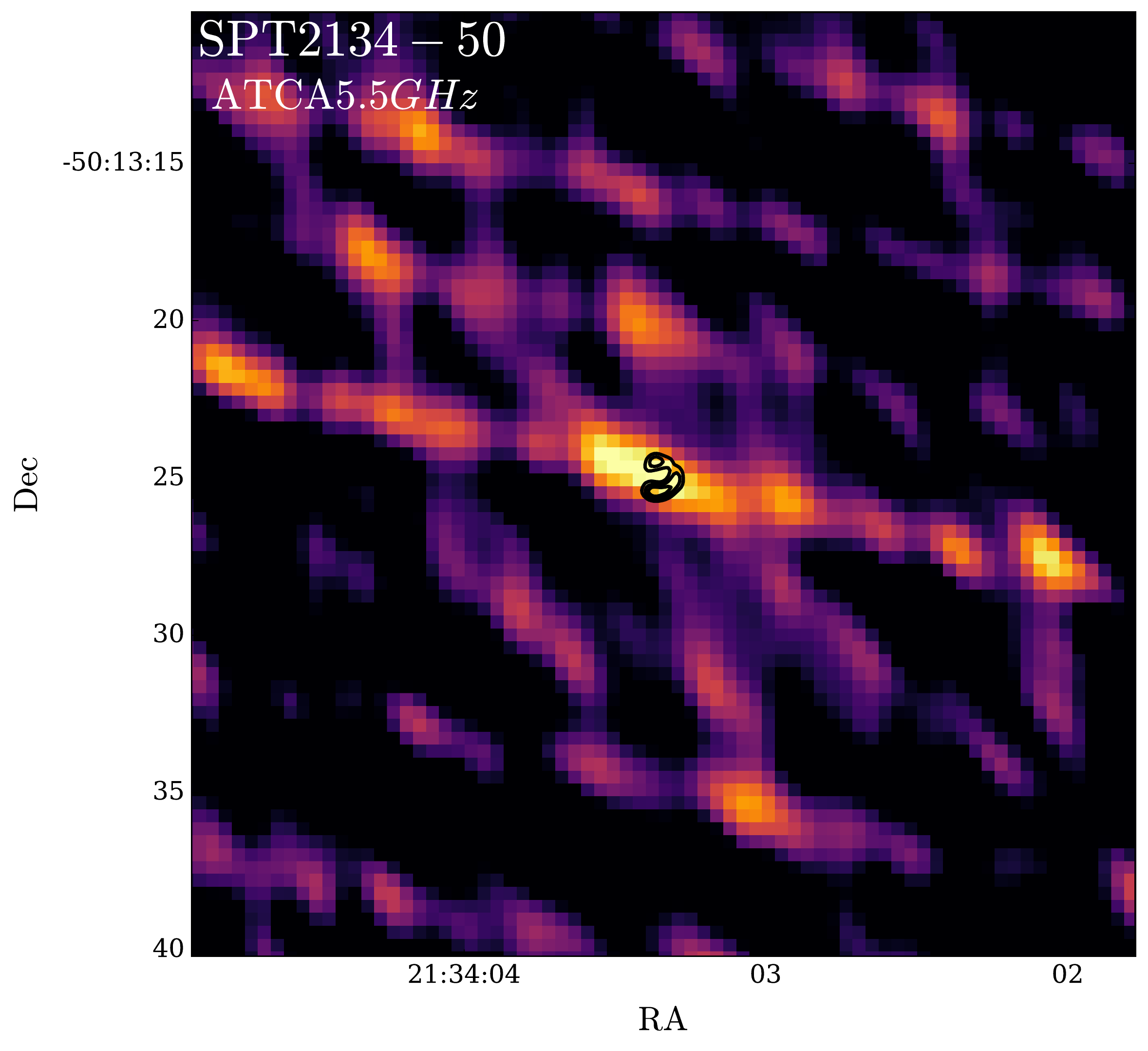}
 \includegraphics[width=0.25\linewidth]{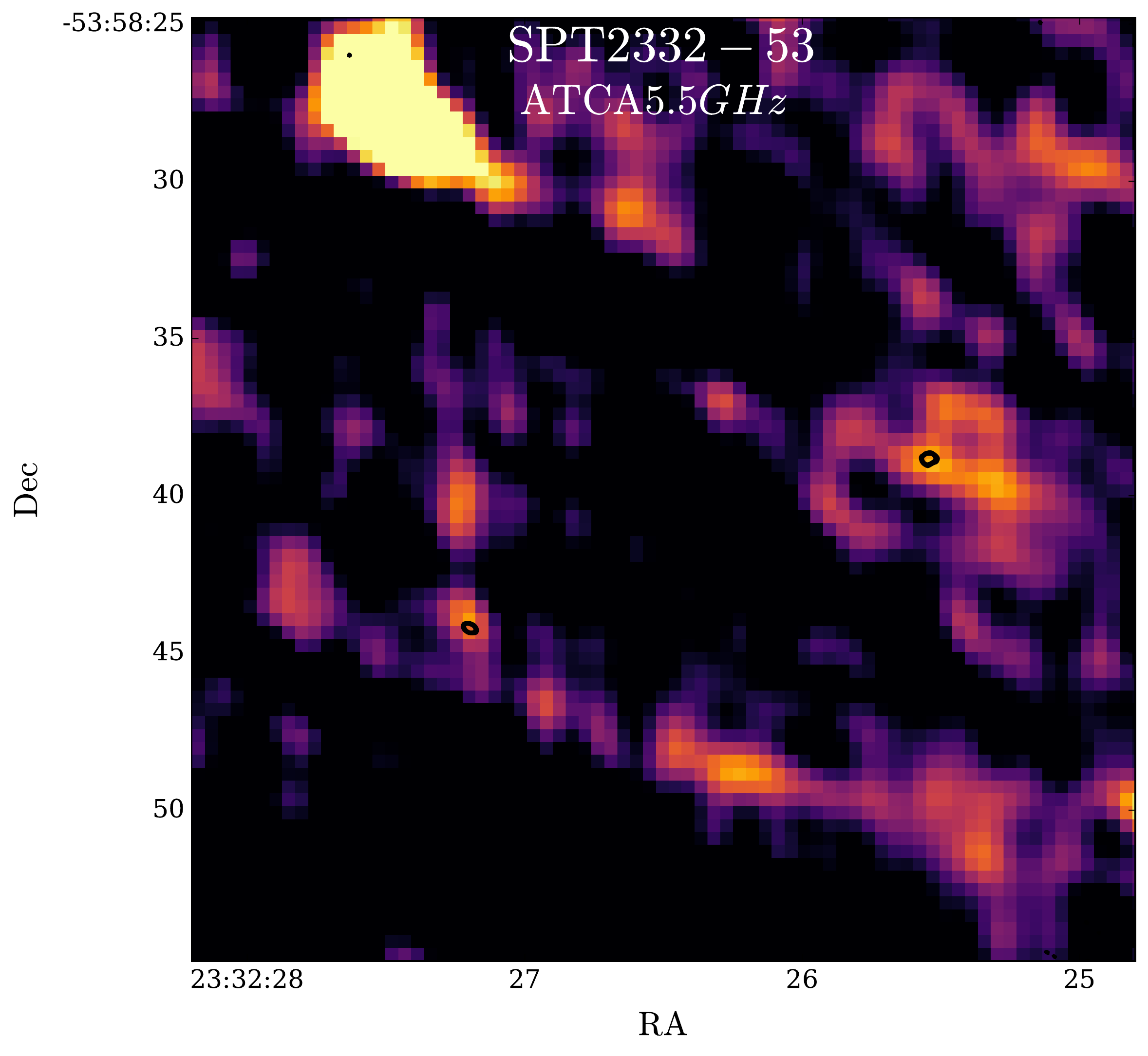}     
 \caption{ATCA 5.5\,GHz images ($30^{\prime\prime}\,{\times}\,30^{\prime\prime}$) of the nine detected SPT-SMGs, with ALMA 850\,$\mu$m contours overlaid.
 }  
    \label{fig:radio_cutouts2}
\end{figure*}

\begin{figure*}
    \centering
 \includegraphics[width=0.244\linewidth]{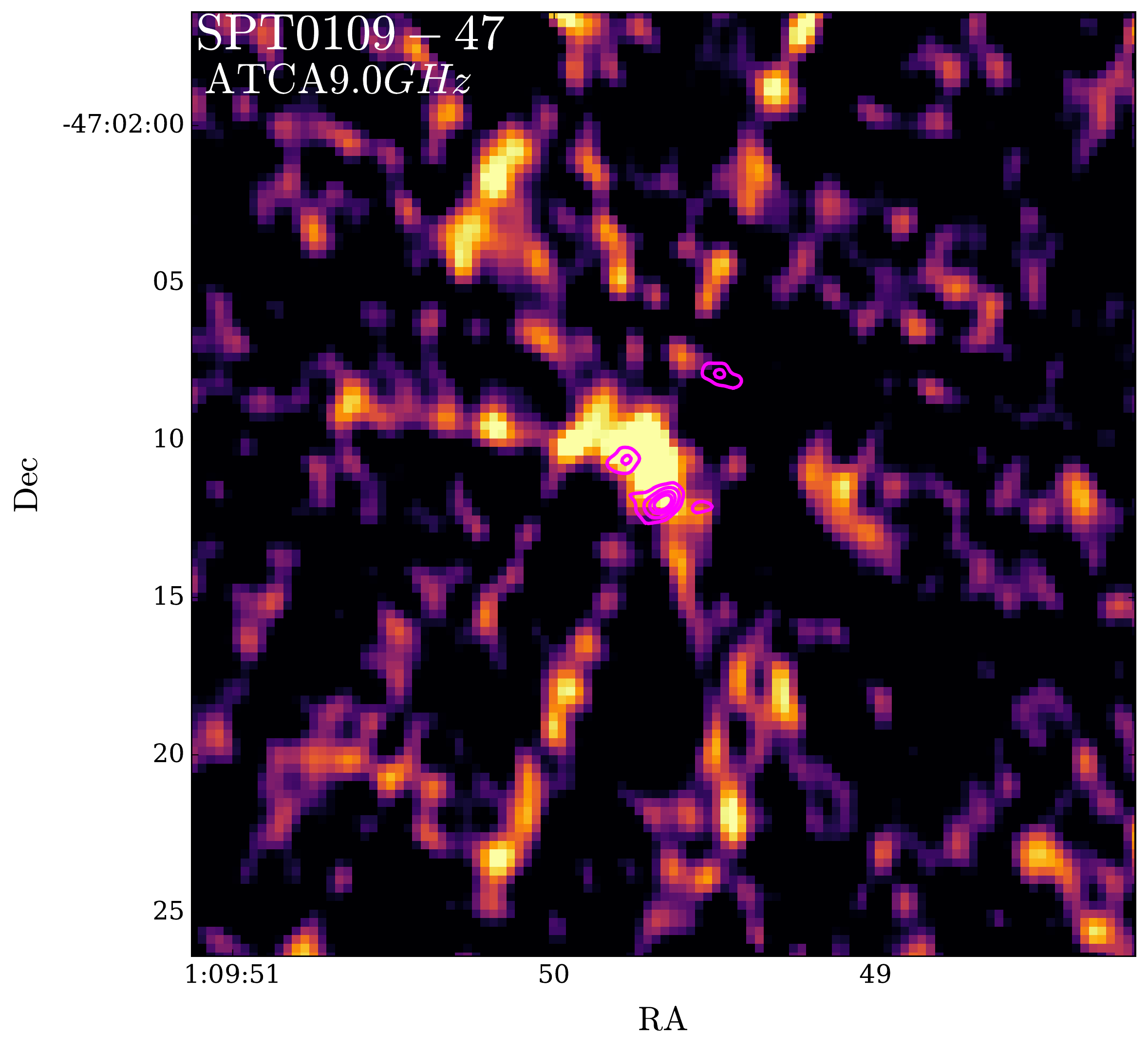}
 \includegraphics[width=0.244\linewidth]{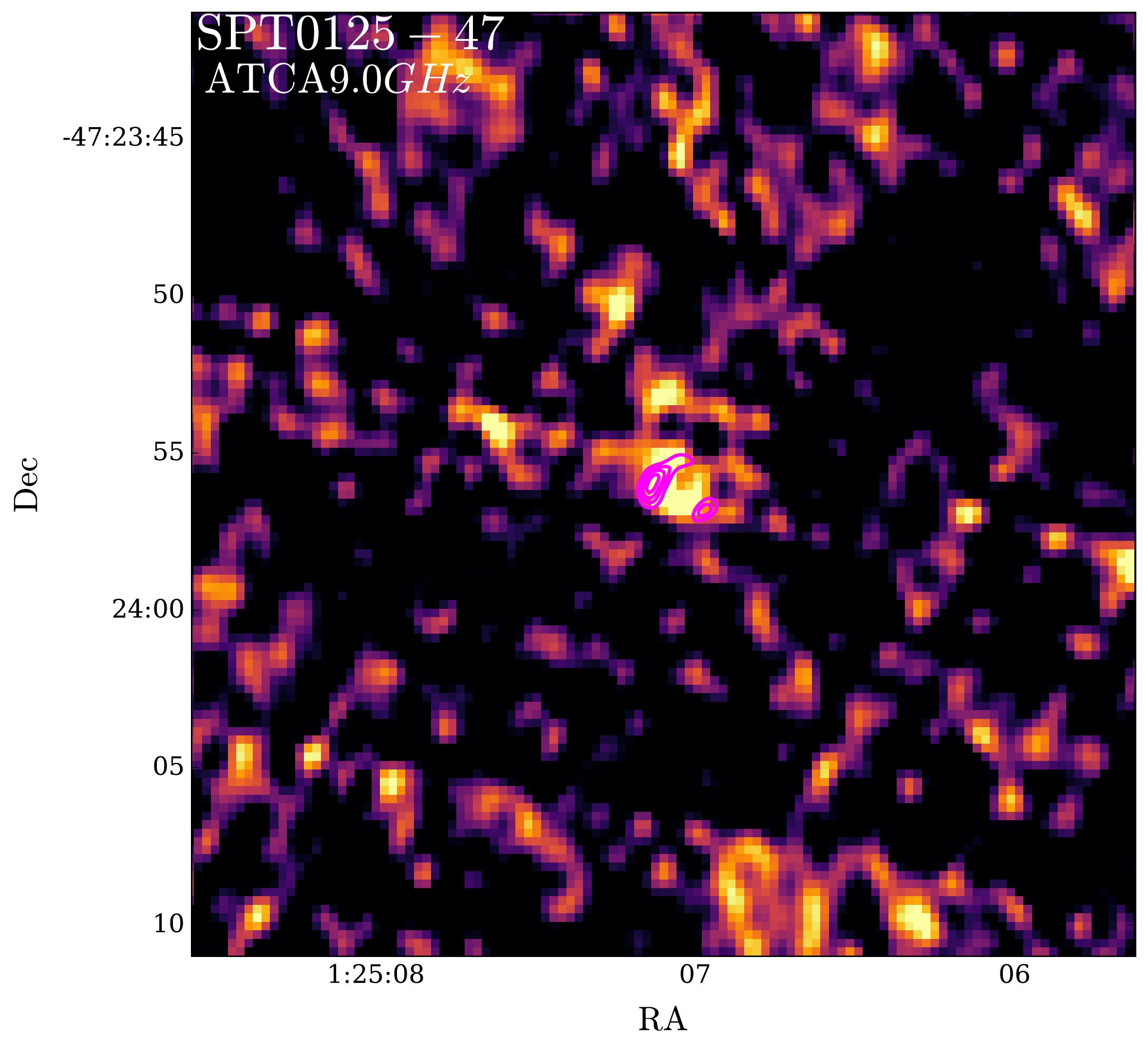}
 \includegraphics[width=0.244\linewidth]{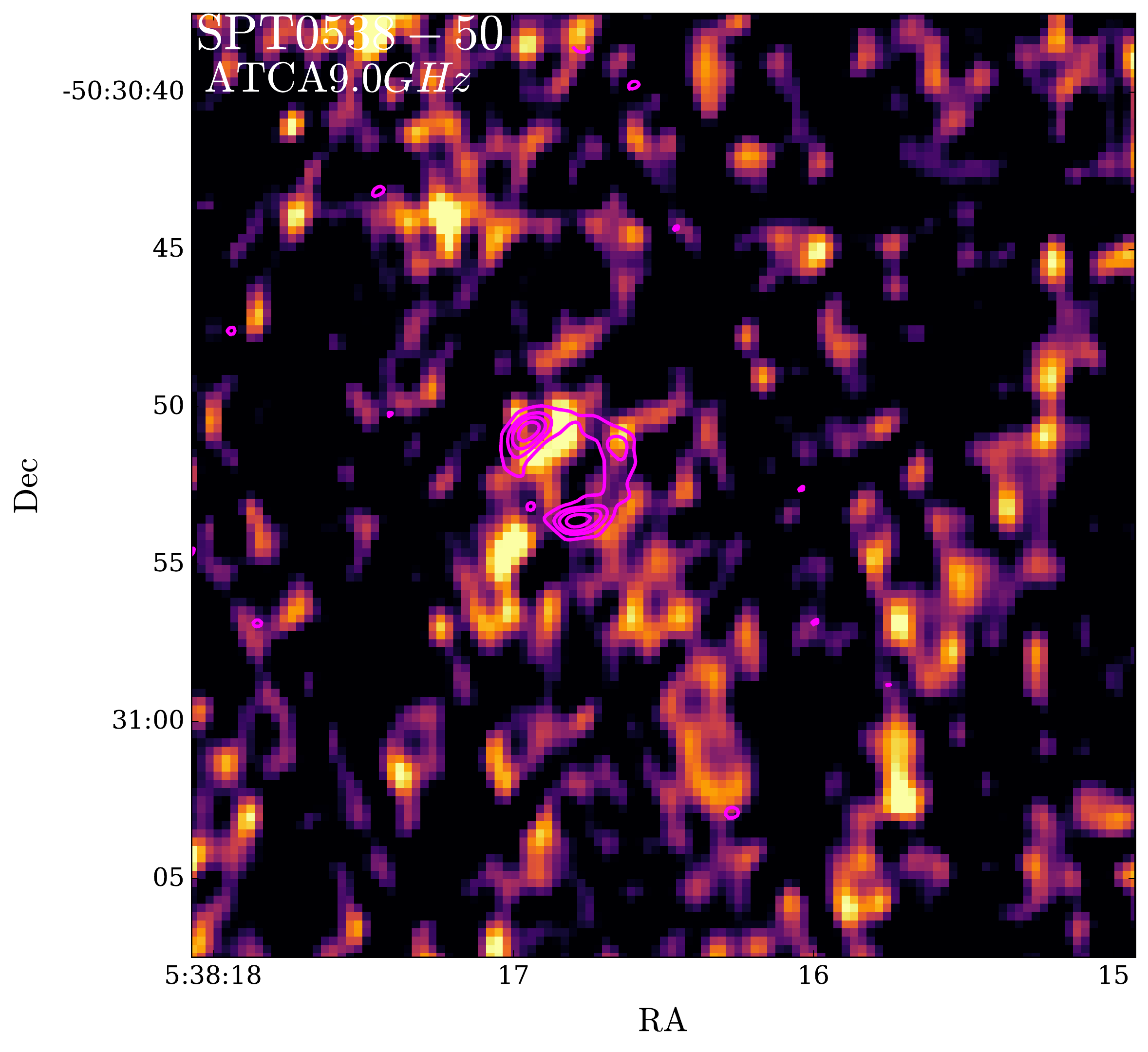}    
 \includegraphics[width=0.244\linewidth]{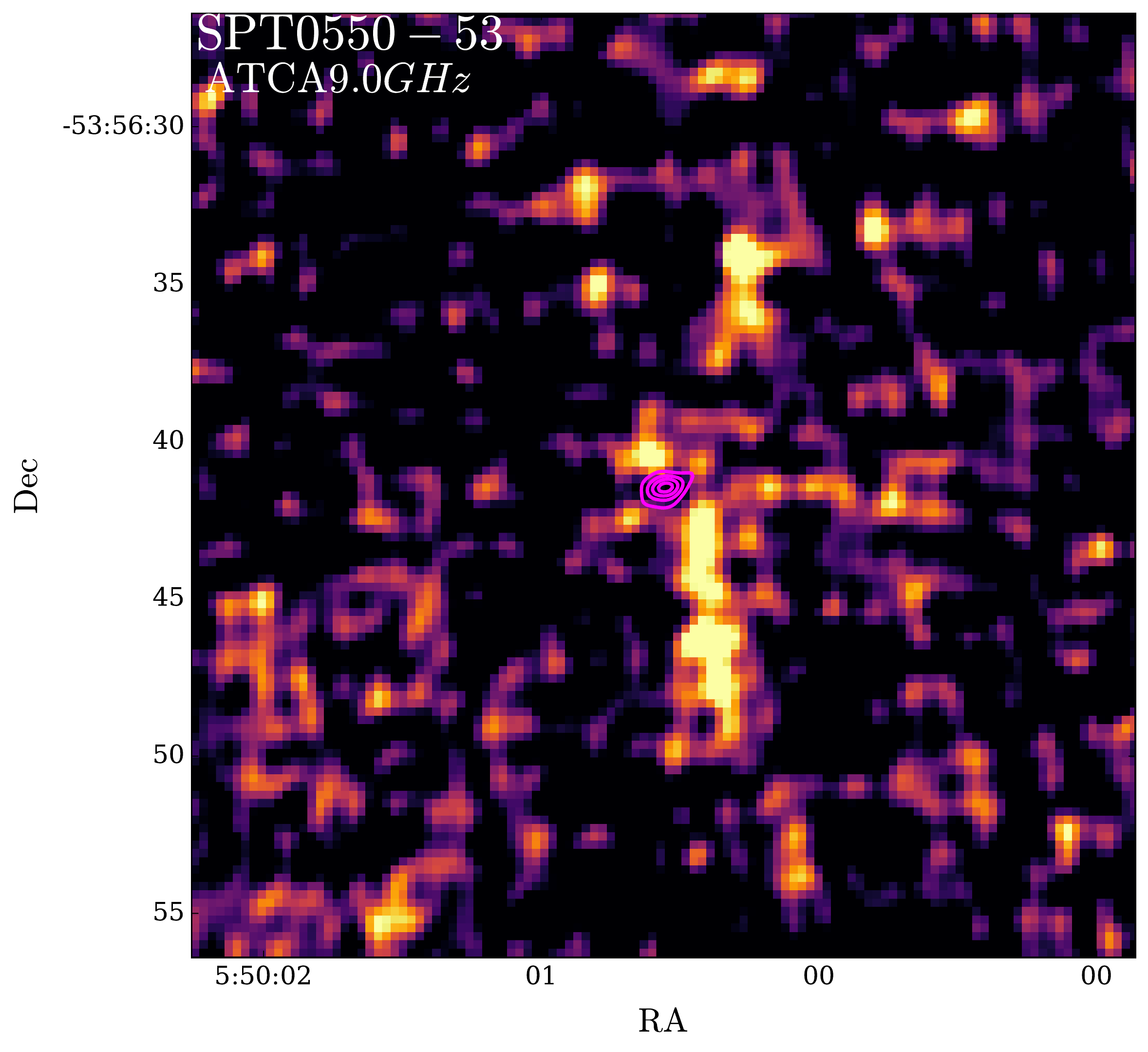}
 \caption{ATCA 9.0\,GHz images ($30^{\prime\prime}\,{\times}\,30^{\prime\prime}$) of the four detected SPT-SMGs, with ALMA 850\,$\mu$m contours overlaid.
 }  
    \label{fig:radio_cutouts3}
\end{figure*}

Most sources are not detected or only marginally detected at 5.5GHz (nine detections at ${>}\,3\sigma$) and 9.0\,GHz (four detections at ${>}\,3\sigma$). For those sources detected at these higher frequencies with ATCA, we measure flux densities from peak pixels when the source is unresolved, or as aperture measurements when the source is resolved. We show the nine sources detected at 5.5\,GHz in Fig.~\ref{fig:radio_cutouts2}, the four sources detected at 9.0\,GHz in Fig.~\ref{fig:radio_cutouts3}.
We have also searched for detections in the ASKAP 0.888\,MHz RACS survey described in section~2.2, listing their flux densities in Table~\ref{table:radio_full}. We find 15 of the 23 sources are significantly detected by ASKAP.

\begin{figure*}
    \centering
 \includegraphics[width=0.288\linewidth]{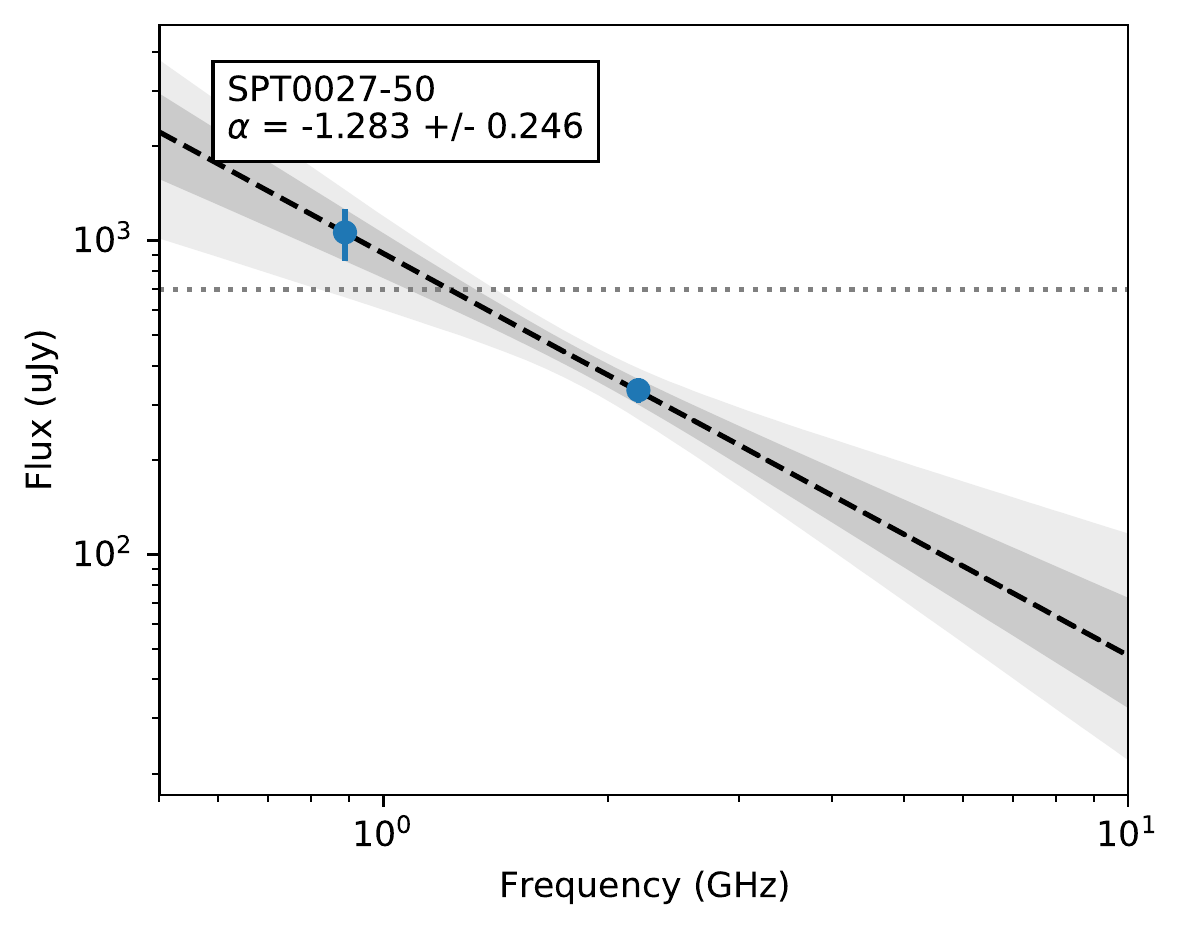}
 \includegraphics[width=0.288\linewidth]{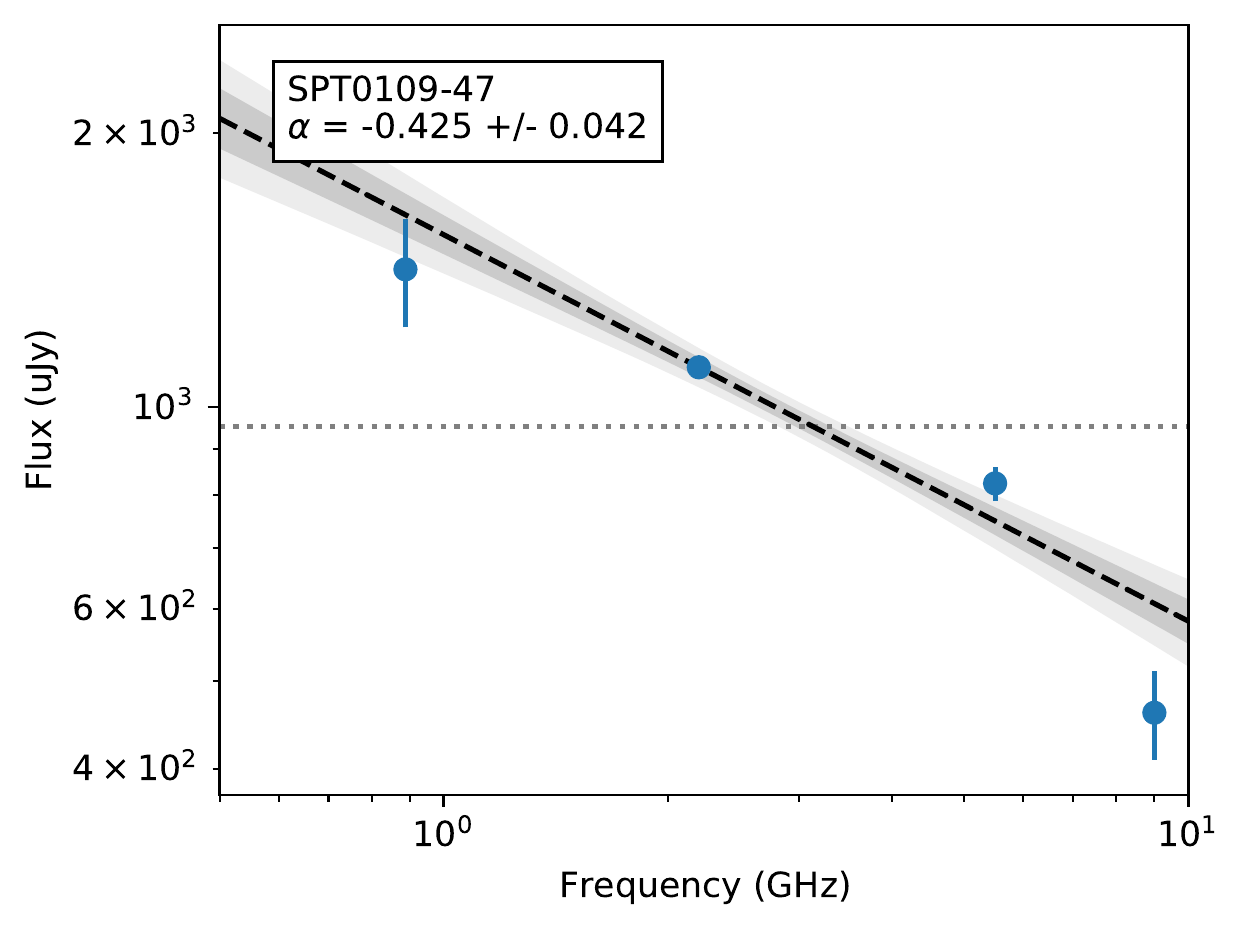}
 \includegraphics[width=0.288\linewidth]{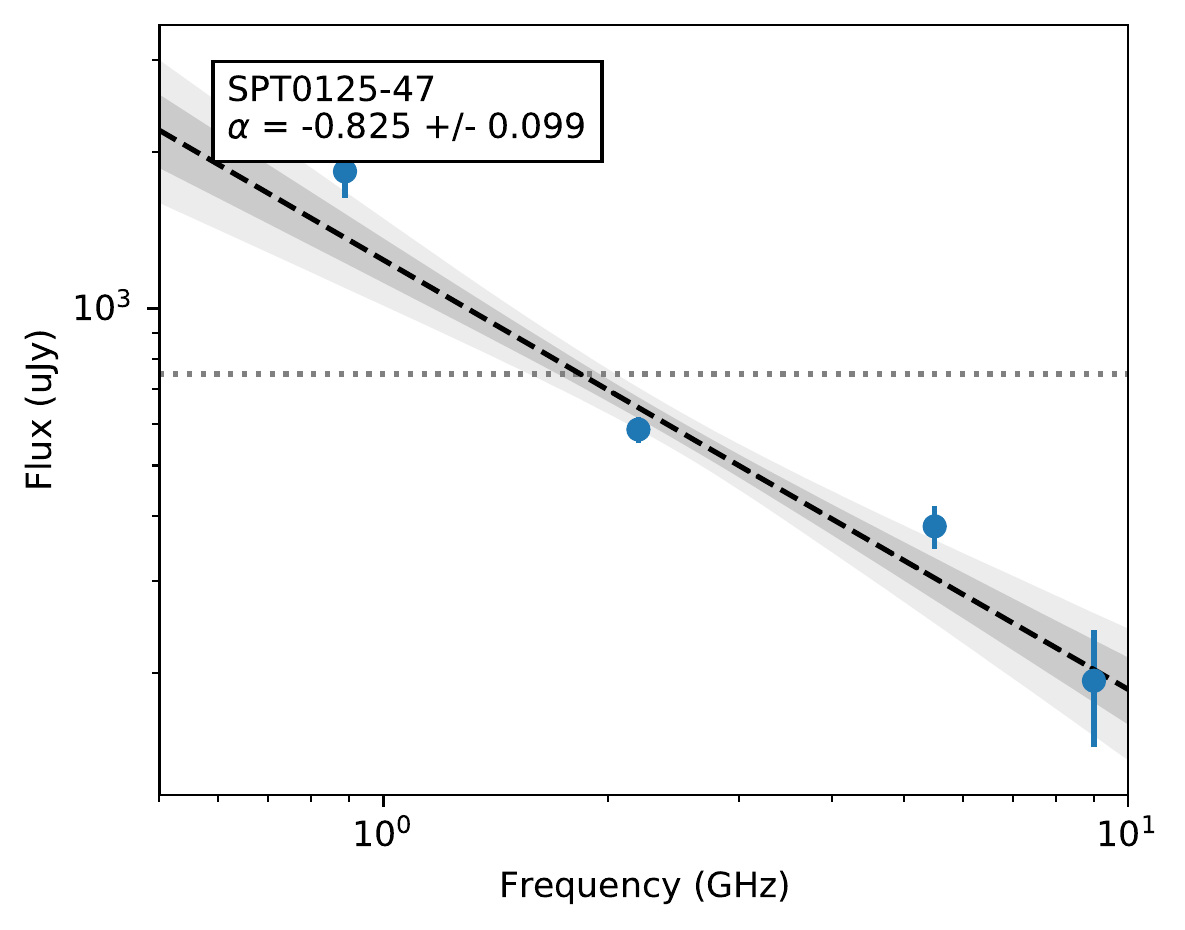}
 \includegraphics[width=0.288\linewidth]{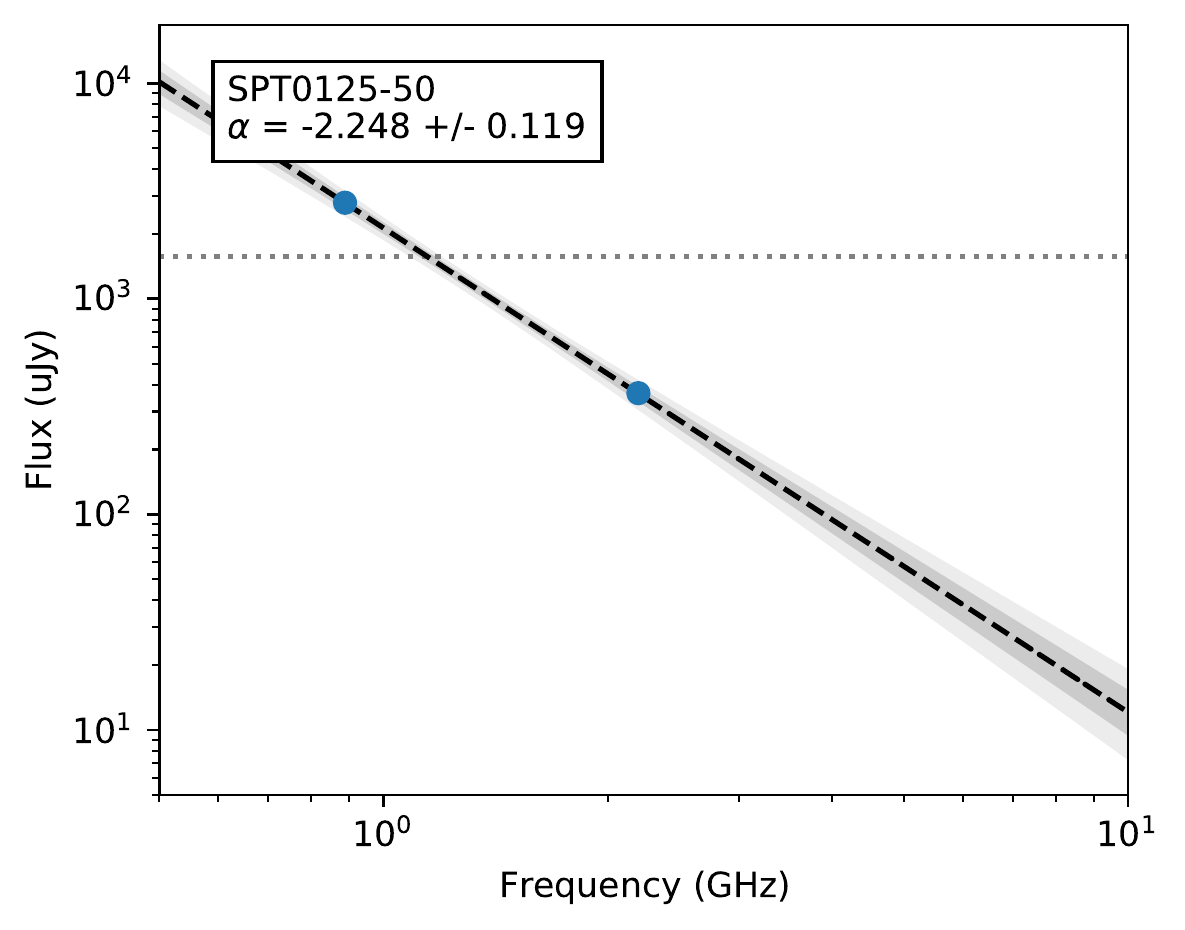}
 \includegraphics[width=0.288\linewidth]{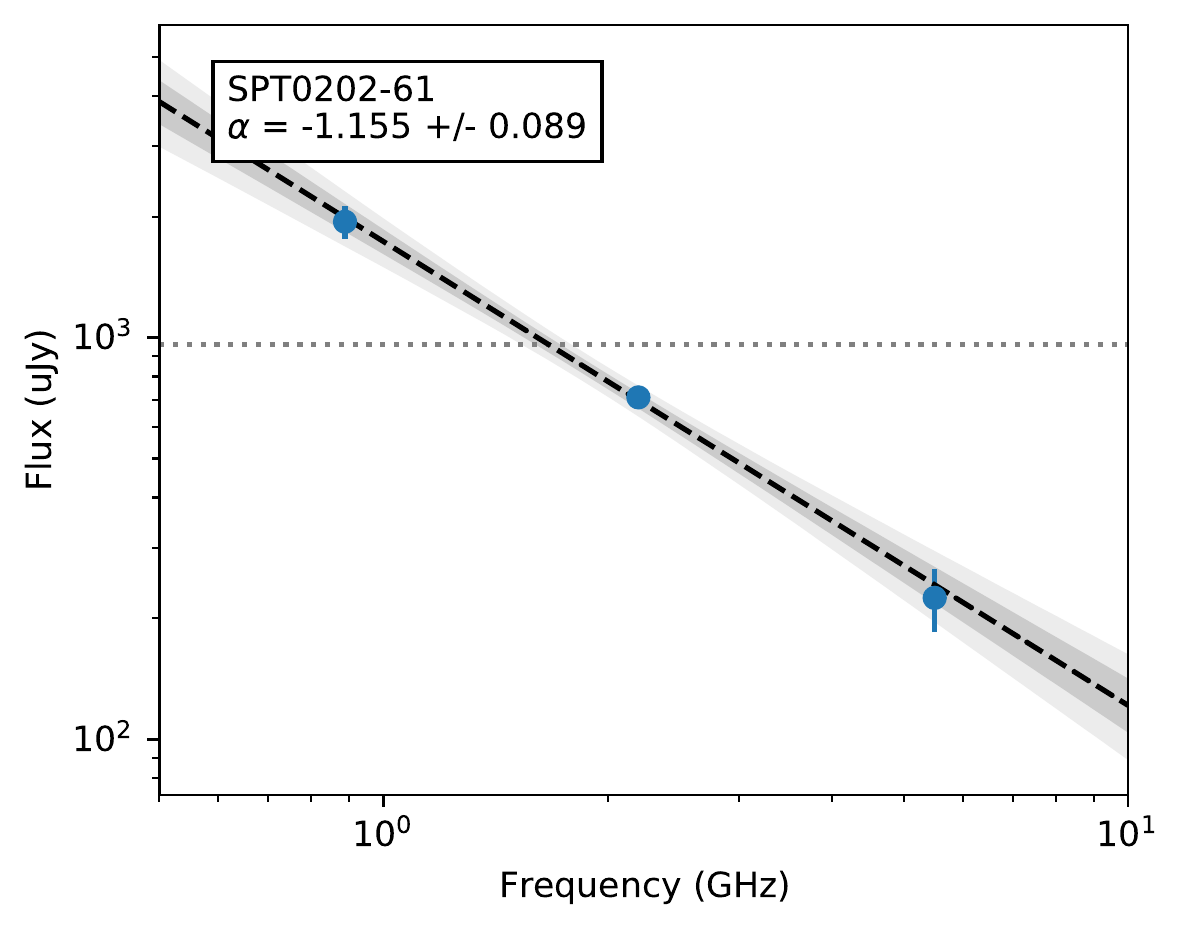}
 \includegraphics[width=0.288\linewidth]{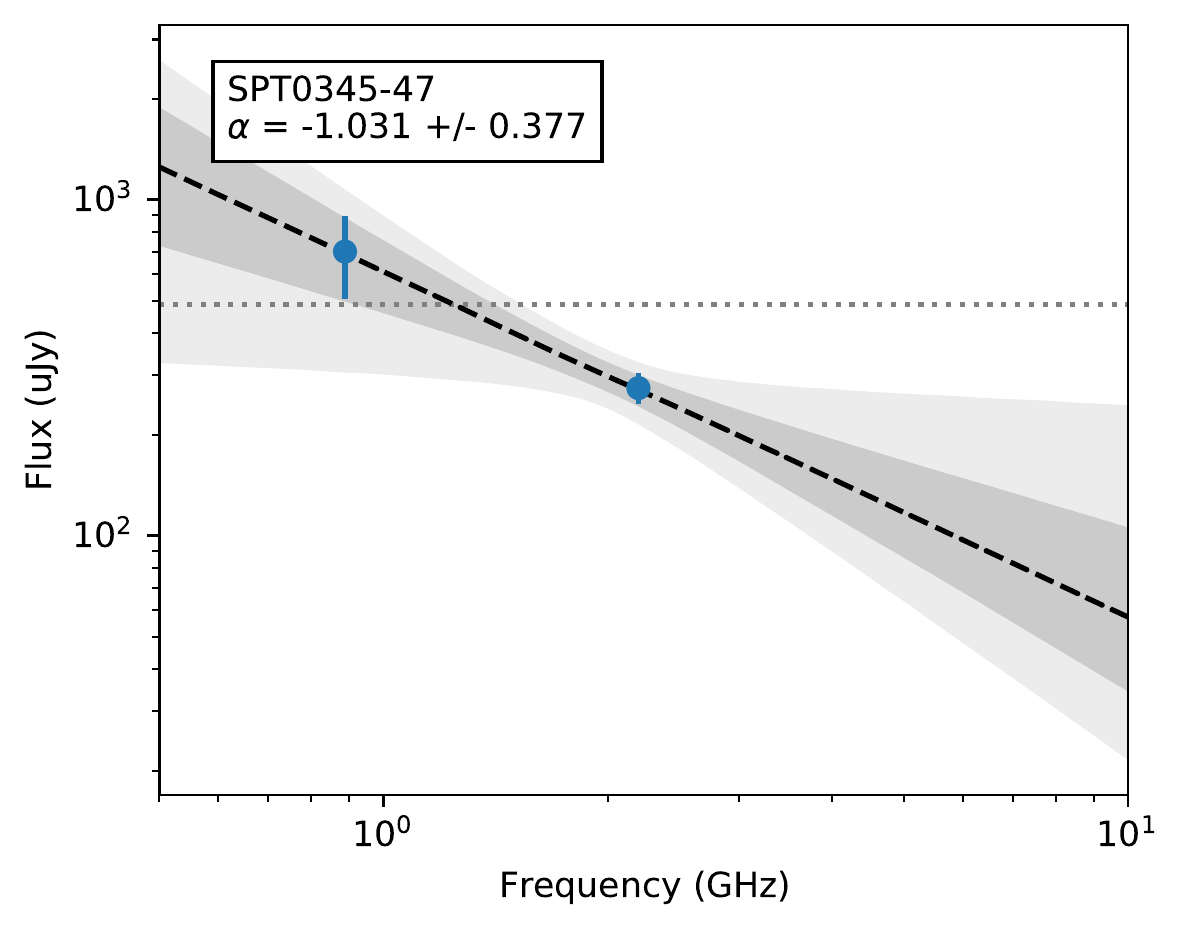}
 \includegraphics[width=0.288\linewidth]{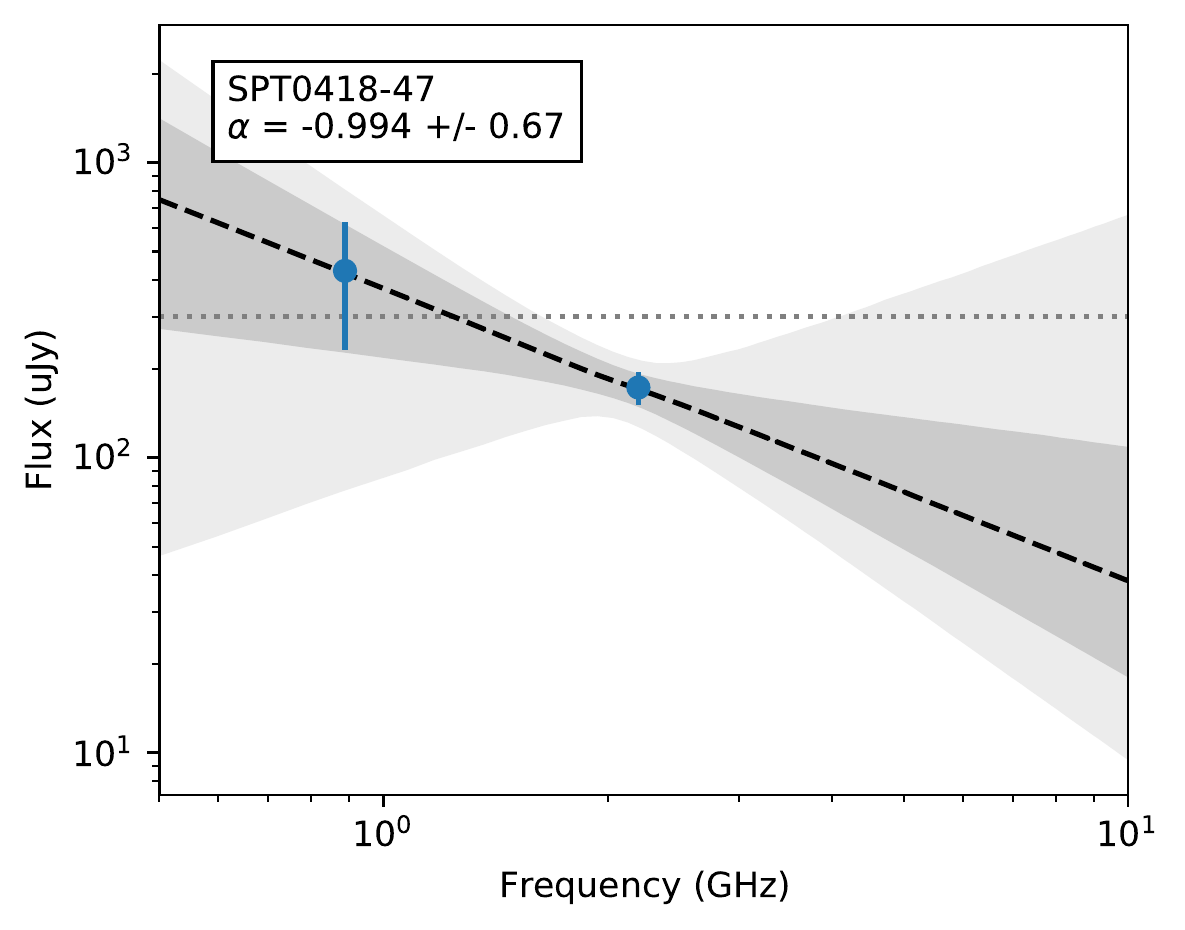}
 \includegraphics[width=0.288\linewidth]{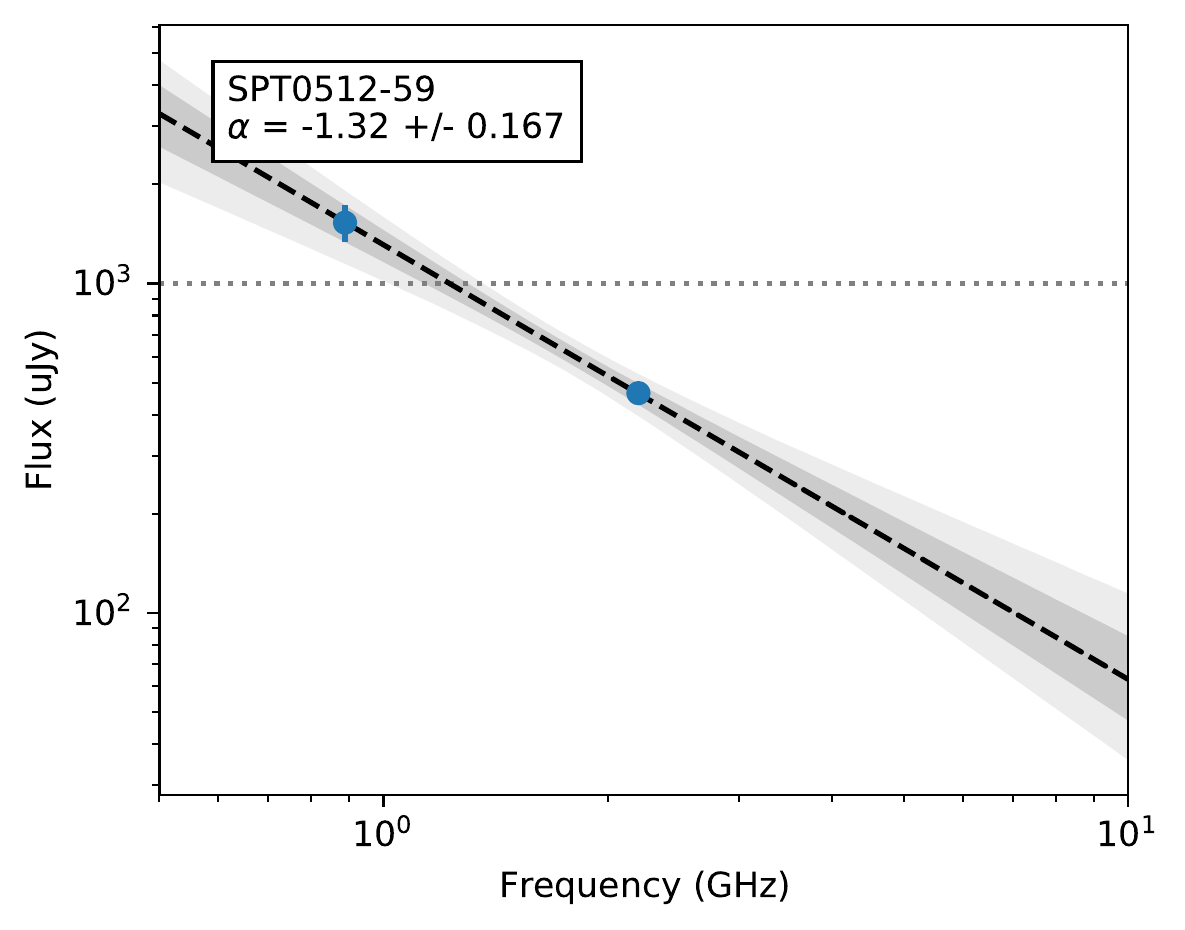}
 \includegraphics[width=0.288\linewidth]{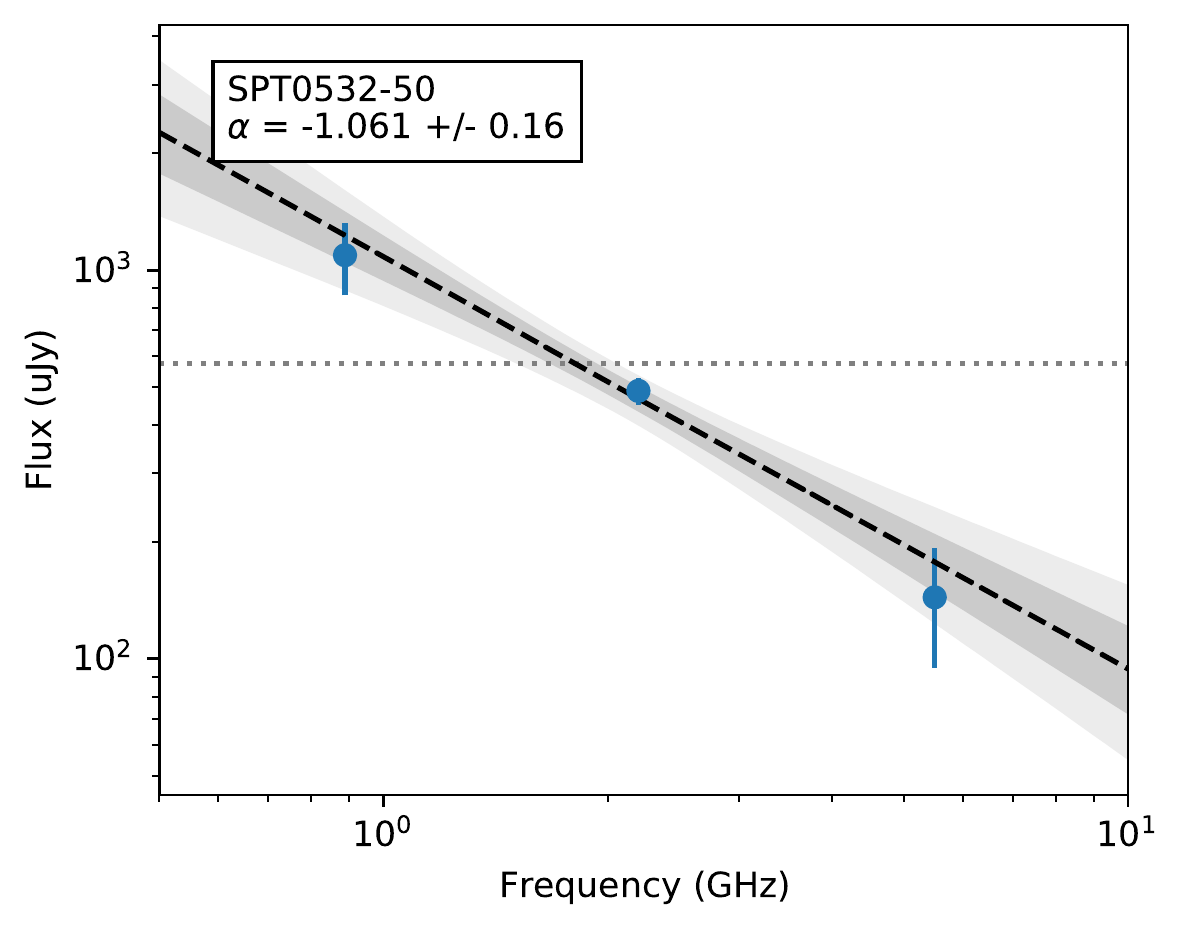}
 \includegraphics[width=0.288\linewidth]{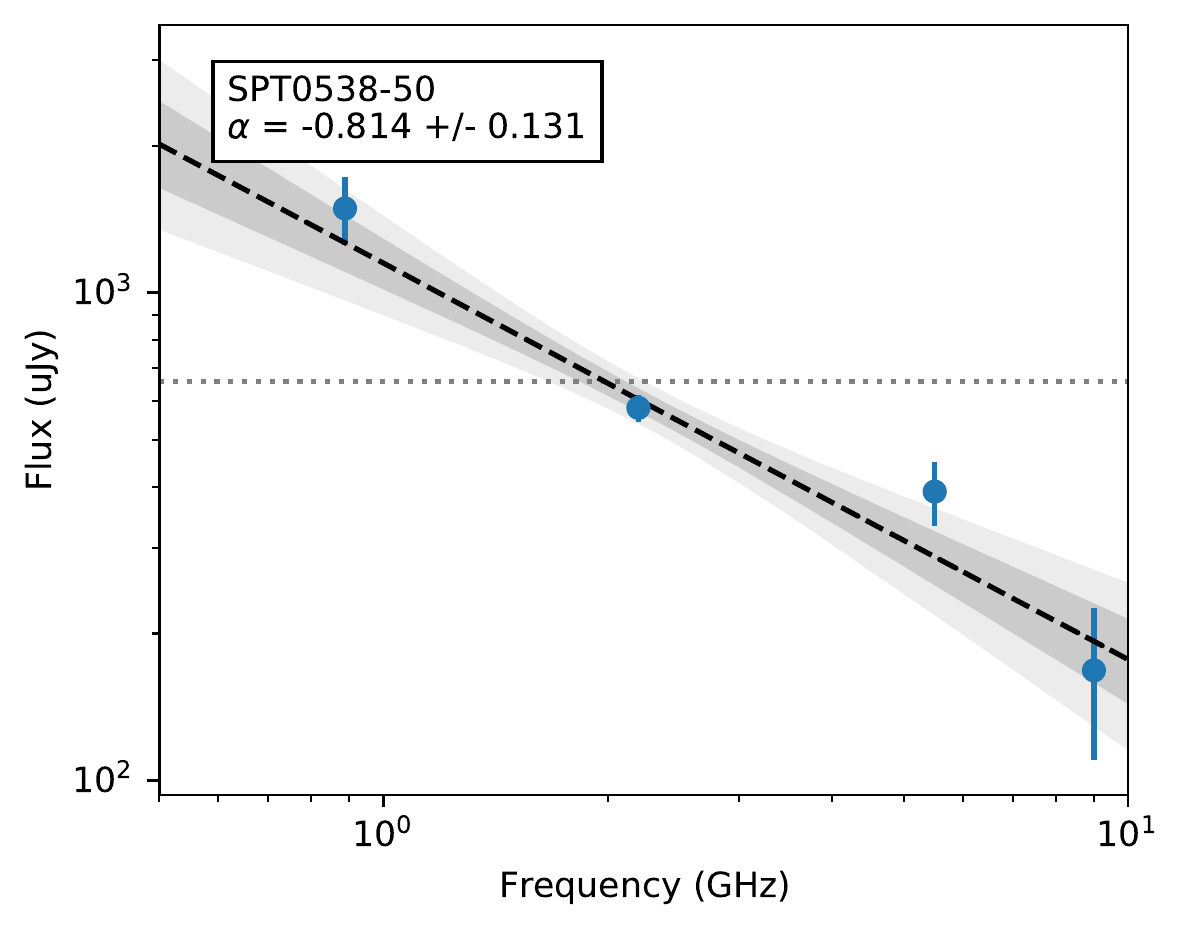}
 \includegraphics[width=0.288\linewidth]{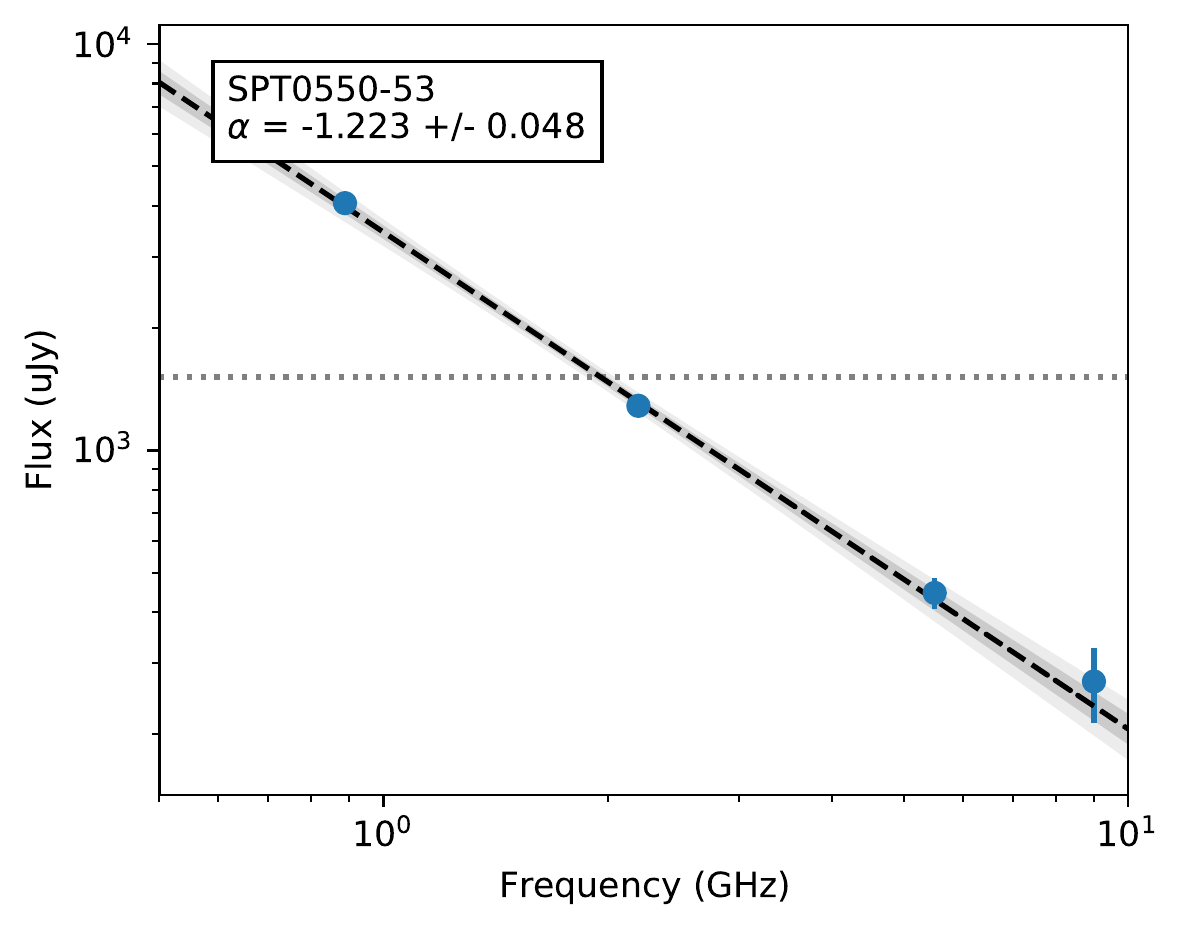}
 \includegraphics[width=0.288\linewidth]{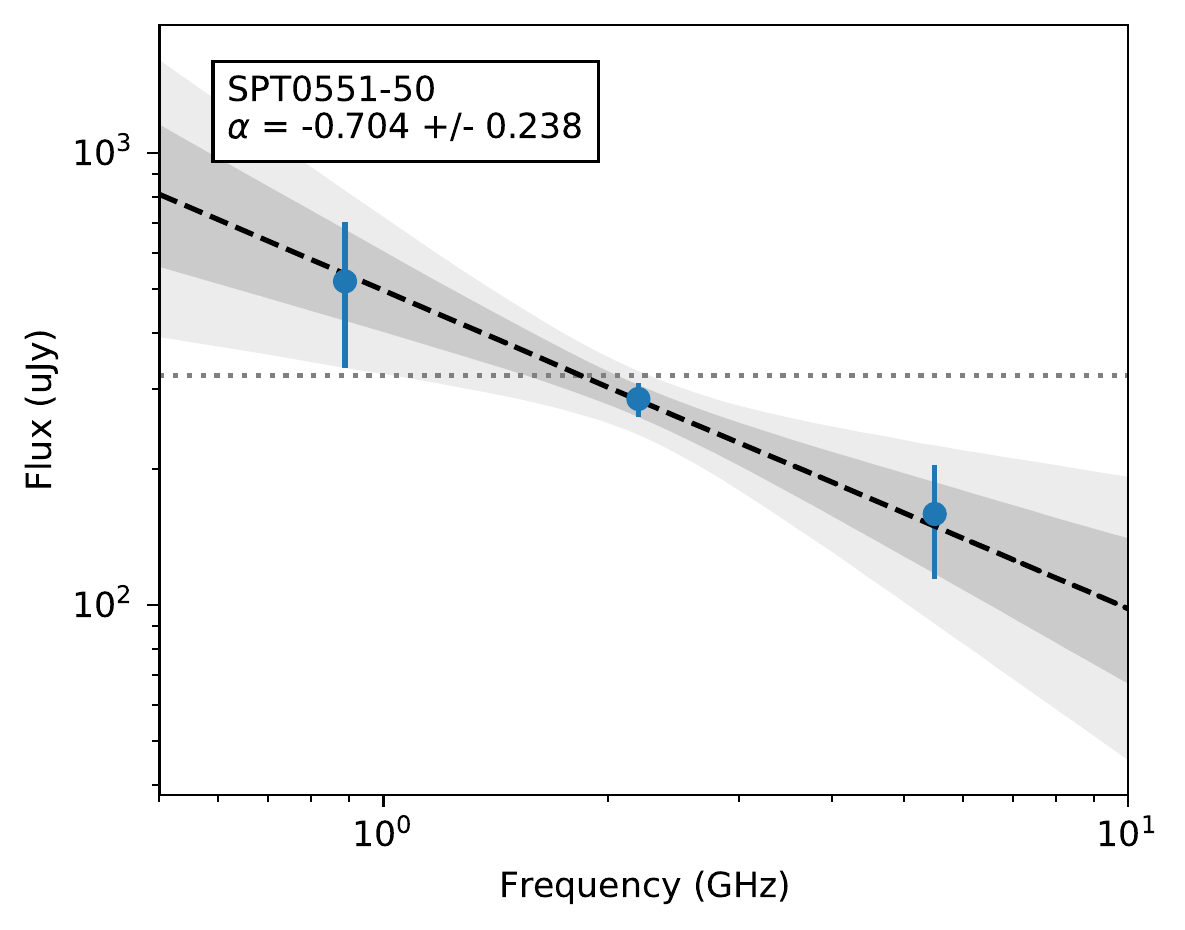}
 \includegraphics[width=0.288\linewidth]{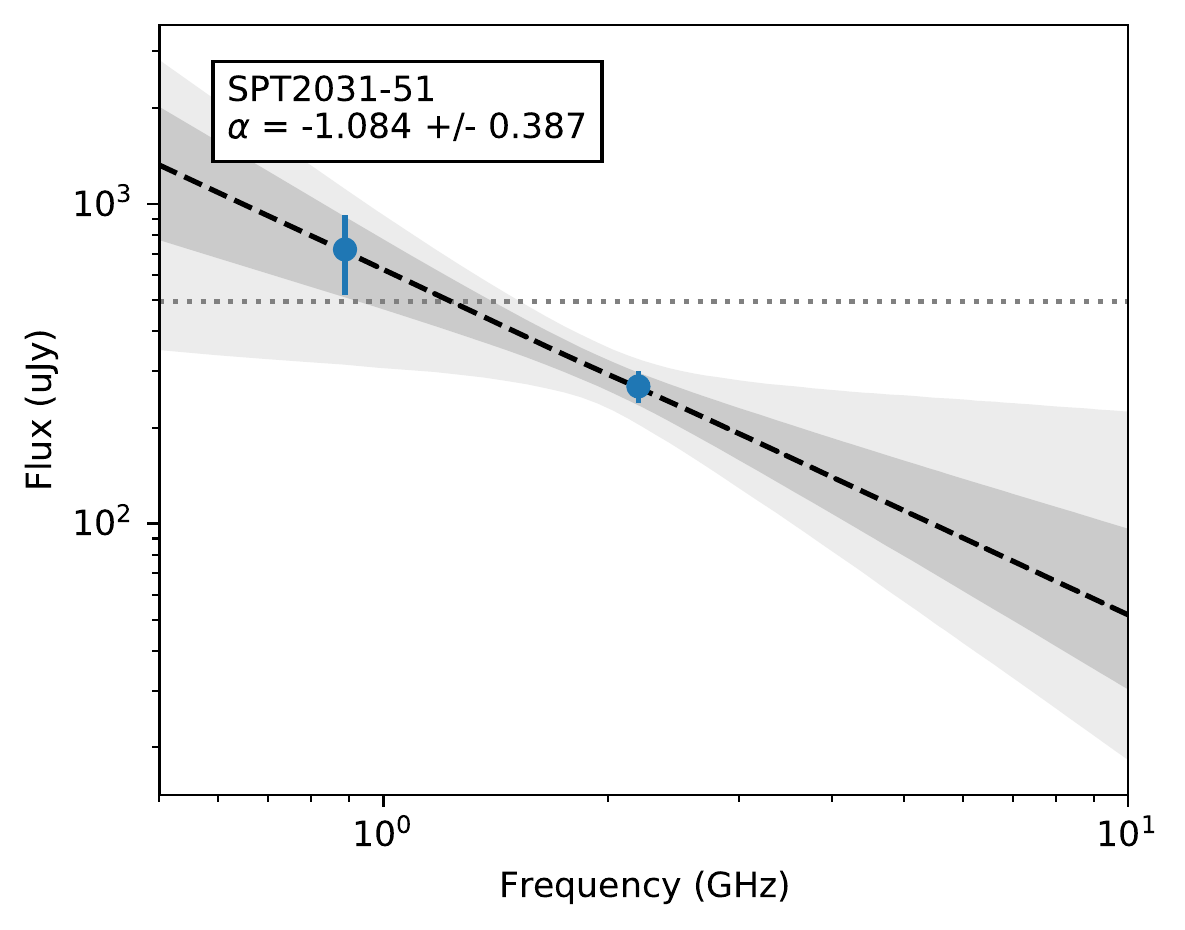}
 \includegraphics[width=0.288\linewidth]{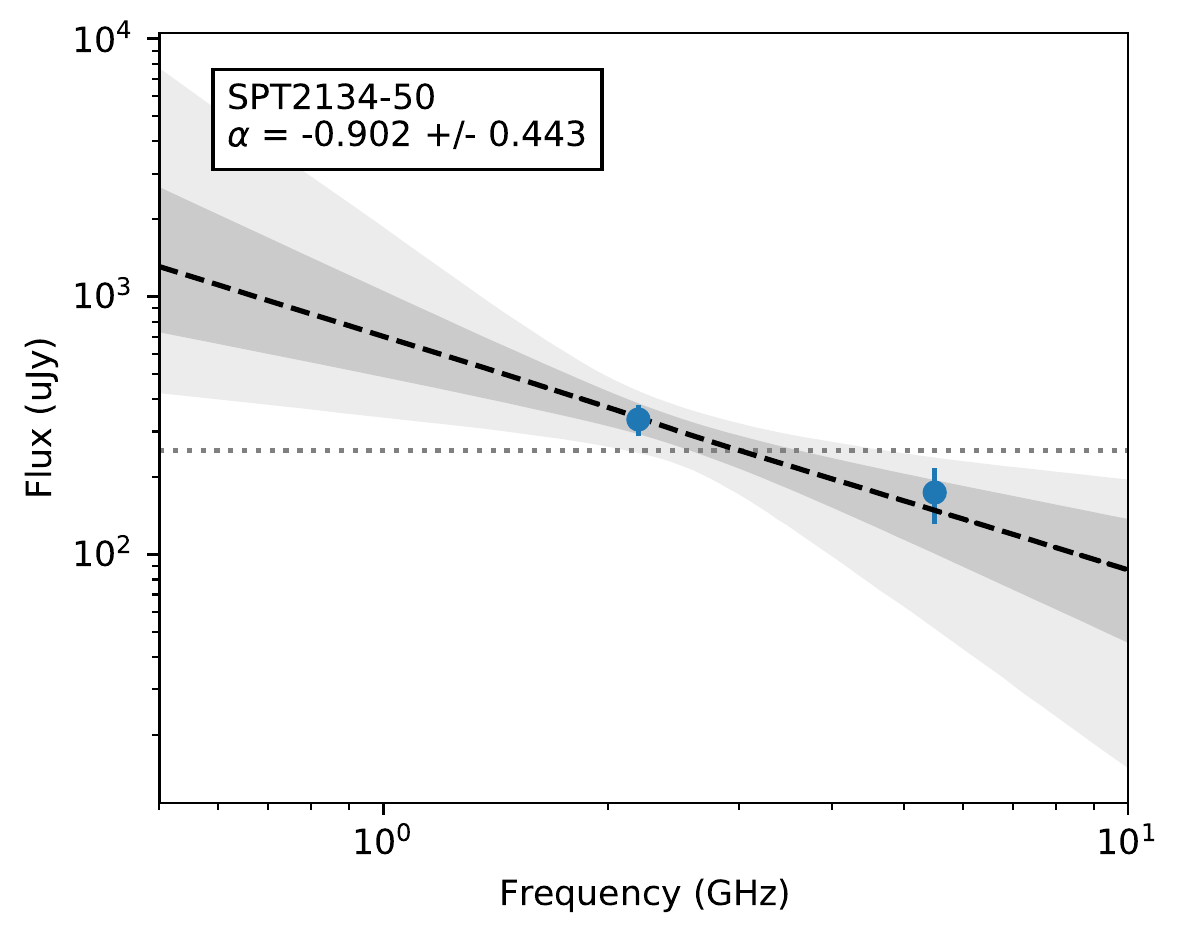}
 \includegraphics[width=0.288\linewidth]{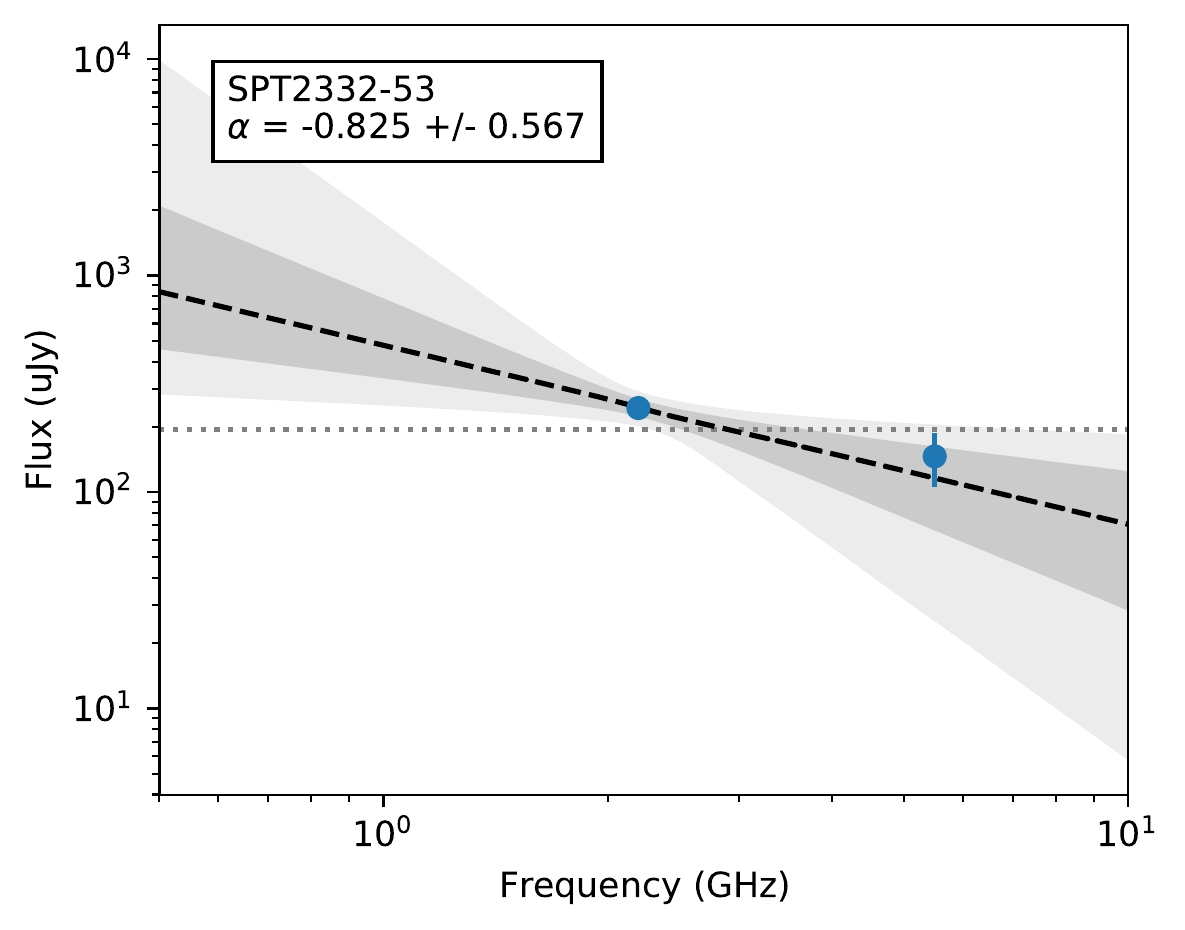}
 \caption{Radio spectral indices with errors constrained by MCMC modelling for gravitatinally lensed SPT-SMGs having at least two detections between the ATCA and ASKAP followup. Grey shaded regions show the 1 and 2$\sigma$ uncertainties on $\alpha$ derived from the ATCA data.
The brighter sources with steeper spectra are generally detected by the ASKAP RACS survey. 
The fit for \spt\ is shown in Figure~\ref{fig:sed}.
}  
    \label{fig:radio_spectra}
\end{figure*}

We derive radio spectral indices directly for all sources with at least two radio detections, and list these together with flux densities in Table~\ref{table:radio_full}. The data was fit according to a linear function using a Markov Chain Monte Carlo algorithm (MCMC) implemented by the \texttt{emcee} package~\citep{foreman2013}.  This MCMC package samples the posterior probability function, and is used to determine the error contours shown in Fig.~\ref{fig:sed}, as well as the uncertainties on $\alpha$ in Table \ref{table:radio_full}.  
 We summarize their radio spectral indices in in Fig.~\ref{fig:radio_spectra} and Table \ref{table:radio_full}. 

The lensed SMGs are shown in Fig.~\ref{fig:distribution}, where we estimate their rest 1.4\,GHz luminosities directly using the measured $\alpha$, or with $\alpha\,{=}\,-0.8$ if only detected at a single radio frequency. In general these SPT-SMGs follow the same FIR-radio correlation as the other field samples shown.
However, three sources are highly significant outliers from the FIR-radio correlation: SPT0125$-$50 at $z\,{=}$3.96; SPT0202$-$61 at $z\,{=}\,$5.02; and SPT0550$-$53 at $z\,{=}\,$3.13. Given how rare such strong outliers are in the field SMG samples (only one of 76 SMGs shows anywhere near this level of radio excess in the ALESS SMG sample -- \citealt{thomson2014}), we propose that the lensing galaxy rather than SMG may be  the more likely radio-AGN in these three cases. 
These radio excess sources exhibit steeper radio indices than typical star-forming galaxies, comparable to or exceeding \spt. Without knowing if the lens or source redshift is correct, we cannot reasonably apply the radio K-correction to estimate the rest 1.4\,GHz luminosity, and therefore we do not include these three in figure~\ref{fig:distribution}.

In particular, SPT0550$-$53 shows an extended radio morphology/jet, well resolved in all three ATCA frequencies, which is more naturally explained by a lower redshift radio-loud galaxy. Further, the optical spectrum of the lens SPT0550$-$53 shows AGN emission lines. 
Neither SPT0125$-$50 nor SPT0202$-$62 show AGN signatures in their optical spectra.
SPT0125$-$50 is curious as the ATCA 2\,GHz flux density is very close to that expected from the FIR-radio relation, however ASKAP reveals a 3\,mJy source well centered on SPT0125$-$50, implying an incredibly steep $\alpha=-2.24$.
Thus any K-correction to lower rest frame frequencies than that probed by 2.2\,GHz observations quickly places SPT0125$-$50 significantly above the FIR-radio correlation. 

These results also beg the question of whether the radio emission in other lensed SPT-SMGs might be contaminated from the often massive lens galaxy \citep{Rotermund20}. 
While in some examples, especially in SPT0538$-$50, the radio emission is directly identified as coming from the ALMA-detected lensed SMG components, in others the Einstein radius of the lensed source \citep{Spilker} is too small to be detected offset from the lens galaxy itself, even at 9\,GHz. 
The distribution in $\alpha$ constrained by the fits in figure~\ref{fig:radio_spectra} show a mean of $-0.93\pm0.14$, offset steeper, but still consistent, with the $\alpha=-0.8$ found in  samples of unlensed SMGs reported in section~\ref{sec:results:3.2} (e.g., \citealt{thomson2014}).
Several of the higher redshift sources in figure~\ref{fig:distribution} do in fact show a marginal excess over that expected from the FIR-radio correlation. This excess sometimes appears only due to the comparison shown at rest 1.4\,GHz, accentuating the K-correction from their steeper than average $\alpha$ we measure. Given the large uncertainties from the often two-point $\alpha$ estimates, this may be inconsequential.
Generally, optically identified AGN are relatively rare in the {\it distant red galaxies} that are often found to lens these SMGs \citep{Rotermund20}, and the gravitationally boosted radio signal associated with the high-SFR SMG is a more probable source of the strong radio emission we see in these 19 SPT-SMGs. Their radio emission is not obviously contaminated by their foreground lens.




\bibliographystyle{aasjournal}
\bibliography{SPT-PC}

\begin{table*}
 \centering{
  \caption{ATCA observations at 2.2, 5.5, and 9.0\,GHz, as well as ASKAP at 888\,MHz, of 23 SPT SMGs. We also include unresolved LABOCA 850\,$\mu$m flux densities ($S_{850}$), ALMA redshifts, and best-fit spectral indices.}
\label{table:radio_full}
\begin{tabular}{lcccccccccccc}
\hline
Name &  $t_{\rm int}$ & $S_{2.2}$  & Beam$^a$  &  PA$^b$ &   $S_{5.5}^\dagger$  &  $S_{9.0}^\dagger$  & $S_{0.9}^{\dagger\dagger}$ & $S_{850}$  &  $z$ & $\alpha^c$ & Comment$^{\dagger\dagger\dagger}$ \\ 
{} &  (hrs) & ($\mu$Jy) & ($^{\prime\prime \times \prime\prime}$) & (deg) &($\mu$Jy) & ($\mu$Jy) & ($\mu$Jy)  & (mJy) & & \\
\hline
SPT\,0027$-$50 & 0.58  &  334$\pm$31  & $8\times4$& 106 &   $<$105 & $<$150 & 1064$\pm$201 & 48   &  3.444 & $-$1.28$\pm$0.25 & n n\\ 
SPT\,0103$-$45 & 0.54 & 229$\pm$29  & $8\times4$& 16   &  $<$111  & $<$159 & $<589$ &  125  &  3.092 & -- & n n\\
SPT\,0109$-$47 &  0.56 & 1106$\pm$31 &  $9\times5$& 11   & 824$\pm$36 & 461$\pm$52 & 1417$\pm$194 & 109  &  3.614 & $-$0.42$\pm$0.04 & yR yR\\
SPT\,0125$-$47 &  0.59 & 586$\pm$35  &  $9\times5$& 112  & 382$\pm$36 & 193$\pm$49 & 1835$\pm$201  &  144  &  2.515& $-$0.82$\pm$0.10  & y mR\\
SPT\,0125$-$50 &  0.56 & 365$\pm$29  &  $5\times7$& 110  &   $<$102 & $<$156 & 2792$\pm$194  & 109  &  3.959& $-$2.25$\pm$0.12 & n n\\ 
SPT\,0202$-$61 & 0.68 & 710$\pm$35  & $9\times5$& 6  & 225$\pm$40 & $<$153 & 1943$\pm$179 & 109  &  5.018 & $-$1.16$\pm$0.09 & y n\\ 
SPT\,0245$-$63 & 0.71 & 94$\pm$33   &  $9\times4$& 77  &   $<$99 & $<$135 & $<598$ & 61   &  5.626 & -- & n n\\ 
SPT\,0345$-$47 &  0.68 & 275$\pm$29  &  $8\times6$& 174  &  $<$99 & $<$132 & 701$\pm$194  & 89   &  4.296 & $-$1.03$\pm$0.38 & n n\\ 
SPT\,0346$-$52 & 0.72 & 162$\pm$38  & $5\times9$ & 68 &  $<$105 & $<$132 & $<603$ & 131  &  5.656 & -- & n n\\ 
SPT\,0418$-$47 &  0.70 & 173$\pm$22  &  $8\times5$& 163  &  $<$93 & $<$135 & 430$\pm$198 & 108  &  4.224 & $-$0.99$\pm$0.67 & n n\\ 
SPT\,0512$-$59 &  0.56 & 465$\pm$34  &  $7\times5$& 153  &   $<$177 & $<$231 & 1531$\pm$197 & 75   &  2.233 & $-$1.32$\pm$0.17 & n n\\ 
SPT\,0529$-$54 &  0.54 & 260$\pm$50  &  $8\times5$& 153  &  $<$120 & $<$180 & $<586$ & 118  &  3.369 & -- & n n \\ 
SPT\,0532$-$50 &  0.57 & 489$\pm$39  &  $8\times5$& 154  & 144$\pm$49 & $<$177 & 1093$\pm$231 & 118  &  3.399& $-$1.06$\pm$0.16 & m n\\ 
SPT\,0538$-$50 &  0.56 & 581$\pm$36  &  $8\times5$& 157  & 341$\pm$59 & 168$\pm$58 & 1490$\pm$237 & 125  &  2.786& $-$0.81$\pm$0.13 & yR mR\\ 
SPT\,0550$-$53 & 0.55 & 1288$\pm$48 & $9\times6$& 169  & 446$\pm$39 & 270$\pm$56 & 4060$\pm$187 & 53   &  3.128 & $-$1.22$\pm$0.05 & yR yR \\ 
SPT\,0551$-$50 &  0.56 & 286$\pm$25  &  $8\times6$& 160  &  159$\pm$45 & $<$171 & 520$\pm$185 & 74& 3.164 & $-$0.70$\pm$0.24 & m n\\ 
SPT\,2031$-$51 & 0.48 & 269$\pm$31  & $9\times5$ & 51 & $<$123 & $<$186 & 721$\pm$203 & 65   &  2.452& $-$1.08$\pm$0.39 & n n\\ 
SPT\,2134$-$50 & 0.53 & 334$\pm$47  & $8\times4$&  33  &  174$\pm$43 & $<$162 & 804$\pm$196 & 101  &  2.780& $-$0.90$\pm$0.44 & y n\\
SPT\,2319$-$55 & 0.54 & 75$\pm$44   & $8\times4$& 47  &   $<$126 & $<$174 & $<600$ & 38   &  5.293& -- & n n\\
SPT\,2332$-$53 & 0.56 & 244$\pm$23  & $9\times5$& 36 &  146$\pm$41 & $<$162 & $<581$ &  57   &  2.756& $-$0.82$\pm$0.56 & m n\\ 
SPT\,2349$-$56 & 0.56 & 215$\pm$27  & $8\times4$  & 30 &  $<$120 & $<$162 & 867$\pm$189 &  106$^{d}$    &  4.303& $-$1.58$\pm$0.31 & n n\\ 
SPT\,2353$-$50 &  0.56  & 24$\pm$53   & $9\times5$ & 30   &  $<$138 & $<$159 & $<594$  & 41   &  5.576& -- & n n\\
SPT\,2357$-$51 &  0.55 & 131$\pm$19  &  $9\times5$ & 31  &    $<$108 & $<$156 & $<589$ & 53   &  3.070& -- & n n\\ 
\hline
\hline
\end{tabular}
}\\
\flushleft{
$^a$ The 2.2\,GHz beam is quoted as x and y FHWM. 
The 5.5\,GHz beam is typically $3.6''\times2.2''$. 
The 9.0\,GHz beam is typically $2.2''\times1.2''$. \\
$^b$ The PA of the 2.2\,GHz beam is the angle East of North. \\ 
$^c$ The radio spectral index $\alpha$, defined as $S_\nu\,{\propto}\,\nu^\alpha$.\\
$^d$ In \spt\ we have assumed source C with $S_{850\,\mu{\rm m}}\,{=}\,$4.7\,mJy is the host of the AGN, although it could be B or G as described in the text; here we still provide the unresolved LABOCA flux density. \\
$^\dagger$ At 5.5 and 9.0\,GHz, the 3$\sigma$ upper limit is listed unless there is a detection at $>3\sigma$ at the ALMA position.\\ 
$^{\dagger\dagger}$ The 888\,MHz measurements are from the ASKAP RACS survey, described in section~2.2. Sources with $<3\sigma$ positive signal are listed at these limits.\\ 
$^{\dagger\dagger\dagger}$ Comments list whether 5.5\,GHz and 9.0\,GHz data show detections $>4\sigma$ (y), marginal detections (m) where $<4\sigma$ flux density is measured at the ALMA position, or no detection (n). We indicate the four sources with resolved radio morphologies (R), in SPT0538$-$50 clearly following the ALMA emission, although in SPT0109$-$47 and especially SPT0550$-$53, the resolved emission appears to be an extended lobe or jet.
}
\end{table*}

\end{document}